\documentclass[reprint,
    superscriptaddress,
    nolongbibliography,
nofootinbib,
    amsmath, amssymb,
    aps, prd,
]{revtex4-2}

\usepackage{enumitem}              \interfootnotelinepenalty = 10000  

\usepackage[dvipsnames]{xcolor}
\definecolor{dodgerblue}{HTML}{1E90FF}
\definecolor{viennared}{HTML}{DA0A14}
\definecolor{ctorange}{HTML}{FF6C0C}
\definecolor{granadagreen}{HTML}{078931}
\definecolor{wales}{HTML}{ff0038}
\definecolor{valenciacfred}{HTML}{ee3524}
\definecolor{barcelonafcgold}{HTML}{edbb00}
\definecolor{jam}{HTML}{A50B5E}
\definecolor{austriawien}{HTML}{441678}

\usepackage{graphicx}
\usepackage[caption=false]{subfig}  \usepackage{placeins}               

\usepackage{tabularray}   \UseTblrLibrary{booktabs} \UseTblrLibrary{siunitx}  

\usepackage{gensymb, upgreek, bm}
\usepackage{commath, tensor, braket}
\usepackage{calc}

\usepackage{mathtools, accents, cancel}
\usepackage{siunitx, xfrac}

\usepackage{stackengine, scalerel, xargs}
\newcommandx\mysqrt[3][1=\phantom{2}, 2=0pt]{\stretchrel{\sqrt[#1]{}}{\addstackgap[#2]{$\displaystyle\overline{#3}$}}}

\usepackage{hyperref}
\usepackage{aas_macros}
\hypersetup{
    colorlinks = true,
    citecolor = dodgerblue,
    linkcolor = dodgerblue,
    urlcolor = dodgerblue,
    linkbordercolor = white
}

\usepackage{orcidlink} \usepackage[capitalize]{cleveref}  

\crefname{section}{Sec.}{Secs.}
\Crefname{section}{Section}{Sections}

\usepackage{tikz}
\tikzset{baseline,
    inner sep=2pt,
    minimum height=12pt,
    rounded corners=2pt  
}

\usepackage{lipsum}          \usepackage[normalem]{ulem}  \usepackage{capt-of}         \usepackage{xspace}          

\usepackage{todonotes}

\usepackage{soul}

\soulregister\cref7
\soulregister\ref7
\soulregister\cite7
\soulregister\alias7

\definecolor{BerlinU1}{HTML}{62AD2D}

\newcommand{\UIB}{\affiliation{Departament de F\'isica, Universitat de les Illes Balears, IAC3 -- IEEC, Crta. Valldemossa km 7.5, E-07122 Palma, Spain}}
\newcommand{\AEI}{\affiliation{Max Planck Institute for Gravitational Physics (Albert Einstein Institute), Am M\"uhlenberg 1, Potsdam 14476, Germany}}
\newcommand{\UP}{\affiliation{Institut f\"{u}r Physik und Astronomie, Universit\"{a}t Potsdam, Haus 28, Karl-Liebknecht-Str. 24/25, 14476, Potsdam, Germany}}
\newcommand{\UoC}{\affiliation{School of Physical \& Chemical Sciences, University of Canterbury, Private Bag 4800, Christchurch 8140, New Zealand}}
\newcommand{\LPC}{\affiliation{CNRS/IN2P3, Laboratoire de Physique Corpusculaire (LPC) UMR6534, 14050 Caen, France}}

\newcommand{\MBH}{M_\mathrm{BH}}
\newcommand{\MNS}{M_\mathrm{NS}}
\newcommand{\chiBH}{\chi_{\mathrm{BH}}}
\newcommand{\chiBHz}{\chi_{\mathrm{BH}z}}
\newcommand{\chiBHvec}{\vec{\chi}_\mathrm{BH}}
\newcommand{\chiNS}{\chi_{\mathrm{NS}}}
\newcommand{\chiNSz}{\chi_{\mathrm{NS}z}}
\newcommand{\chiNSvec}{\vec{\chi}_\mathrm{NS}}
\newcommand{\lambdaNS}{\Lambda} \newcommand{\Msun}{\,\mathrm{M_\odot}}
\newcommand{\chieff}{\chi_\mathrm{eff}}
\newcommand{\chip}{\chi_\mathrm{p}}

\newcommand{\modelname}[1]{\textsc{#1}\xspace} \newcommand{\Undr}{\rule{1ex}{0.5pt}} \newcommand{\undr}{\penalty0\hskip0pt\rule{1ex}{0.5pt}\hskip0pt\penalty0} 

\usepackage{expl3}
\ExplSyntaxOn
\prop_new:N \g_model_full_prop
\prop_new:N \g_model_alias_prop

\prop_put:Nnn \g_model_full_prop {ph}{Phenom}
\prop_put:Nnn \g_model_full_prop {phD}{IMRPhenomD}
\prop_put:Nnn \g_model_full_prop {phC}{IMRPhenomC}
\prop_put:Nnn \g_model_full_prop {phX}{IMRPhenomX}
\prop_put:Nnn \g_model_full_prop {XAS}{IMRPhenomXAS}
\prop_put:Nnn \g_model_full_prop {XHM}{IMRPhenomXHM}
\prop_put:Nnn \g_model_full_prop {phNSBH}{IMRPhenomNSBH}
\prop_put:Nnn \g_model_full_prop {XHM_NRT2}{IMRPhenomXHM{\undr}NRTidalv2}
\prop_put:Nnn \g_model_full_prop {XHM_NRT3}{IMRPhenomXHM{\undr}NRTidalv3}
\prop_put:Nnn \g_model_full_prop {XNSBH}{IMRPhenomXHM{\undr}NSBH}
\prop_put:Nnn \g_model_full_prop {XP}{IMRPhenomXP}
\prop_put:Nnn \g_model_full_prop {XPHM}{IMRPhenomXPHM}
\prop_put:Nnn \g_model_full_prop {XPNSBH}{IMRPhenomXPHM{\undr}NSBH}

\prop_put:Nnn \g_model_full_prop {v4ROM}{SEOBNRv4{\undr}ROM}
\prop_put:Nnn \g_model_full_prop {v5ROM}{SEOBNRv5{\undr}ROM}
\prop_put:Nnn \g_model_full_prop {v4ROM_NRT2}{SEOBNRv4{\undr}ROM{\undr}NRTidalv2}
\prop_put:Nnn \g_model_full_prop {v4ROM_NSBH}{SEOBNRv4{\undr}ROM{\undr}NRTidalv2{\undr}NSBH}
\prop_put:Nnn \g_model_full_prop {v5ROM_NRT2}{SEOBNRv5{\undr}ROM{\undr}NRTidalv2}
\prop_put:Nnn \g_model_full_prop {v5ROM_NRT3}{SEOBNRv5{\undr}ROM{\undr}NRTidalv3}
\prop_put:Nnn \g_model_full_prop {v5HMROM}{SEOBNRv5HM{\undr}ROM}
\prop_put:Nnn \g_model_full_prop {v5HMROM_NRT2}{SEOBNRv5HM{\undr}ROM{\undr}NRTidalv2}
\prop_put:Nnn \g_model_full_prop {v5HMROM_NRT3}{SEOBNRv5HM{\undr}ROM{\undr}NRTidalv3}
\prop_put:Nnn \g_model_full_prop {v5HMROM_NSBH}{SEOBNRv5HM{\undr}ROM{\undr}NRTidalv3{\undr}NSBH}

\prop_put:Nnn \g_model_full_prop {TEOB}{TEOBResumS}
\prop_put:Nnn \g_model_full_prop {DALI}{TEOBResumS-Dalí}
\prop_put:Nnn \g_model_full_prop {GIOTTO}{TEOBResumS-GIOTTO}

\prop_put:Nnn \g_model_full_prop {NRT}{NRTidal}
\prop_put:Nnn \g_model_full_prop {NRT2}{NRTidalv2}
\prop_put:Nnn \g_model_full_prop {NRT3}{NRTidalv3}
\prop_put:Nnn \g_model_full_prop {NRSurTidal}{NRHybSur3dq8Tidal}

\prop_put:Nnn \g_model_alias_prop {phNSBH}{PhenomNSBH}

\prop_put:Nnn \g_model_alias_prop {XHM}{XHM}
\prop_put:Nnn \g_model_alias_prop {XPHM}{XPHM}
\prop_put:Nnn \g_model_alias_prop {XNSBH}{XHM{\undr}NSBH}
\prop_put:Nnn \g_model_alias_prop {XPNSBH}{XPHM{\undr}NSBH}

\prop_put:Nnn \g_model_alias_prop {v4ROM_NSBH}{SEOBv4{\undr}NSBH}
\prop_put:Nnn \g_model_alias_prop {v5HMROM}{SEOBHMv5}
\prop_put:Nnn \g_model_alias_prop {v5HMROM_NSBH}{SEOBHMv5{\undr}NSBH}

\prop_put:Nnn \g_model_alias_prop {DALI}{DALI}
\prop_put:Nnn \g_model_alias_prop {GIOTTO}{GIOTTO}

\newcommand{\model}[1]{
    \modelname{\prop_item:Nn \g_model_full_prop {#1}}
}

\newcommand{\alias}[1]{
  \prop_if_in:NnTF \g_model_alias_prop {#1}
    {\modelname{\prop_item:Nn \g_model_alias_prop {#1}}}
    {\model{#1}}
}
\ExplSyntaxOff

\newcommand{\lm}{{\ell m}}
\newcommand{\Mf}{\mathit{Mf}}
\newcommand{\fring}[1][]{\Mf_{\!\mathrm{ring}}^{#1}}
\newcommand{\fdamp}[1][]{\Mf_{\!\mathrm{damp}}^{#1}}
\newcommand{\fmeco}[1][]{\Mf_{\!\mathrm{MECO}}^{#1}}
\newcommand{\ftide}[1][]{\Mf_{\!\mathrm{tide}}^{#1}}
\newcommand{\fzero}[1][]{\Mf_{\!0}^{#1}}
\newcommand{\ftransI}[1][]{\Mf_{\!\mathrm{transI}}^{#1}}
\newcommand{\ftransM}[1][]{\Mf_{\!\mathrm{transM}}^{#1}}
\newcommand{\fcutRD}[1][]{\Mf_{\!\mathrm{cutRD}}^{#1}}
\newcommand{\fmrd}[2][]{\Mf_{\!#2\mathrm{MRD}}^{#1}}

\newcommand{\ABBH}[1][]{A_\mathrm{BBH}^{#1}}
\newcommand{\ANSBH}[1][]{A_\mathrm{NSBH}^{#1}}
\newcommand{\Ains}{A_\mathrm{INS}}
\newcommand{\Amrd}[2][]{A_{#2\mathrm{MRD}}^{#1}}
\newcommand{\Ymrd}[2][]{Y_{#2\mathrm{MRD}}^{#1}}
\newcommand{\Yfit}[2][]{Y_{#2\mathrm{FIT}}^{#1}}

\begin{document}

\title{Fast gravitational waveform models for quasi-circular coalescences of neutron star--black hole binaries}

\author{Felip A. Ramis Vidal}
\UIB
\email{f.ramis@uib.cat}

\author{Adrian Abac}
\AEI \UP \LPC

\author{Marta Colleoni}
\UIB

\author{Tim Dietrich}
\UP \AEI

\author{Pierre Mourier}
\UIB \UoC

\author{Alejandra Gonzalez}
\UIB

\author{Ivan Markin}
\UP

\author{Anna Puecher}
\UP

\date{\today}

\begin{abstract}
    We present \model{XNSBH} and \model{v5HMROM_NSBH}, the first two frequency-domain models for gravitational-wave signals from quasi-circular, aligned-spin neutron star--black hole (NSBH) binaries including higher-order modes beyond the dominant quadrupole. We also present \model{XPNSBH}, an extension of the former model to the spin-precessing case. These models incorporate tidal effects in the gravitational-wave phasing and amplitude using a higher-mode extension of the \model{NRT3} model, as well as dedicated amplitude models calibrated to numerical relativity (NR) simulations of NSBH mergers. We test the performance and validity of the new models by comparing them to NR simulations and other existing models for these systems. Finally, we perform parameter estimation studies. The new models show clear improvements over their predecessors in analyses of simulated signals, while yielding results consistent with the literature when applied to real events from the GWTC-3 and GWTC-4 catalogs.
\end{abstract}

\maketitle

\section{Introduction}\label{sec:intro}
In 2020, the LIGO--Virgo--KAGRA (LVK) collaboration announced the first observations of neutron star--black hole (NSBH) binaries, GW200105\_162426 and GW200115\_042309~\cite{LIGOScientific:2021qlt}. The associated gravitational-wave (GW) signals were detected by the Advanced LIGO and Advanced Virgo detectors~\cite{LIGOScientific:2014pky,VIRGO:2014yos}, with KAGRA joining the network during the later phase of the same observing run~\cite{KAGRA:2018plz}. Though no signatures of tidal disruption could be detected for either of them, the secondary masses of both events were found to lie below the maximal mass of a neutron star (NS) with high probability, after accounting for uncertainties in the astrophysical mass priors and equation of state (EOS)~\cite{Essick:2020ghc}. Hence, these signals were recognized as the first confident detections of coalescing NSBHs through GWs.

While still much rarer than binary black-hole (BBH) signals, observations of NSBH mergers are on the rise and have outnumbered the detections of binary neutron star (BNS) mergers. Several other candidates detected during the third and fourth observing runs have been regarded as potential NSBHs, though with large uncertainties due to either their marginal significance~\cite{LIGOScientific:2021usb, Pillas:2025pfc} or heavier secondary object, e.g., GW190814~\cite{LIGOScientific:2020zkf}. The latter event has been interpreted as either the heaviest NS observed in a binary (possibly fast spinning~\cite{Most:2020bba}, or containing exotic matter~\cite{Dexheimer:2020rlp}) or, more likely, containing a light black hole (BH)~\cite{Tews:2020ylw}. 

More recently, during the fourth observing run O4a~\cite{LIGOScientific:2025slb}, two additional significant NSBH candidates have been reported: GW230518\_125908, detected in the engineering run preceding O4a, representing the loudest NSBH candidate to date with a network signal-to-noise ratio (SNR) of ${\,\sim\,}14$, and GW230529\_181500 (hereafter GW230529)~\cite{LIGOScientific:2024elc}. The latter emerged as a particularly interesting NSBH candidate, as its primary mass lies in the so-called lower mass gap, i.e., the interval between approximately 2.5 and 5 solar masses, expected to be scarcely populated based on observations of galactic low-mass X-ray binaries~\cite{2010ApJ...725.1918O}, radio pulsars~\cite{Ye:2024wqj}, and some supernovae core-collapse models~\cite{2012ApJ...749...91F, 2012ApJ...757...91B}. In addition, thanks to the low mass of the primary, the system was inferred to have a relatively small mass ratio of $Q \simeq 2.5$, making it the most comparable-mass NSBH candidates reported so far. The detection of GW230529 adds to other recent observations of mass-gap objects~\cite{2019Sci...366..637T, 2021MNRAS.504.2577J, Barr:2024wwl} and could have deep implications for our understanding of the supernova engine~\cite{Boccioli:2024kvw, Fryer:2022lla, Olejak:2022zee} and stellar binary evolution~\cite{Mandel:2020cig,Xing:2024ydg}. Although most dynamical and isolated binary formation models predict predominantly unequal-mass NSBH systems~\cite{Sedda:2020wzl, Chattopadhyay:2020lff}, uncertainties in core-collapse supernova physics, including convection growth, explosion asymmetries, and fallback mechanisms, leave room for the formation of more comparable-mass systems~\cite{Olejak:2022zee,Vigna-Gomez:2021oqy}. It was found that including GW230529 in the population inference modestly increases the 90\% credible upper limit on the fraction of electro-magnetic bright NSBH mergers that can be detected through GWs~\cite{LIGOScientific:2024elc}. Hence, this event could provide a glimpse of a population of comparable-mass (and potentially bright) NSBH systems, though conclusions remain tentative given the current number of observations.

NSBHs are not necessarily associated with tidal disruption, which only occurs when tidal forces overcome the self-gravity of the NS during the inspiral. When this condition is not met, the NS is swallowed by the BH, and the resulting GW signal closely resembles that of a BBH merger, making it hard to distinguish between the two types of events. When disruption occurs, on the other hand, the merger-ringdown GW signal is sharply suppressed, due to the incoherent excitation of the remnant's quasinormal modes by infalling matter~\cite{1982ApJ...260..838S,Lackey:2013axa,Topolski:2024jva,Steppohn:2025kbh}. 

Tidal disruption is favoured by comparable mass ratios, high prograde BH spins aligned with the binary's orbital angular momentum, and large NS tidal deformabilities~\cite{Foucart:2012nc,Kyutoku:2015gda}. In disruptive mergers, radioactive decays in the ejecta can drive an observable kilonova~\cite{Fernandez:2016sbf}; furthermore, a highly magnetized accretion disk can form around the remnant BH and power an ultrarelativistic jet, potentially prompting a gamma-ray burst~\cite{Paschalidis:2014qra,Hayashi:2022cdq,Gottlieb:2023est}. 

While non-disruptive NSBH mergers are generally not expected to produce significant EM counterparts due to the absence of substantial matter ejection, several studies have explored scenarios leading to detectable EM signals even in such cases. Examples include NS crust shattering due to tidal deformation~\cite{Tsang:2012PhysRevLett.108.011102}, reconnection-driven emission arising from the interaction between the BH and NS magnetosphere~\cite{East:2021spd}, or the so-called black-hole battery mechanism~\cite{McWilliams:2011zi}, which could be amplified if the BH accumulates a significant electric charge while spinning in the magnetic field of its companion~\cite{PhysRevD.10.1680,Levin:2018mzg}. 

The growing number of observations of NSBH mergers, together with their potential role as multimessenger sources, call for drastic improvements to the GW models employed to study their properties. Several existing inspiral-merger-ringdown (IMR) GW models for NSBH coalescences, such as \model{phNSBH}~\cite{Thompson:2020nei} and \model{v4ROM_NSBH}~\cite{Matas:2020wab}, only capture the dominant quadrupolar mode of GW radiation, and are restricted to aligned-spin binaries. While potential biases due to the neglect of higher-order harmonic modes (HMs) and precession have been explored through the use of BBH templates accounting for these effects~\cite{LIGOScientific:2021qlt,Huang:2020pba}, models with a complete physics content are highly desirable to deliver fully self-consistent analyses. The \model{NRSurTidal} model~\cite{Barkett:2019tus} incorporates post-Newtonian (PN) tidal corrections into several harmonics of a spin-aligned BBH surrogate model, but its validity is limited to the inspiral regime, since these additional terms diverge close to the merger. More recently, the time-domain (TD) model \model{GIOTTO}~\cite{Gonzalez:2022prs} provided the IMR waveforms for NSBHs, including HMs and precession. This was subsequently improved by \model{DALI}~\cite{Gonzalez:2025xba}, which individually fits tidal corrections to higher harmonics against NR simulations. The computational cost of these TD models, however, is significantly higher than that of frequency-domain (FD) models, which are ideally suited for low-latency applications and can be efficiently combined with a number of techniques to accelerate parameter estimation (PE), such as likelihood multibanding~\cite{Morisaki:2021ngj} and heterodyning~\cite{Zackay:2018qdy}.

In this paper, we present \model{XNSBH} and \model{v5HMROM_NSBH}, two new IMR GW models for quasi-circular, spin-aligned NSBH binaries that combine accuracy and computational efficiency in an unprecedented way. The new approximants build upon two consolidated families of GW templates for compact binary coalescences, namely spinning effective-one-body (SEOB)~\cite{Buonanno:1998gg,Pompili:2023tna} and phenomenological (Phenom) models~\cite{Khan:2015jqa,Pratten:2020fqn,Garcia-Quiros:2020qpx,Garcia-Quiros:2020qlt}. To maximize accuracy, tidal information is encoded in both the phasing and the amplitude of several GW harmonics beyond the $(2,2)$ mode. The phase leverages the closed-form representation provided by the \model{NRT3} model~\cite{Abac:2023ujg} for the inspiral-merger signal, as well as fits for the remnant properties of NSBH mergers to approximate the quasinormal ringing of the remnant BH~\cite{Gonzalez:2022prs}. As for the GW strain amplitudes, we propose a new fitting strategy augmenting PN information~\cite{Dones:2024odv} with direct tuning to NR simulations. Computational efficiency is achieved through optimized FD representations of the GW signal: for our new SEOB model, we build upon the reduced-order model \model{v5HMROM}~\cite{Pompili:2023tna}, while our phenomenological model is based on \model{XHM}~\cite{Garcia-Quiros:2020qpx}.

In addition to the two new spin-aligned models introduced above, we present \model{XPNSBH}, an extension of \model{XNSBH} to the spin-precessing case. This model is built upon the \model{XPHM}~\cite{Pratten:2020ceb} BBH baseline, in which precessing waveforms are constructed by ``twisting up'' an aligned-spin approximation of the signal in a frame that is co-precessing with the binary's orbital angular momentum.

The paper is structured as follows. In \cref{sec:nr_dataset}, we will describe the dataset of numerical-relativity (NR) simulations employed for the calibration of the GW amplitudes. In \cref{sec:model}, we will summarize our amplitude and phasing models and explain our fitting procedure. After clarifying these technical aspects, we will provide details about the performance of the models in \cref{sec:validation}, including TD comparisons against NR waveforms (\cref{subsec:td_comparisons}), matches (\cref{subsec:mismatches}), and benchmarks (\cref{subsec:benchmarks}). \Cref{sec:pe} will present some PE studies of real and simulated GW signals, followed by our conclusions in \cref{sec:conclusions}.

\paragraph*{Notation.} Throughout this work, we use geometrized units $G = c = 1$ unless otherwise stated.
We denote the BH mass by $\MBH$, the NS mass by $\MNS$, and the total mass by $M = \MBH + \MNS$. The mass ratio is defined as $Q = \MBH / \MNS$, the inverse mass ratio as $q = 1/Q$, the symmetric mass ratio as $\eta = Q / (1 + Q)^2$, and the chirp mass as $\mathcal{M} = \eta^{3/5}M$.
The dimensionless tidal deformability of the NS is denoted by
\[\lambdaNS = \dfrac{2}{3}\dfrac{k_2}{C_\mathrm{NS}^5}\,,\]
where $k_2$ is the quadrupolar gravito-electric Love number and $C_\mathrm{NS} = \MNS/R_\mathrm{NS}$ is the compactness of the NS.
The dimensionless spins of the objects are denoted by $\vec{\chi}_i$
and their magnitudes by $\chi_i = \norm[0]{\vec{\chi}_i}$. Decomposing these spins with respect to the direction $\hat{L}$ of the orbital angular momentum, we define the magnitudes of the aligned (parallel) components as $\chi_{iz} = \vec{\chi}_i \cdot \hat{L}$, and the magnitudes of the in-plane (perpendicular) contributions as $\chi_i^\perp = \norm[0]{\vec{\chi}_i \times \hat{L}}$. From these, the effective spin parameter is defined as $\chieff = (\MBH \chiBHz + \MNS \chiNSz) / M$, and the effective precession parameter as
\[\chip = \max\!{\cbr[3]{\chi_\mathrm{BH}^\perp,\: \frac{A_\mathrm{NS} M_\mathrm{NS}^2}{A_\mathrm{BH} M_\mathrm{BH}^2} \chi_\mathrm{NS}^\perp}}\,,\]
where the coefficients are given by $A_\mathrm{BH} = 2 + \frac{3}{2Q}$ and $A_\mathrm{NS} = 2 + \frac{3Q}{2}$, assuming $M_\mathrm{BH} > M_\mathrm{NS}$.

\section{Calibration Data}\label{sec:nr_dataset}
The NR dataset employed in the calibration of the amplitude models constructed in this work comprises simulations performed with three different codes: SpEC~\cite{SpEC}, SACRA~\cite{Yamamoto:2008js}, and BAM~\cite{Bruegmann:2006ulg,Thierfelder:2011yi}. SpEC employs a multidomain pseudospectral method to evolve the spacetime metric in the generalized harmonic gauge~\cite{Lindblom:2005qh}, and excision to remove the BH interior from the computational domain. The hydrodynamic evolution is instead handled via high-resolution shock-capturing methods on a separate finite-difference grid, with metric and fluid variables interpolated to and from the two grids at each time step in the evolution~\cite{Foucart:2020xkt}. On the other hand, SACRA and BAM employ the moving puncture gauge~\cite{Campanelli:2005dd} and finite differences for the spacetime evolution and finite volumes for matter evolution. 

The NR waveforms publicly released by the SXS collaboration~\cite{SXS:catalog,Foucart:2018lhe,Foucart:2020xkt} only comprise 10 waveforms available in 3 different resolutions, with a number of simulated orbits spanning from approximately 10 to 16. In these simulations, the NS is modelled considering an ideal gas EOS with $\Gamma=2$, except for SXS:BHNS:0003, which uses a piecewise EOS with segments fitted to the H1 EOS. Two of these configurations, SXS:BHNS:0008 and SXS:BHNS:0009, feature a highly spinning BH ($\chiBHz = 0.9$). Additionally, two of these configurations, SXS:BHNS:0005 and SXS:BHNS:0007, include a spinning NS with $\chiNSz=-0.2$, and one, SXS:BHNS:0010, features spin-precession~\cite{Foucart:2020xkt}. These last three simulations are the only public simulations including spin on the NS or BH spin components misaligned with the orbital angular momentum. Given this scarcity, it is not feasible to calibrate our model to these parameters, and we therefore exclude these configurations from our calibration dataset and use them only for validation.

We also employ 162 simulations produced with the SACRA code~\cite{Kyutoku:2010zd,Kyutoku:2011vz,Lackey:2013axa,Kyutoku:2013wxa,Kyutoku:2015gda}, which have already been exploited in the calibration of several GW models for NSBHs~\cite{Pannarale:2015jka,Thompson:2020nei,Matas:2020wab,Gonzalez:2022prs,Gonzalez:2025xba}. The configurations spanned by this dataset include several values of mass ratio $Q \in \{2,3,4,5\}$ and BH spins $\chiBHz \in \{-0.5, 0, 0.25, 0.5, 0.75\}$, as well as several piecewise polytropic EOS for the NS with two or four segments, allowing to probe both disruptive and non-disruptive mergers. These simulations are relatively short and do not consistently include multiple resolutions; furthermore, only the $(2,2)$ mode is available.

Additionally, we consider a set of 51 NSBH simulations produced with the code BAM~\cite{Gonzalez:2025xba}. This dataset comprises three different piecewise polytropic EOS for the NS, with a variety of configurations close to the tidal disruption regime. This is the largest dataset available, including subdominant modes up to $\ell=4$. Most of these simulations are rather short, with a typical length of 3--4 orbits, and are not eccentricity-reduced, with an upper bound on residual eccentricity estimated around $e \sim 0.02$, i.e., 1 or 2 orders of magnitude higher than the reference eccentricity achieved by SpEC waveforms. For some configurations, we find evidence of mode mixing of the $(2,2)$ mode into the subdominant harmonics on the orbital timescale, likely due to a non-negligible centre-of-mass drift\footnote{Simulations from this dataset are continuously being corrected against these effects and are available at the CoRe database~\cite{Gonzalez:2022mgo}.}; hence, we exclude them from the calibration dataset and retain only a subset of 25 simulations. In order to use these short simulations for calibration, we first hybridize them with \model{DALI}~\cite{Gonzalez:2025xba} as detailed in the subsection below. 

Lastly, we employ an additional set of $12$ BAM simulations from a recent study of equal- and near-equal-mass NSBH mergers~\cite{Markin:2026eyc}. These simulations are of systems with mass ratios $Q \in \{1, 2, 3\}$, non-spinning components, and two different EOSs describing the NS matter. These simulations last for around nine orbits, and have residual eccentricity of the initial data reduced to the values below $10^{-3}$. The waveforms from these simulations include modes up to $\ell=4$, and are corrected for the center-of-mass drift~\cite{Woodford:2019tlo}, cf.\ Ref~\cite{Markin:2026eyc} for details\footnote{While preparing this manuscript, we identified an EOS inconsistency between the initial data and evolution for the simulations with the DD2 EOS (seven simulations). This causes reconfiguration of the NS for the first milliseconds, and leaves the star in a perturbed state. As a consequence, the tidal deformability changes by a few percent except for BAM:0232, BAM:0233, and BAM:0227, where the change remains at a sub-percent level. In \cref{tab:simulation_table}, we list the original values used in the calibration of the amplitude models and refer the reader to Ref.~\cite{Markin:2026eyc} for the corrected parameter values. In all cases, however, the differences remain within the uncertainties of the amplitude calibration procedure.}. One of these simulations, BAM:0238, was finished in a later stage and was only included for validation. 

In summary, the NR dataset used in the calibration of the amplitude models detailed in \cref{sec:model} comprises 162 SACRA waveforms for the $(2,2)$ mode, together with the HM waveforms listed in \cref{tab:simulation_table}; some of which were hybridized with \model{DALI} for calibration, while all were hybridized with \model{NRSurTidal} for validation.

\begin{table}
    \caption{NR simulations including HMs employed in the calibration of the amplitude models. The columns report the simulation tag, the mass ratio $Q$, the dimensionless BH spin $\chiBHz$, the NS tidal deformability $\lambdaNS$, and the NS mass $\MNS$ in solar masses. All of these simulations feature a non-spinning NS. The simulations above the dashed line were hybridized with \model{DALI} prior to calibration.}
    \label{tab:simulation_table}
    \vspace{0.5mm}
    \begin{tblr}{
        rowsep = 0.25mm,
        colsep = 4pt,
        colspec = {c S[table-format=1.2] S[table-format=-1.2] S[table-format=5.0] S[table-format=1.2]},
column{1} = {rightsep=12pt},
row{2} = {abovesep=1mm},
    }
\SetCell{c} \text{Tag} & \text{$Q$} & \text{$\chiBHz$} & \text{$\lambdaNS$} & \text{$\MNS\,[\Msun]$} \\
        \midrule
        BAM:0181      & 2.25 & -0.30 & 494    & 1.44  \\
        BAM:0182      & 2.35 & -0.60 & 494    & 1.44  \\
        BAM:0183      & 3.33 & 0.00  & 494    & 1.44  \\
        BAM:0184      & 3.38 & 0.30  & 494    & 1.44  \\
        BAM:0185      & 3.38 & -0.30 & 494    & 1.44  \\
        BAM:0187      & 3.53 & 0.60  & 494    & 1.44  \\
        BAM:0190      & 1.92 & 0.00  & 1010   & 1.46  \\
        BAM:0194      & 2.03 & -0.60 & 1010   & 1.46  \\
        BAM:0196      & 2.22 & -0.30 & 1010   & 1.46  \\
        BAM:0197      & 2.22 & 0.30  & 1010   & 1.46  \\
        BAM:0198      & 2.32 & -0.59 & 1010   & 1.46  \\
        BAM:0199      & 3.29 & 0.00  & 1010   & 1.46  \\
        BAM:0200      & 3.33 & -0.30 & 1010   & 1.46  \\
        BAM:0201      & 3.33 & 0.30  & 1010   & 1.46  \\
        BAM:0202      & 3.48 & 0.60  & 1010   & 1.46  \\
        BAM:0204      & 1.96 & 0.00  & 276    & 1.43  \\
        BAM:0205      & 1.98 & 0.30  & 276    & 1.43  \\
        BAM:0206      & 2.24 & 0.00  & 276    & 1.43  \\
        BAM:0209      & 2.27 & 0.30  & 276    & 1.43  \\
        BAM:0210      & 2.27 & -0.30 & 276    & 1.43  \\
        BAM:0213      & 2.36 & 0.60  & 276    & 1.43  \\
        BAM:0214      & 3.36 & 0.00  & 276    & 1.43  \\
        BAM:0215      & 3.40 & 0.30  & 276    & 1.43  \\
        BAM:0216      & 3.40 & -0.30 & 276    & 1.43  \\
        BAM:0220      & 3.55 & -0.60 & 276    & 1.43  \\
        \hline[dashed] BAM:0227      & 2.00 & 0.00  & 12530  & 0.80  \\
        BAM:0228      & 1.00 & 0.00  & 1633   & 1.20  \\
        BAM:0229      & 2.00 & 0.00  & 1633   & 1.20  \\
        BAM:0230      & 1.00 & 0.00  & 701    & 1.40  \\
        BAM:0231      & 2.00 & 0.00  & 701    & 1.40  \\
        BAM:0232      & 1.00 & 0.00  & 29     & 2.20  \\
        BAM:0233      & 2.00 & 0.00  & 29     & 2.20  \\
        BAM:0234      & 1.00 & 0.00  & 811    & 1.20  \\
        BAM:0235      & 2.00 & 0.00  & 811    & 1.20  \\
        BAM:0236      & 1.00 & 0.00  & 307    & 1.40  \\
        BAM:0237      & 2.00 & 0.00  & 307    & 1.40  \\
        SXS:BHNS:0001 & 6.00 & 0.00  & 525    & 1.40  \\
        SXS:BHNS:0002 & 2.00 & 0.00  & 791    & 1.40  \\
        SXS:BHNS:0003 & 3.00 & 0.00  & 607    & 1.35  \\
        SXS:BHNS:0004 & 1.00 & 0.00  & 791    & 1.40  \\
        SXS:BHNS:0006 & 1.50 & 0.00  & 791    & 1.40  \\
        SXS:BHNS:0008 & 3.00 & 0.90  & 792    & 1.40  \\
        SXS:BHNS:0009 & 4.00 & 0.90  & 793    & 1.40  \\
        \bottomrule
    \end{tblr}
\end{table}

\subsection{Hybrid construction}\label{subsec:hybrids}
Due to the short length of NR waveforms, it is common practice to build \textit{hybrid} waveforms, where an NR waveform is smoothly connected to a GW model covering the early inspiral regime. In this work, we employ hybrid waveforms both in the construction and in the validation of GW models. For the low frequency part of the signal, we consider two different waveform models, \model{DALI}, where extreme matter effects have been explicitly tuned to NR waveforms, and \model{NRSurTidal}, which implements instead the PN tidal slicing method~\cite{Barkett:2015wia}\footnote{In the PN tidal splicing method, the BBH contribution to the inspiral dynamics is extracted from NR waveforms; analytic PN tidal corrections are then linearly added to the flux-balance equations to modify the rate of the adiabatic inspiral.}. For the tuning of the amplitudes, we rely on hybrids constructed with \model{DALI}, since this model provides a better behaved approximation throughout merger-ringdown, allowing some NR waveforms that cover only a few orbits before merger to be included in the calibration. For the computation of mismatches and TD comparisons presented in \cref{sec:validation}, as well as for our PE studies in \cref{sec:pe}, we instead employ hybrids constructed with \model{NRSurTidal}. As a surrogate model, \model{NRSurTidal} provides an accurate description of the inspiral over the relevant region of parameter space while remaining independent of the \model{DALI} waveforms used for the amplitude calibration, thereby reducing the possibility of introducing favourable biases in the validation. Furthermore, this choice follows the hybridization strategy adopted in Ref.~\cite{Huang:2020pba}, enabling a direct comparison with previous studies of NSBH waveform systematics.

For each simulation, we select an alignment window, $\sbr{t_1, t_2}$, after discarding the portion of the NR waveform contaminated by the initial junk radiation. The start of the window, $t_1$, is placed sufficiently after the discarded segment to ensure that transient effects associated with the initial data have fully subsided. The end of the window, $t_2$, is chosen such that the alignment spans several orbital periods while retaining an adequate buffer before merger, whenever the simulation length permits. The alignment is validated by verifying that the frequency difference between the model and NR waveform exhibits no secular trend across the alignment window, that the accumulated phase difference remains comparable to or smaller than the estimated NR error, and that the fitted time and phase shifts vary smoothly under small perturbations of $t_1$ and $t_2$.

Assuming that the NR and model waveforms follow the same tetrad convention, they can be aligned by finding the time- $t_\mathrm{c}$ and phase-shift $\phi_\mathrm{c}$ minimizing the integrated difference between their phasings over a subset of $(\ell, m)$ modes:
\begin{equation}\label{eq:wav_alignment}
    \mathcal{I} = \sum_{\lm}{
        w_\lm \! \int_{t_1}^{t_2}{
            \abs{\underset{m\pi}{\bmod}{\del[2]{
                \phi^{\mathrm{NR}}_{\lm}(t) - \sbr[1]{\phi_{\lm}(t - t_\mathrm{c}) + {m \phi_\mathrm{c}}}
                    }}
                }
            dt}
        }\,,
\end{equation}
where $w_\lm$ are weights assigned to each mode according to their relative contribution to the signal.

Once the optimal phase and time shifts have been determined, the individual spin-weighted spherical harmonic modes of a TD hybrid waveform can be constructed through a piecewise function. Assuming one is interested in the hybrid GW strain $h$, then:
\begin{equation}
    h^{\mathrm{hyb}}_{\lm} =
    \begin{cases}
        h_{\lm}(t-t_c)\, e^{-i \phi_c} &t < t_1 \\
        \begin{aligned}
        & w(t)\, h_{\lm}(t-t_c)\, e^{-i \phi_c} \\
        & \quad + [1-w(t)]\, h^{\mathrm{NR}}_{\lm}(t)
        \end{aligned} &t_1 \le t \le t_2 \\
        h^{\mathrm{NR}}_{\lm}(t) &t > t_2
    \end{cases},
\end{equation}
where we take $w(t)$ to be a Hann window
\begin{equation}
     w(t)=\frac{1}{2}\sbr{1+\cos\del{\frac{t-t_1}{t_2-t_1}}}.
\end{equation}
A similar construction can be applied to the individual modes of the Newman-Penrose scalar $\psi_{4}$, which is our preferred strategy when constructing hybrid BAM waveforms for our calibration dataset, since this avoids the need of performing a fixed frequency integration. 

\section{Construction of the Models}\label{sec:model}
In this section, we present a summary of the construction of the \model{XNSBH}, \model{XPNSBH}, and \model{v5HMROM_NSBH} models presented in this paper.

These models are constructed in the FD by modelling the spherical harmonic modes $h_{\lm}$ of the multipolar decomposition of the GW strain $h$,
\begin{equation}
    h(f,\theta,\phi;\vec{\sigma}) = \sum_{\ell \ge 2}\sum_{m = -\ell}^{\ell} {}_{-2}Y_{\lm}(\theta, \phi) h_{\lm}(f;\vec{\sigma})\,,
\end{equation}
where ${}_{-2}Y_{\lm}$ are the spin-weighted spherical harmonics of spin weight $-2$, $\theta$ and $\phi$ denote the polar and azimuthal angles of the observer in the source frame, and $\vec{\sigma}$ stands for the set of intrinsic parameters of the system. These complex modes can be written in polar form in terms of real amplitude $A_{\lm}$ and phase $\psi_{\lm}$ functions:
\begin{equation}
    h_{\lm}(f;\vec{\sigma}) = A_{\lm}(f;\vec{\sigma}) e^{-i\psi_{\lm}(f;\vec{\sigma})}\,.
\end{equation}
In practice, these functions are only modelled for the negative $m < 0$ modes, which, under the Fourier transform convention used in the LVK Algorithm Library Suite (LALSuite)~\cite{lalsuite, swiglal},
\begin{equation}
    h(f) = \int_{-\infty}^{\infty} h(t) e^{-i 2\pi\!f t} dt\,,
\end{equation}
have support over positive frequencies $f > 0$. The positive $m > 0$ modes are then included by exploiting the equatorial symmetry of non-precessing binaries, which implies
\begin{equation}
    h_{\lm}(f;\vec{\sigma}) = (-1)^{\ell} h_{\ell -m}^*(-f;\vec{\sigma})\,.
\end{equation}
Modes with $m=0$ are not included in the aligned-spin models presented in this work. 

\subsection{Amplitude}\label{subsec:amplitudes}
\subsubsection{\model{XNSBH}}
The IMR amplitude model of \model{XNSBH} builds upon \model{XHM} by incorporating tidal effects and disruption physics. The model employs a piecewise approach with three distinct components:

\begin{enumerate}
    \item \emph{Inspiral region}: The amplitude of \model{XHM} is augmented with PN tidal corrections that capture the deformation of the NS due to the companion's gravitational field.
    \item \emph{Merger-ringdown region}: The amplitude of \model{XHM} is multiplied by a suppressing function calibrated to NR simulations, which models the amplitude reduction with respect to the BBH model caused by tidal effects.
    \item \emph{Transition region}: A smooth windowing function connects the inspiral and merger-ringdown regions, ensuring continuity and differentiability.
\end{enumerate}

The calibration of the merger-ringdown ansatz is performed independently for each mode in the ``collocation points'' approach introduced in \model{phX}~\cite{Pratten:2020fqn}. In this procedure, the amplitude ratio of NR simulations with respect to the underlying BBH model is fitted at a number of frequencies spanning the region of interest across the space of intrinsic parameters of the system, and these fits are then used to inform the merger–ringdown ansatz. Rather than using the raw NR amplitudes directly, each waveform is first matched to the merger–ringdown ansatz, yielding an intermediate, mode-specific representation that closely approximates the ideal behaviour of the model. This procedure acts as a smoothing operation, reducing the impact of residual numerical noise in the NR data and providing a more robust input for the calibration of the model. Further details are provided in \cref{app:phenomxnsbh-amplitude}.

\subsubsection{\model{v5HMROM_NSBH}}
In the case of \model{v5HMROM_NSBH}, the amplitude is constructed by extending the corrections used in \model{v4ROM_NSBH} describing tidal disruption effects. As in \model{XNSBH}, these corrections are applied multiplicatively to the amplitude of the underlying BBH model---here \model{v5HMROM}---and are divided into three regions: (i) an inspiral region where the amplitude of the BBH model is left unchanged; (ii) a merger-ringown region where the amplitude is suppressed relative to the BBH case; and (iii) a smooth transition between the two. The boundaries of these regions are defined in terms of a pivot frequency $\Mf_\mathrm{\!piv}$, determined as a function of the effective tidal deformability $\tilde{\Lambda}$.
The multiplicative correction to the $(2,2)$-mode is adapted from Ref.~\cite{Matas:2020wab} and fitted to the simulations of \cref{tab:simulation_table} choosing arbitrary collocation points. The HM corrections are then obtained by evaluating the $(2,2)$-mode amplitude corrections at frequencies scaled with a mode-dependent factor also fitted to NR. Further details are provided in \cref{app:seobv5nsbh-amplitude}.

\subsection{Phase}\label{subsec:phases}
The phase model used in \model{XNSBH} and \model{v5HMROM_NSBH} is constructed by augmenting the BBH models \model{XHM} and \model{v5HMROM} with the tidal phase contributions provided by \modelname{NRTidalv3}~\cite{Abac:2023ujg}, including its recent extension to HMs implemented for several BNS models in Ref.~\cite{Abac:2025brd}.

The \modelname{NRTidalv3} model~\cite{Abac:2023ujg} provides the contributions to the phase of the GW modes arising from the matter interactions, which can be linearly added to the phase of the corresponding BBH waveform, i.e.,
\begin{equation}
    \psi_{\lm}(f) = \psi_{\lm}^{\rm BBH}(f) + \psi_{\lm}^T(f)\,,
\end{equation}
where $\psi_{\lm}^{\rm BBH}$ is the phase of the underlying BBH model, and $\psi_{\lm}^T$ is the tidal phase provided by \model{NRT3}. In particular, this model provides the tidal contribution to the phase of the dominant mode through the following closed-form expression
\begin{equation}\label{eq:psi_T}
    \psi_{22}^T(x) = -c_{\rm Newt}\kappa(\lambdaNS;x)x^{5/2}P(x)\,,
\end{equation}
where $c_{\rm Newt}$ is the leading-order PN constant, $\kappa(\lambdaNS;x)$ is the dynamical tidal parameter, and $P(x)$ is a rational function (a polynomial in the original PN representation, and a Padé approximant in \model{NRT3}) of the PN parameter $x = (\pi \Mf)^{2/3}$. This expression is constructed such that it reduces to the 7.5PN tidal phase at low frequencies, and is fitted to 55 NR simulations of BNS systems containing mass ratios up to $Q=2$ across a wide range of EOSs.
Since the calibration is based exclusively on BNS simulations, the lack of dedicated NSBH physics becomes a limiting factor in the accuracy of \model{NRT3} from the late inspiral through the merger and post-merger regimes of NSBH systems, especially during tidal disruption. Nevertheless, the validation in \cref{sec:validation} shows that these models achieve an accuracy comparable to other state-of-the-art NSBH waveform models, supporting the use of this tidal extension until a dedicated NSBH-based calibration becomes feasible.

The tidal contributions to the HM phases are then included through the following approximated scaling relation
\begin{equation}\label{eq:psi_T_lm}
    \psi_{\lm}^T(f) = \frac{\abs{m}}{2} \psi_{22}^T\del{\frac{2f}{\abs{m}}},
\end{equation}
as was done in Ref.~\cite{Abac:2025brd} for BNS models. Furthermore, the EOS-dependent spin-squared terms up to 3.5PN order and the leading-order spin-cubed terms entering at 3.5PN order, which were introduced in Ref.~\cite{Dietrich:2019kaq} and are present in \model{NRT3}, are also included and appropriately scaled as in \cref{eq:psi_T_lm}. These terms are included separately, since \model{NRT} phasings were tuned to non-spinning NR simulations and so do not include spin-dependent effects by construction.

Due to the nature of the rational function $P(x)$ used in \cref{eq:psi_T}, divergences (poles) or spurious inflection points may arise in the post-merger regime, outside the calibration region of \model{NRT3}. To prevent such unphysical behaviours from entering the waveform, we implement a simple algorithm that checks for these features and replaces the problematic region with a second-order Taylor expansion, ensuring continuity up to the second derivative. Specifically, we locate potential pathological behaviour by identifying the poles of the rational function $P(x)$, and the zeros of the second derivative ${\rm d}^2 \psi_{22}^T(x) / {\rm d}x^2$. If either of these features occurs before the termination of the waveform, we define the corresponding limiting frequencies $\Mf_{\!\mathrm{pole}}$ and $\Mf_{\!\mathrm{infl}}$, and switch to a second-order Taylor extrapolation at
\begin{equation}
    \Mf_{\!\mathrm{ext}} = 0.95\min(\Mf_{\!\mathrm{pole}}, \Mf_{\!\mathrm{infl}})\,,
\end{equation}
providing a buffer from the onset of the first singular or non-concave feature. The resulting phase is twice continuously differentiable in the FD (the relevant domain for PE).\footnote{When computing the inverse Fourier transform to obtain TD waveforms, some oscillatory features can occasionally appear shortly before merger or during the ringdown for high dimensionless tidal deformabilities  ($\Lambda\approx O(10^3)$). These features arise from the use of high-frequency amplitude tapering in the FD and their magnitude generally depends on the details of the post-inspiral phasing.}

In the case of \model{XNSBH}, on top of the tidal contributions from \modelname{NRTidalv3}, the phase of the underlying BBH baseline is also modified by using calibrated remnant properties in the computation of the ringdown's frequencies and damping times (see \cref{app:xnsbh_fring_fdamp}), improving accuracy in the post-merger.

\subsection{Precession}\label{subsec:precession}
In \model{XPNSBH}, precession effects are included through the standard twisting-up procedure~\cite{Schmidt:2010it} inherited from its baseline model \model{XPHM}. Within this framework, the signal in a frame that is co-precessing with the binary's orbital angular momentum is approximated by the aligned-spin waveform provided by \model{XNSBH}. This procedure is performed using a time-dependent set of Euler angles that map the co-precessing frame to the inertial frame, propagating the tidal corrections included in the aligned-spin model to the precessing regime.

The Euler angles are derived in closed form using a multiple scale analysis of the orbit-averaged PN spin-precession equations. Exploiting the hierarchy between the precession and radiation-reaction timescales, the dynamics are treated perturbatively, yielding a solution composed of a secular (slowly varying) term and oscillatory corrections on the precession timescale~\cite{Chatziioannou:2016ezg, Chatziioannou:2017tdw, Pratten:2020ceb}. Alternatively, the user can activate another prescription for the Euler angles~\cite{Colleoni:2025aoh} based on the numerical integration of the orbit-averaged SpinTaylorT4 PN equations of Ref.~\cite{SpinTaylor_TechNote}. This prescription is, however, computationally more expensive, especially for low-mass systems. %

\section{Performance and Model Validation}\label{sec:validation}
In this section, we validate the computational efficiency and accuracy of the new models by presenting timings, mismatches, and TD comparisons against other waveform models and NR simulations. For convenience, from this point onward we resort to shorthand names for all the approximants used in this study, which are summarized in \cref{table:model_abbreviations}.

\begin{table*}
  \caption{Shorthand names used for the models considered in \cref{sec:validation,sec:pe}, along with their respective BBH baselines, included modes, domain of implementation, and whether they can generate spin-precessing signals.}
  \label{table:model_abbreviations}
  \centering
  \scriptsize
  \begin{tblr}{
    width   = \textwidth,
    colspec = {l l l X[l] c c},
    rowsep  = 2pt,
    colsep  = 3.25pt,
    row{1}  = {font=\bfseries, belowsep=0.5mm},
    row{2}  = {abovesep=1mm},
    row{5}  = {abovesep=3pt},
    row{7}  = {abovesep=3pt},
    column{1} = {leftsep=2pt},
    column{4} = {rightsep=2pt},
    column{6} = {rightsep=2pt},
    cells = {valign = m} }
Shorthand Name   & Waveform Model     & BBH Baseline   & \SetCell{c} Modes  & Domain & Precession \\
\midrule
    \alias{phNSBH}        &  \model{phNSBH}~\cite{Thompson:2020nei}   &  {\model{phC}~\cite{Santamaria:2010yb} (amplitude)\\
                                                                          \model{phD}~\cite{Khan:2015jqa} (phase)}       & 22                         & FD & --         \\
    \alias{XNSBH}         &  \model{XNSBH}                            &  \model{XHM}~\cite{Garcia-Quiros:2020qpx}        & 22, 21, 33, 32, 44         & FD & --         \\
    \alias{XPNSBH}        &  \model{XPNSBH}                           &  \model{XPHM}~\cite{Pratten:2020ceb}             & 22, 21, 33, 32, 44         & FD & \checkmark \\
    \hline[dashed]     
    \alias{v4ROM_NSBH}    &  \model{v4ROM_NSBH}~\cite{Matas:2020wab}  &  \model{v4ROM}~\cite{Bohe:2016gbl}               & 22                         & FD & --         \\
    \alias{v5HMROM_NSBH}  &  \model{v5HMROM_NSBH}                     &  \model{v5HMROM}~\cite{Pompili:2023tna}          & 22, 21, 33, 32, 44, 43, 55 & FD & --         \\
    \hline[dashed]     
    \alias{GIOTTO}        &  \model{GIOTTO}~\cite{Gonzalez:2022prs}   &  ---                                             & 22, 21, 33, 32, 44, 43, 55 & TD & \checkmark \\
    \alias{DALI}          &  \model{DALI}~\cite{Gonzalez:2025xba}     &  ---                                             & 22, 21, 33, 32, 44         & TD & \checkmark \\
    \bottomrule
\end{tblr}
\end{table*}

\subsection{Time-domain comparisons}\label{subsec:td_comparisons}
In this subsection, we compare the waveforms produced by \alias{XNSBH} and \alias{v5HMROM_NSBH} converted to the TD, with those produced by NR simulations and \alias{DALI}, the most recent TD model including HMs. In particular, we consider the simulations SXS:BHNS:0001 and SXS:BHNS:0002 from \cref{tab:simulation_table}, with mass ratios $Q=6$ and $Q=2$, respectively.

In order to perform these comparisons, each model is aligned with the NR simulation using \cref{eq:wav_alignment}. For consistency, we restrict the sum over the $(\ell, m)$ modes over those common between \alias{XNSBH} and \alias{v5HMROM_NSBH}, i.e., $(\ell, \abs[0]{m}) \in \cbr[0]{(2,2), (2,1), (3,2), (3,3), (4,4)}$ for unequal mass systems, and $(\ell, \abs[0]{m}) \in \cbr[0]{(2,2), (3,2), (4,4)}$ for equal-mass systems (where the odd $m$-modes are suppressed due to symmetries). The integration window $[t_1,t_2]$ is chosen between $-1200 \Msun$ and $-800 \Msun$ before the peak of the $(2,2)$ mode. 

\begin{figure*}[htp]
	\begin{center}
		\includegraphics[width=1\linewidth]{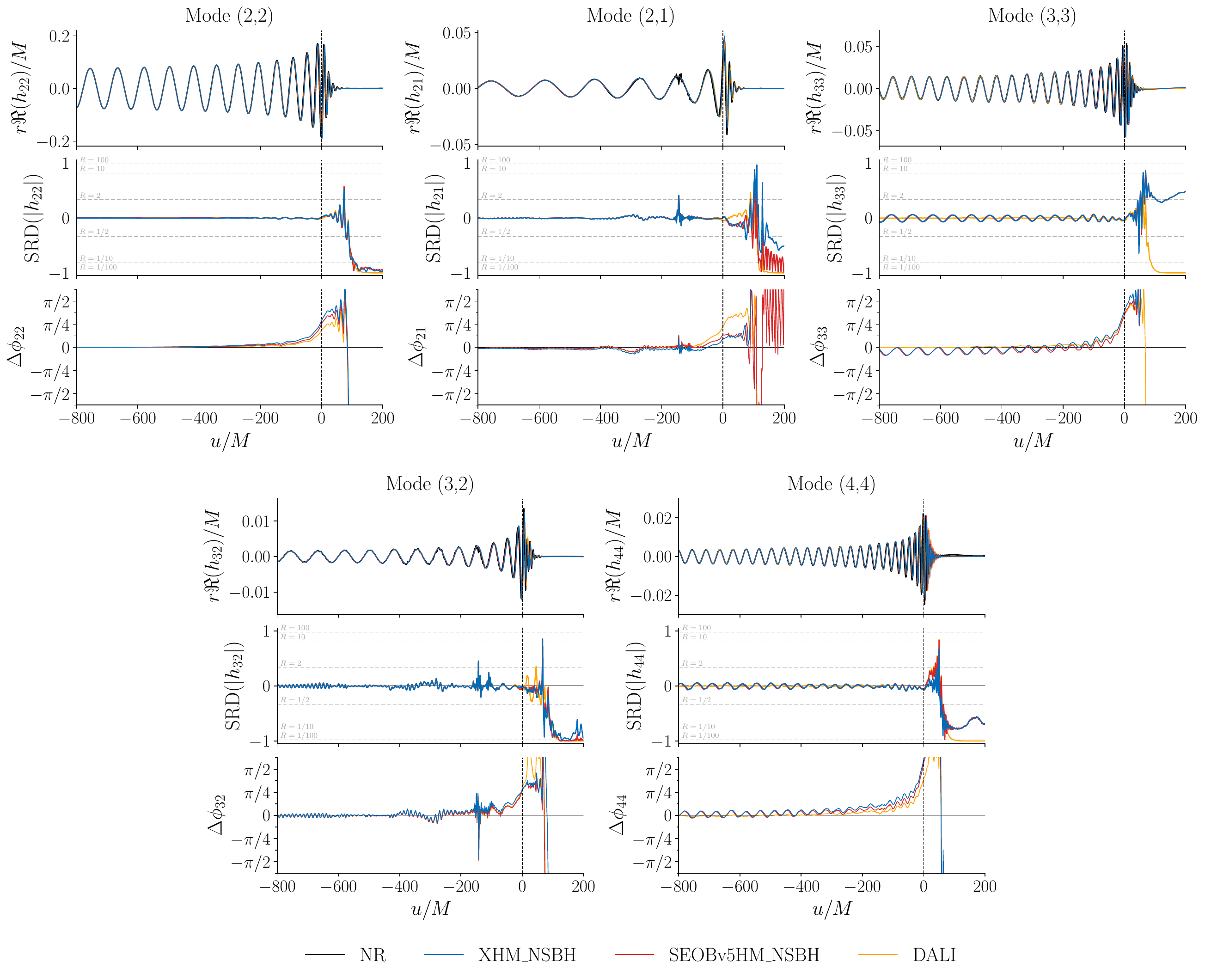}
        \caption{TD comparison between the non-disruptive NR simulation SXS:BHNS:0001 ($Q=6$, $\chiBH=0$, $\lambdaNS=525$) and the waveforms produced by \alias{XNSBH}, \alias{v5HMROM_NSBH}, and \alias{DALI}. Each column shows the real part, phase difference, and amplitude SRD for a given mode with respect to the NR simulation, as a function of the retarded time $u = t - r_*$, where $r_*$ is the tortoise coordinate. The horizontal lines in the amplitude panels correspond to values of the ratio $R = |h^{\vphantom{\mathrm{NR}}}_{\lm}|/|h^\mathrm{NR}_{\lm}|$. We do not show error bands for the NR data, since this configuration does not have multiple resolutions.}
		\label{fig:TDComparison_SXS0001}
	\end{center}
\end{figure*}

\begin{figure*}[htp]
	\begin{center}
		\includegraphics[width=1\linewidth]{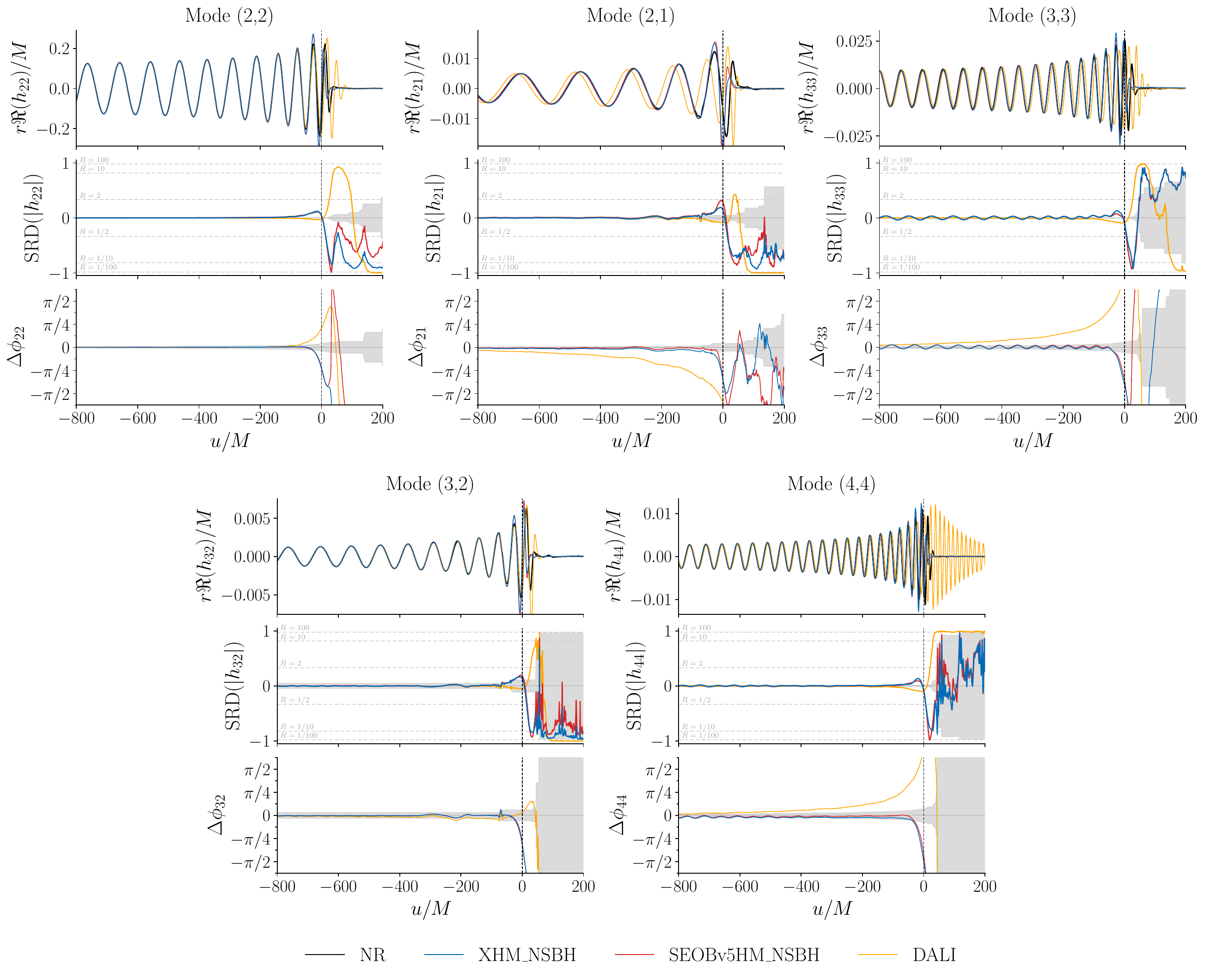}
		\caption{TD comparison between the disruptive NR simulation SXS:BHNS:0002 ($Q=2$, $\chiBH=0$, $\lambdaNS=791$) and the waveforms produced by \alias{XNSBH}, \alias{v5HMROM_NSBH}, and \alias{DALI}. Each column shows the real part, phase difference, and amplitude SRD for a given mode with respect to the NR simulation, as a function of the retarded time $u$.}
		\label{fig:TDComparison_SXS0002}
	\end{center}
\end{figure*}

The results of these comparisons are presented in \cref{fig:TDComparison_SXS0001,fig:TDComparison_SXS0002}, where we show the real part of each mode as a function of the retarded time, along with its phase difference and amplitude symmetric relative difference (SRD)\footnote{We define the symmetric relative difference of the amplitude with respect to NR as $\mathrm{SRD}\big(\big|h_{\ell m}\big|\big)=\big(\big|h_{\ell m}^{\vphantom{\mathrm{NR}}}\big| - \big|h_{\ell m}^{\mathrm{NR}}\big|\big)/\big(\big|h_{\ell m}^{\vphantom{\mathrm{NR}}}\big| + \big|h_{\ell m}^{\mathrm{NR}}\big|\big)$, which takes values between $-1$ and $1$.} with respect to the NR simulation. When multiple resolutions are available, we use shaded regions to indicate the estimated uncertainty of the NR simulation with respect to the second-highest resolution.

Against the non-disruptive SXS:BHNS:0001 simulation (\cref{fig:TDComparison_SXS0001}), all models show good agreement until close to merger in terms of both amplitude and phase, indicating a good recovery of the underlying BBH models in this configuration where tidal effects are subdominant given the asymmetric masses.

\begin{figure}[htp]
	\begin{center}
		\includegraphics[width=\columnwidth]{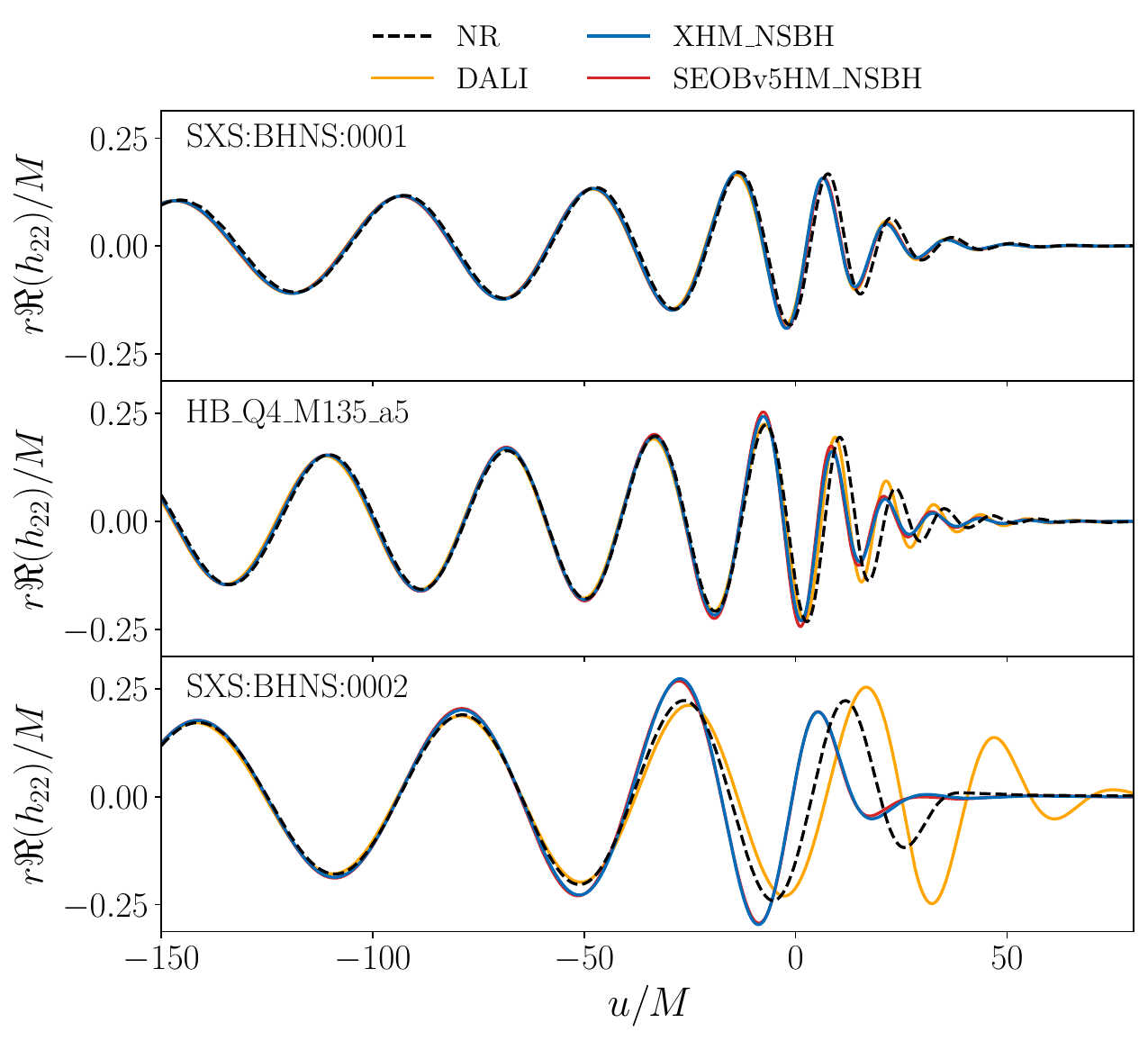}
        \caption{TD comparison of the real part of the $(2,2)$ mode around the merger region for the non-disruptive SXS:BHNS:0001 ($Q=6$, $\chiBHz=0$, $\lambdaNS=525$), mildly disruptive HB\_Q4\_M135\_a5 ($Q=4$, $\chiBHz=0.5$, $\lambdaNS=422$), and disruptive SXS:BHNS:0002 ($Q=2$, $\chiBHz=0$, $\lambdaNS=791$) NSBH simulations, compared against \alias{XNSBH}, \alias{v5HMROM_NSBH}, and \alias{DALI}.}
        \label{fig:TD_NDMDD}
	\end{center}
\end{figure}

In the case of SXS:BHNS:0002 (\cref{fig:TDComparison_SXS0002}), which has a lower mass ratio and is in the disruptive merger regime, we observe better agreement than for SXS:BHNS:0001 throughout the inspiral and close to merger, after which the agreement rapidly deteriorates. In this case, both \alias{XNSBH} and \alias{v5HMROM_NSBH} exhibit larger disagreements with NR near and after merger, likely due to the tidal phase entering its extrapolation regime. \alias{DALI} shows large dephasings, particularly on the $(2,1)$ mode, as well as insufficient amplitude suppression after merger across all of its modes, which is especially pronounced in the case of the $(4,4)$ mode.

A close-up of the merger region showing the real part of the $(2,2)$ mode for these simulations is presented in \cref{fig:TD_NDMDD}, where the mildly disruptive SACRA simulation HB\_Q4\_M135\_a5 ($Q=4$, $\chiBHz=0.5$, $\lambdaNS=422$) is included to illustrate the transition between the three NSBH merger regimes. For this comparison, the waveforms are aligned over a window extending from two orbital periods after the dissipation of the junk radiation to one orbital period before merger. As the degree of tidal disruption increases from the non-disruptive to the mildly disruptive and fully disruptive cases, the waveform exhibits progressively stronger suppression of the post-merger signal. Although all three waveform models reproduce the qualitative transition across the three merger regimes, their agreement with NR gradually worsens from the non-disruptive to the disruptive case. The models also differ in the amount of amplitude suppression they predict, with \alias{DALI} exhibiting the weakest suppression.

\begin{figure}[htp]
	\begin{flushleft}
		\includegraphics[width=\columnwidth]{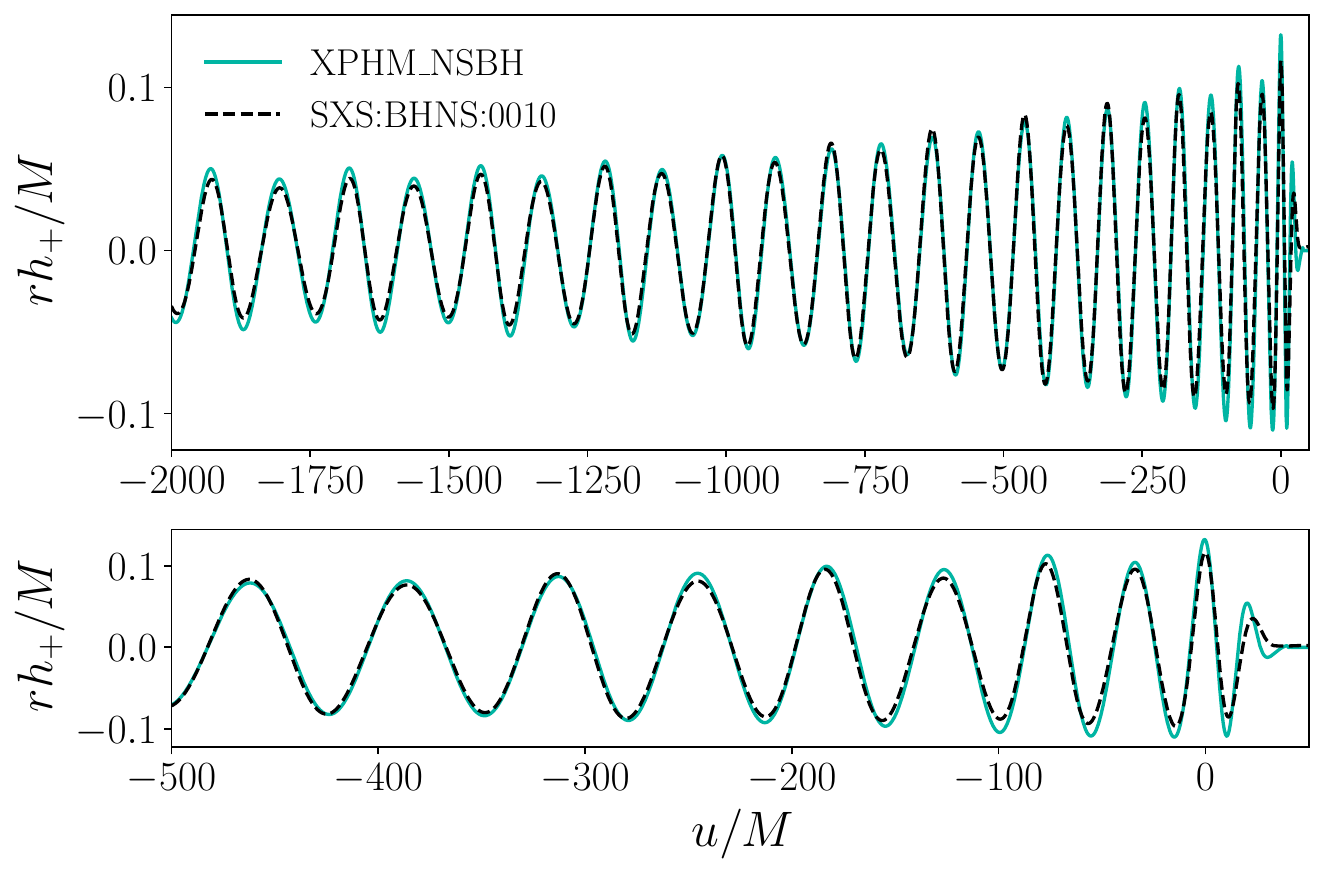}
        \caption{TD comparison between the plus polarization of the GW strain of the disruptive NR simulation SXS:BHNS:0010 ($Q=3$, $\chiBHvec = [0.53, 0.04, 0.53]$, $\lambdaNS=791$) and \alias{XPNSBH} as a function of retarded time at an inclination of $\iota=\pi/6$. The alignment between the NR and model waveform is performed following the optimization described in the main text. The bottom panel corresponds to a zoom into the last cycles before merger.}
		\label{fig:td_precessing}
	\end{flushleft}
\end{figure}

Finally, we compare \alias{XPNSBH} against SXS:BHNS:0010, a $Q=3$ simulation with a primary's reference dimensionless spin of $\chiBHvec=[0.53, 0.04, 0.53]$, and a NS tidal deformability of $\lambdaNS=791$, resulting in a disruptive merger. The alignment is performed by maximizing the normalized noise-weighted overlap between the cross polarization of the NR strain and \alias{XPNSBH} for a source with an inclination of $\iota=\pi/6$. The optimization is performed over the polarization angle, time and phase shifts, and over a rigid (common) rotation of the in-plane spin components of the model template, meaning that the two in-plane spin vectors undergo the same rotation and are not varied independently. This procedure effectively reabsorbs differences in the reference spin definitions arising from gauge effects, as well as ambiguities in the time--frequency mapping when comparing TD and FD waveforms. The resulting alignment and time-domain comparison are shown in \cref{fig:td_precessing}. The waveform produced by \alias{XPNSBH}, obtained via an inverse Fourier transform using LALSuite's \texttt{SimInspiralChooseTDWaveform} interface, shows good agreement with SXS:BHNS:0010 through merger and ringdown.
We note, however, that this validation is based on a single NR simulation, and a more comprehensive assessment of \alias{XPNSBH} should be performed once a larger set of simulations becomes available. Nevertheless, the agreement observed here provides encouraging evidence that the model can accurately reproduce precessing NSBH waveforms.

\subsection{Mismatches}\label{subsec:mismatches}
To assess the accuracy of the new waveform models, we perform a series of mismatch computations against NR simulations and established waveform models.

The mismatch between two complex TD waveforms ${h}_1$ and ${h}_2$ is a measure of their dissimilarity defined as
\begin{equation}\label{eq:mismatch}
    \bar{\mathcal{F}} = 1 -\!\!\max_{\phi_0, t_c, \psi_p}\!\frac{(h_1(\phi_c, t_c, \psi_p)|h_2)}{\sqrt{(h_1|h_1)(h_2|h_2)}}\,,
\end{equation}
where the overlap
\begin{equation}\label{eq:overlap}
    (h_1|h_2) = 4\mathrm{Re}\int_{f_\mathrm{min}}^{f_{\mathrm{max}}}\frac{\tilde{h}_1^{*}(f)\tilde{h}_2(f)}{S_n(f)}df\,,
\end{equation}
represents the inner product between the waveforms in the frequency range $\sbr[0]{f_\mathrm{min}, f_\mathrm{max}}$ for a given power spectral density (PSD) of the noise $S_n(f)$. In the above expressions, the maximization over the reference phase $\phi_0$, coalescence time $t_c$, and polarization angle $\psi_p$ ensures the correct alignment between the two waveforms. Throughout this section, we use the zero-detuned, high-power PSD of Advanced LIGO as provided by PyCBC~\cite{pycbc}.

\subsubsection{Mismatches against NR}\label{subsubsec:NR_mismatches}

We start by computing mismatches between the new waveform models and NR waveforms hybridized with \model{NRSurTidal}, including the previously established NSBH waveform models from \cref{table:model_abbreviations} for comparison, as well as \alias{XNSBH} and \alias{v5HMROM_NSBH} restricted to the leading-order $(2,2)$ mode. In particular, we consider the simulations listed in \cref{tab:hybrid_mismatch_table}, which correspond to the cleanest aligned-spin simulations including HMs available to us, while also excluding a few equal-mass simulations to avoid an over-representation of these systems, in line with current predictions of dynamical and isolated binary formation models.

We note that, although most of the NR simulations considered here were used in the calibration of the amplitude models described in \cref{subsec:amplitudes}, mismatches for long signals such as the hybrids considered in this section are dominated by differences in the phasing, and none of the NSBH simulations considered here were used in the calibration of \model{NRT3}.

In the computation of these mismatches, we include the modes $(2,\pm2)$, $(2,\pm1)$, $(3,\pm2)$, $(3,\pm3)$ and $(4,\pm4)$ in our hybrid waveforms, considering the frequency content between \SI{25}{\Hz} (the minimum frequency common to all hybrids) and \SI{4096}{\Hz}. For each hybrid, the mismatches are computed over a grid of extrinsic parameters including $9$ values of the inclination $\cos{\iota}\in[1, 0)$, and values of either 0 or $\pi/2\;\si{\radian}$ for the reference phase $\phi_0$ and polarization angle $\psi_p$, resulting in a total of $36$ combinations of extrinsic parameters for each simulation.

\begin{figure*}[htp]
    \begin{center}
    \includegraphics[width=0.93\linewidth]{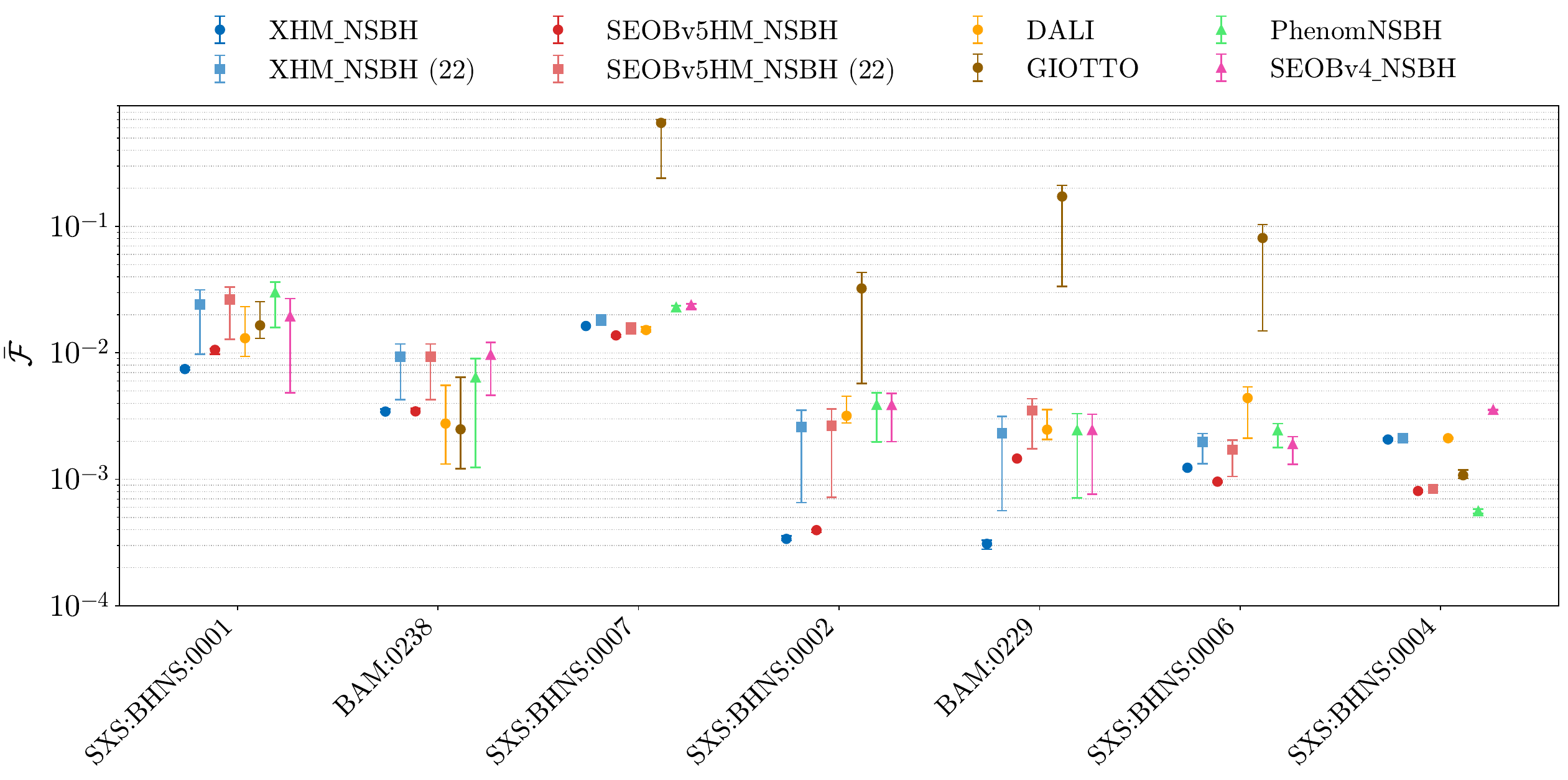}
    \caption{Mismatches between NSBH waveform models and seven NR waveforms from \cref{tab:hybrid_mismatch_table} hybridized with \model{NRSurTidal}. The mismatches are computed over 36 combinations of inclination $\iota$, reference phase $\phi_0$, and polarization angle $\psi_p$. For each model, the marker denotes the median mismatch, while the whiskers extend to the 10th and 90th percentiles. Models including “(22)” in the legend are restricted to the $(\ell, m) = (2,\pm 2)$ modes. The simulations are sorted from least to most disruptive.}
    \label{fig:multimode_mismatches_tags}
\end{center}
\end{figure*}

\begin{figure}[htp]
	\begin{center}
		\includegraphics[width=1\linewidth]{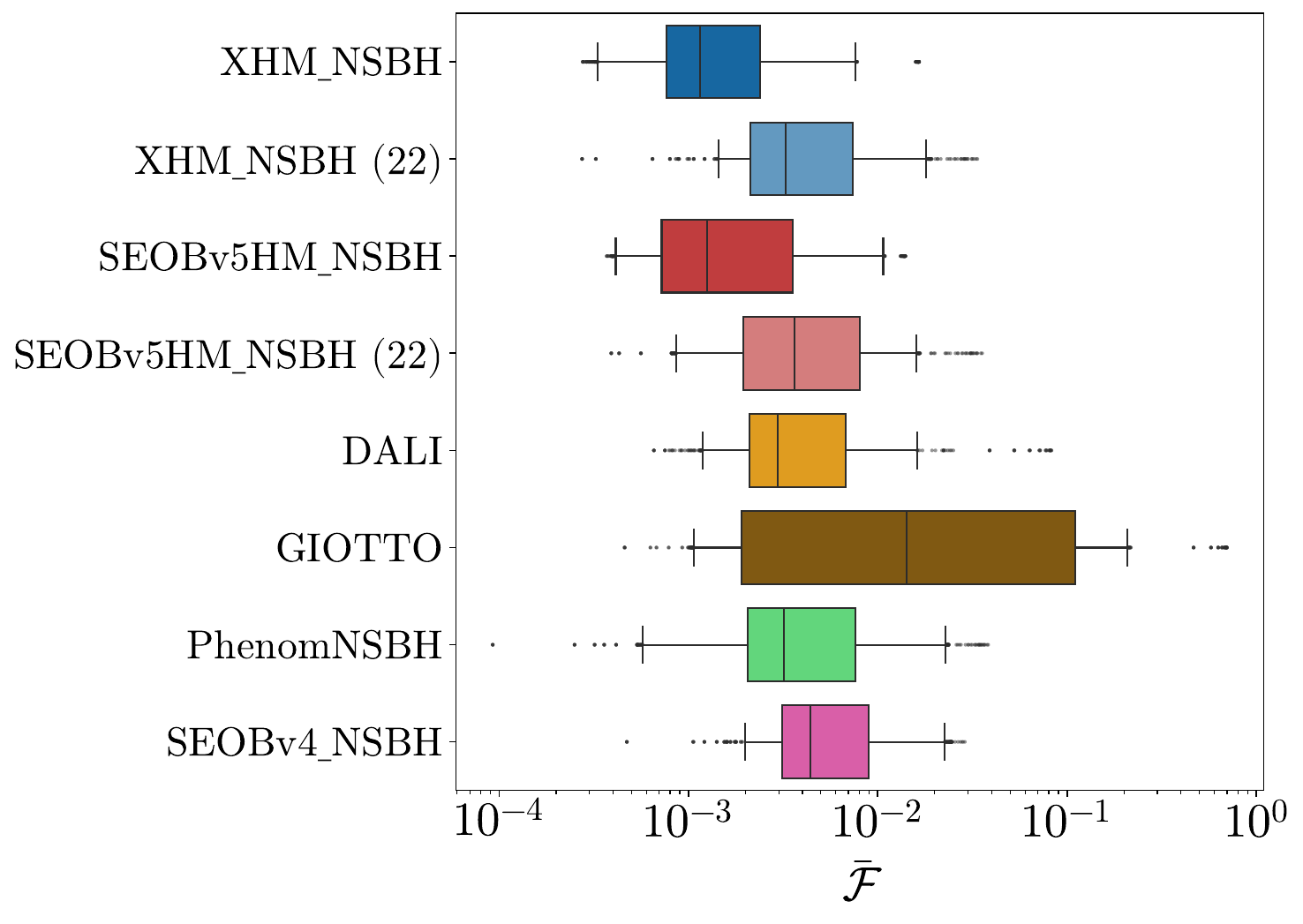}
        \caption{Mismatches between the NSBH waveform models of \cref{table:model_abbreviations} and the NR waveforms of \cref{tab:hybrid_mismatch_table} hybridized with \model{NRSurTidal}, computed across 36 combinations of inclination $\iota$, reference phase $\phi_0$, and polarization angle $\psi_p$. Colored boxes indicate the interquartile range of each distribution, with the central line showing the median, and whiskers extending to the 10th and 90th percentiles. Values beyond this range are individually shown as points. Models including “(22)” in the legend are restricted to the $(\ell, m) = (2,\pm 2)$ modes.}
		\label{fig:multimode_mismatches_boxes}
	\end{center}
\end{figure}

The results of these mismatches are presented in \cref{fig:multimode_mismatches_tags,fig:multimode_mismatches_boxes}. \Cref{fig:multimode_mismatches_tags} compares the median mismatches and 10th--90th percentile ranges obtained by each model for seven increasingly disruptive simulations from \cref{tab:hybrid_mismatch_table}, while \cref{fig:multimode_mismatches_boxes} shows the distribution of mismatches produced by each model when considering all the cases in \cref{tab:hybrid_mismatch_table} presented as box plots indicating the median, quartiles, and 10th and 90th percentiles.

In these figures, we observe how the new models produce lower mismatches with much tighter spreads than both their leading-order mode counterparts and earlier leading-order models, highlighting the increased accuracy and reduced dependency on the extrinsic parameters brought by the inclusion of HMs. This improvement becomes smaller towards equal-mass systems, where HMs are largely suppressed by symmetry in aligned-spin systems, resulting in the mismatch being dominated by the leading-order contribution, as can be seen for SXS:BHNS:0006 ($Q=1.5$) and SXS:BHNS:0004 ($Q=1$) in \cref{fig:multimode_mismatches_tags}. A notable exception to this trend is the case of SXS:BHNS:0007 ($Q=2$), for which all models yield systematically larger mismatches and the improvements provided by the new models over their predecessors are less pronounced. This is because this simulation includes a spinning neutron star, for which the tidal sector of current NSBH waveform models is not calibrated over the merger-ringdown, limiting the achievable agreement with NR irrespective of the treatment of higher harmonics.

The relative performance of the different models for equal-mass systems, where tides are strongest, is largely dependent on the specific simulation, but \alias{phNSBH} showed the best overall agreement in our tests, hinting at possible advantages of its phasing model in this regime.

The mismatches produced by \alias{DALI} and \alias{GIOTTO} against these simulations are in line with those produced by the other models considered in this section. However, \alias{GIOTTO} exhibits significantly larger mismatches than the other models for certain configurations, due to occasional amplitude spikes in some of its higher harmonics that pollute the FD spectrum of the signal. This can be traced back to the fact that, in GIOTTO, tidal corrections to higher harmonics have not been individually fitted to NR. DALI mitigates these issues, and achieves mismatches that are consistent with other models when restricted to the leading order mode; some small differences are visible when including higher harmonics, in line with our time comparisons.

Overall, \alias{XNSBH} and \alias{v5HMROM_NSBH} show an improvement in terms of their mismatches against NR simulations when compared to previously established NSBH waveform models.

As for precessing systems, due to the scarcity of NR simulations, we complement the visual comparison of \cref{subsec:td_comparisons} with a short match study based on SXS:BHNS:0010, where we vary the polarization angle and inclination of the source template over a $10 \times 5$ grid with bounds $[0,\pi] \times [0,\pi/2]$. Matches are computed using the same power spectral density employed for aligned-spin matches over the frequency range $[300,2048]$ Hz, given that the (2,2) mode of this simulation starts just below 250 Hz. The median value of the corresponding mismatch distribution is $0.009$, with the highest mismatch being around $0.03$. Together with the visual comparison of \cref{fig:td_precessing}, these results indicate that \alias{XPNSBH} can reproduce with good accuracy the GW emission from precessing systems, at least in the single-spin case.

\begin{table}
    \caption{Simulations hybridized with \model{NRSurTidal} included in the computation of mismatches against NR (see main text for references describing the input data). The simulations marked with an asterisk were not included in the calibration of the amplitude model. The column $N_\mathrm{cyc}$ represents the number of orbital cycles before merger provided by each simulation previous to hybridization.}
    \label{tab:hybrid_mismatch_table}
    \vspace{0.5mm}
    \begin{tblr}{
        rowsep = 1pt,
        colspec = {l S[table-format=1.2] S[table-format=-1.2] S[table-format=-1.2] S[table-format=5.0] S[table-format=1.2] S[table-format=2]},
        colsep = 3pt,
        column{1} = {rightsep=8pt},
        row{1} = {halign=c},
        row{2} = {abovesep=1mm},
        }
Tag & $Q$ & $\chiBHz$ & $\chiNSz$ & $\lambdaNS$ & $\MNS$ & $N_\mathrm{cyc}$ \\
        \midrule
        BAM:0229            & 2.00 & 0.00 & 0.00  & 1633 & 1.20 & 10  \\
        BAM:0231            & 2.00 & 0.00 & 0.00  & 701  & 1.40 & 9  \\
        BAM:0233            & 2.00 & 0.00 & 0.00  & 29   & 2.20 & 11 \\
        BAM:0235            & 2.00 & 0.00 & 0.00  & 811  & 1.20 & 9  \\
        BAM:0237            & 2.00 & 0.00 & 0.00  & 307  & 1.40 & 10  \\
        BAM:0238\bf{*}      & 3.00 & 0.00 & 0.00  & 307  & 1.40 & 11 \\
        SXS:BHNS:0001       & 6.00 & 0.00 & 0.00  & 525  & 1.40 & 11 \\
        SXS:BHNS:0002       & 2.00 & 0.00 & 0.00  & 791  & 1.40 & 12 \\
        SXS:BHNS:0003       & 3.00 & 0.00 & 0.00  & 607  & 1.35 & 2  \\
        SXS:BHNS:0004       & 1.00 & 0.00 & 0.00  & 791  & 1.40 & 11 \\
        SXS:BHNS:0006       & 1.50 & 0.00 & 0.00  & 791  & 1.40 & 15 \\
        SXS:BHNS:0007\bf{*} & 2.00 & 0.00 & -0.20 & 791  & 1.40 & 11 \\
        \bottomrule
    \end{tblr}
\end{table}

\subsubsection{Mismatches between waveform models}\label{subsubsec:model_mismatches}
In this section, we quantify the agreement among \alias{XNSBH}, \alias{v5HMROM_NSBH}, \alias{GIOTTO}, and \alias{DALI} by computing pairwise mismatches across the parameter space.

In the case of aligned-spin NSBH systems, we compute these mismatches over a sample of 5000 aligned-spin configurations with uniform priors in mass ratios $Q \in [2,20]$, NS masses $\MNS \in [1,3]\,\Msun$, aligned-spin components $\chiBHz \in [-0.9,0.9]$ and $\chiNSz \in [-0.7,0.7]$, and tidal deformabilities of the NS uniformly sampled under the constraint $\lambdaNS \leq 5000$. This sample is not designed to represent an astrophysically likely population: it purposely spans a very broad region of parameter space, allowing for thorough stress-testing of the models. Matches are computed activating each model’s default mode content as specified in \cref{table:model_abbreviations}.

\begin{figure}[htp]
	\begin{center}
		\includegraphics[width=1\linewidth]{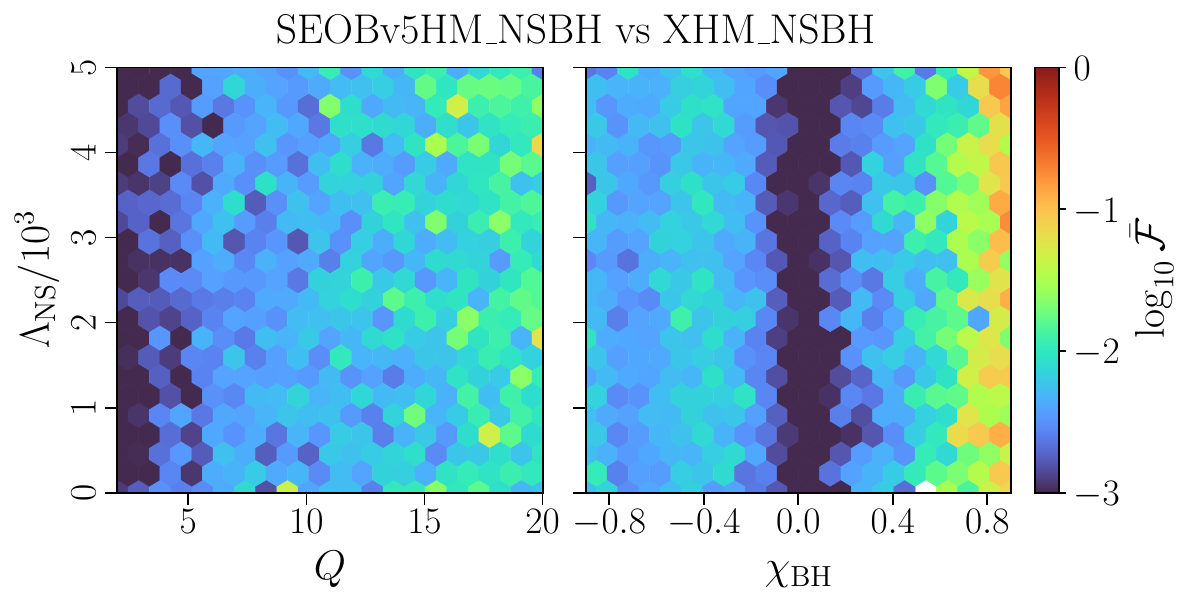}
		\includegraphics[width=1\linewidth]{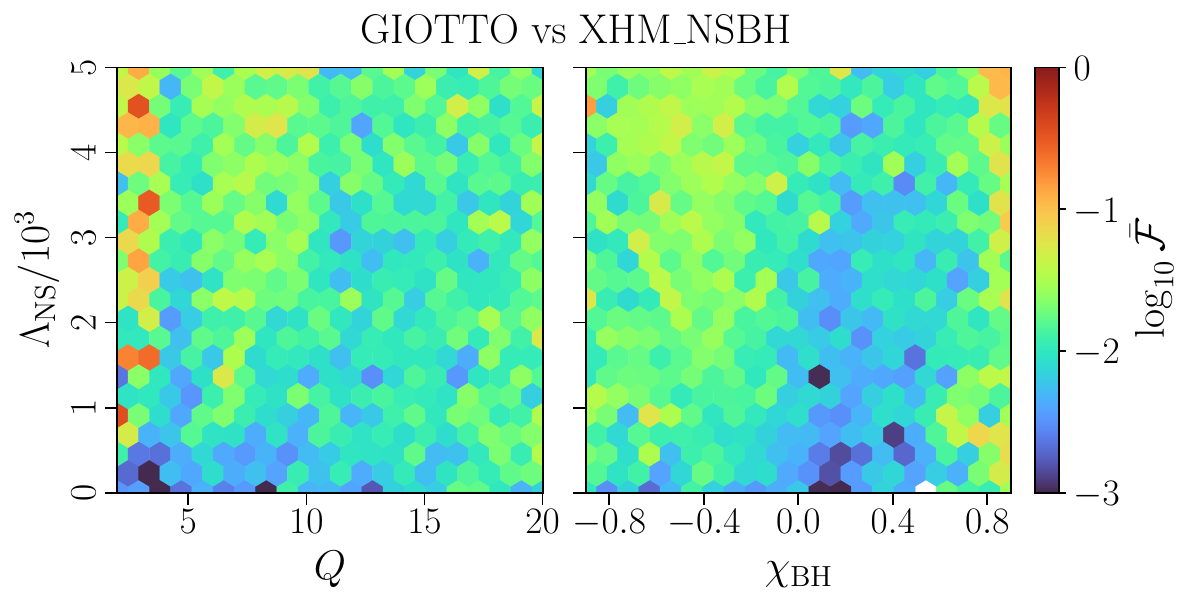}
		\includegraphics[width=1\linewidth]{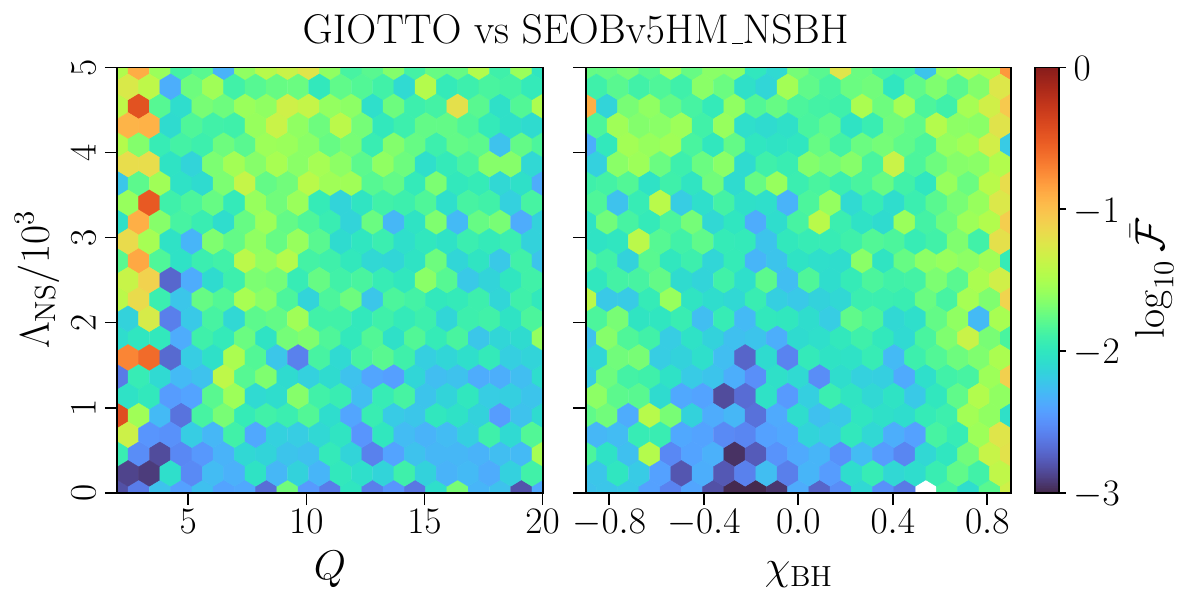}
		\includegraphics[width=1\linewidth]{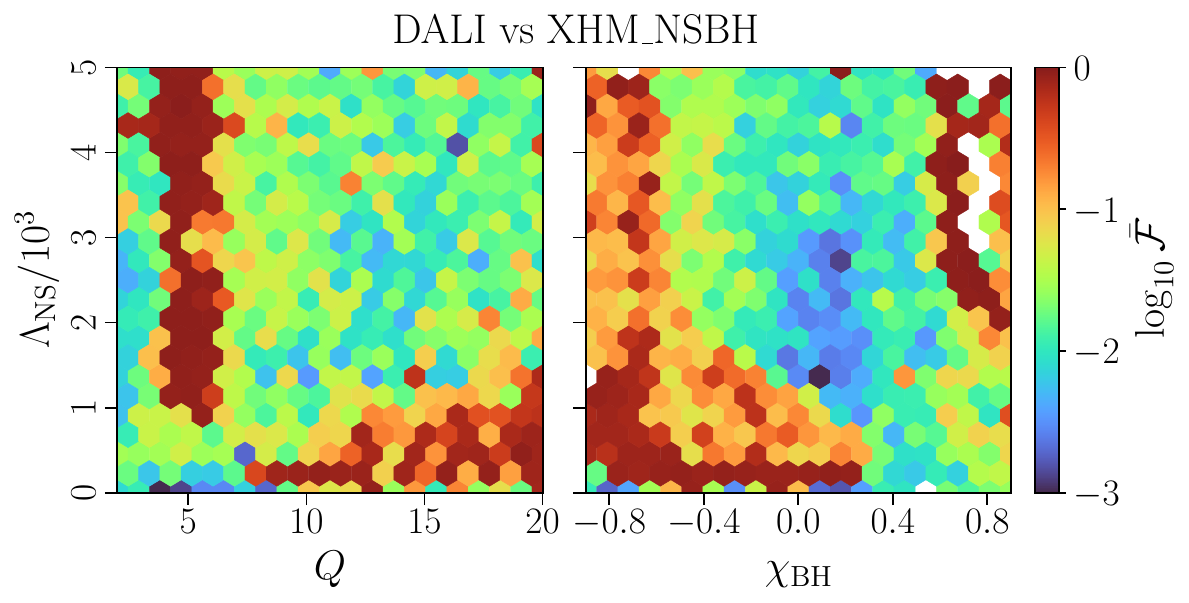}
		\includegraphics[width=1\linewidth]{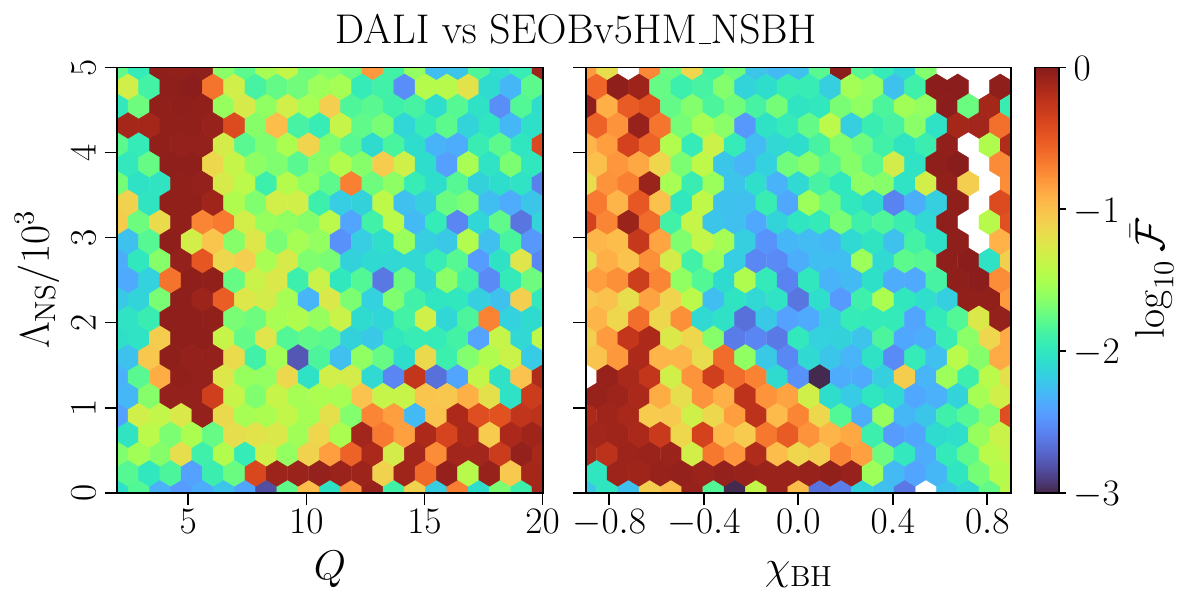}
		\caption{Mismatches between \alias{XNSBH}, \alias{v5HMROM_NSBH}, \alias{GIOTTO} and \alias{DALI}. The mismatches are shown as 2D hexagonal binning plots for the $(Q, \lambdaNS)$ and $(\chiBHz, \lambdaNS)$ pairs, where the colour of each bin indicates the median logarithmic mismatch over the configurations that fall within it. In all cases, the mismatches are computed over a random sample of 5000 configurations (see main text for further details).}
		\label{fig:as_nsbh_model_mismatches}
	\end{center}
\end{figure}

\Cref{fig:as_nsbh_model_mismatches} shows the mismatches between these waveform models as a function of the tidal deformability $\lambdaNS$ and mass ratio $Q$ or BH spin $\chiBHz$. Overall, \alias{XNSBH} and \alias{v5HMROM_NSBH} show good consistency over parameter space, with mismatches generally increasing towards more asymmetric-mass systems and for large prograde spins. The comparisons involving \alias{GIOTTO} follow similar qualitative trends but yield larger mismatches, which tend to worsen towards comparable masses, consistent with the results of \cref{subsubsec:NR_mismatches}. Finally, comparisons involving \alias{DALI} are obtained over a reduced subset of the sample due to the model returning \texttt{NaN} values for about $10\%$ of the configurations, and exhibit localized regions of exceedingly large mismatches; both of these effects can be traced back to pathological behaviours of its HMs in our tests. Indeed, mismatches above 0.5 can be clearly correlated to \alias{DALI} templates having optimal SNRs several orders of magnitudes higher than expected, indicating excess power in one or more harmonics. This behaviour may be linked to the difficulty of accurately modelling relatively sharp transitions at the boundary between different merger types (e.g., disruptive versus non-disruptive). Overall, although \alias{DALI} produced better results than \alias{GIOTTO} over the set of NR simulations considered in \cref{subsubsec:NR_mismatches}, \alias{GIOTTO} appears to more robust across the parameter space typically sampled in PE.

\begin{figure}[htp]
	\begin{center}
        \includegraphics[width=1\linewidth]{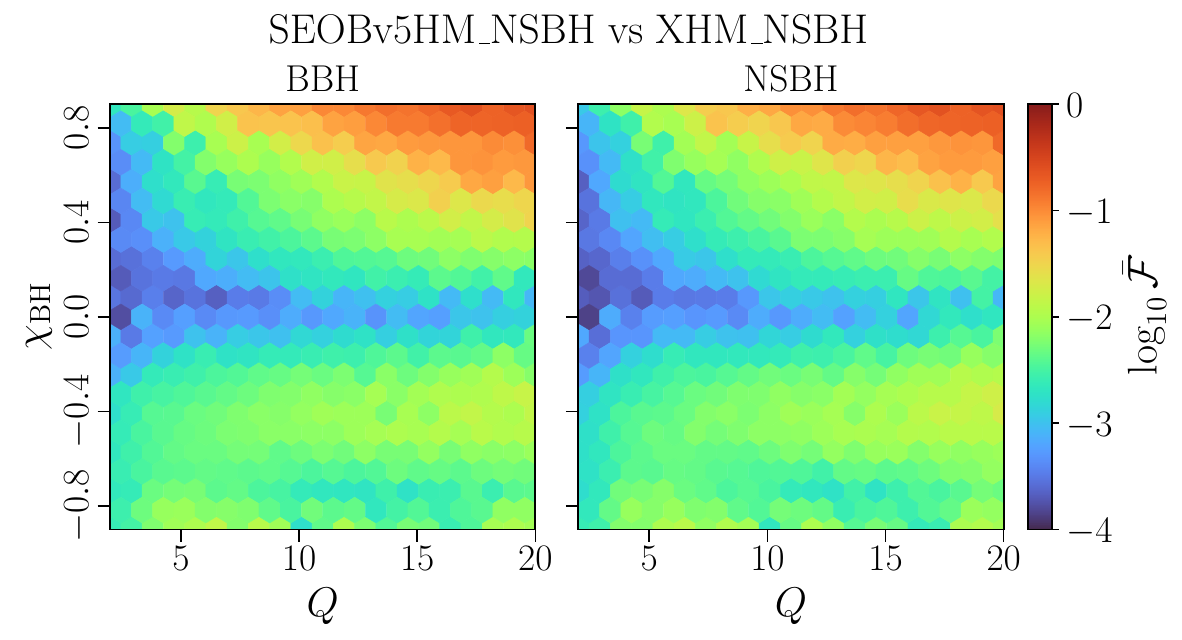}
		\includegraphics[width=1\linewidth]{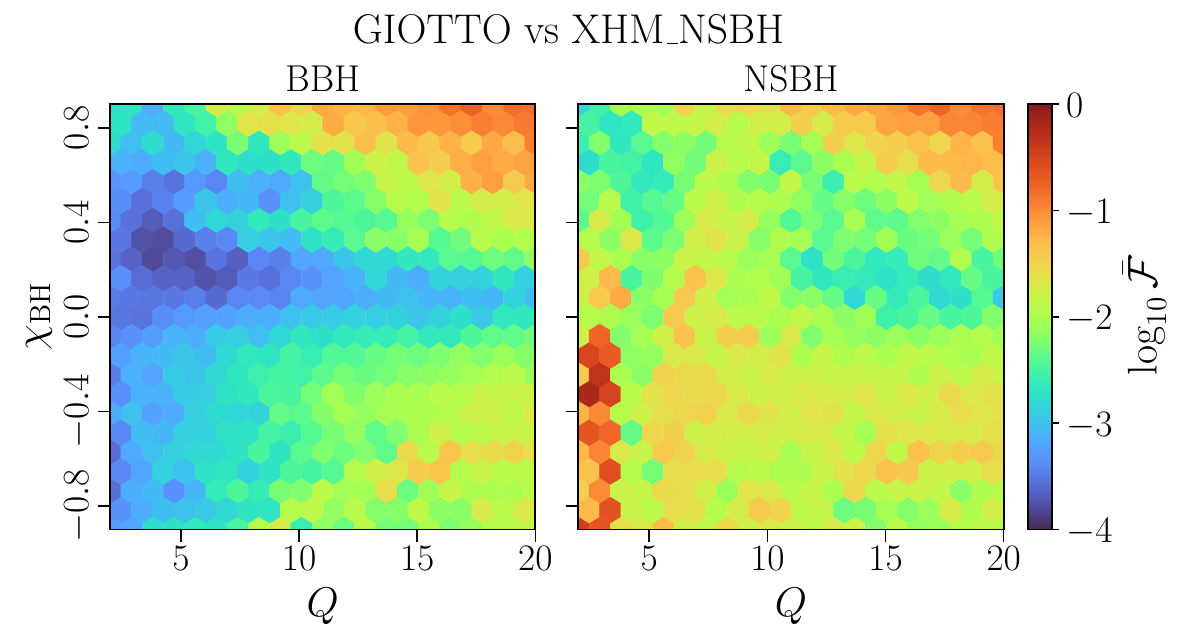}
		\includegraphics[width=1\linewidth]{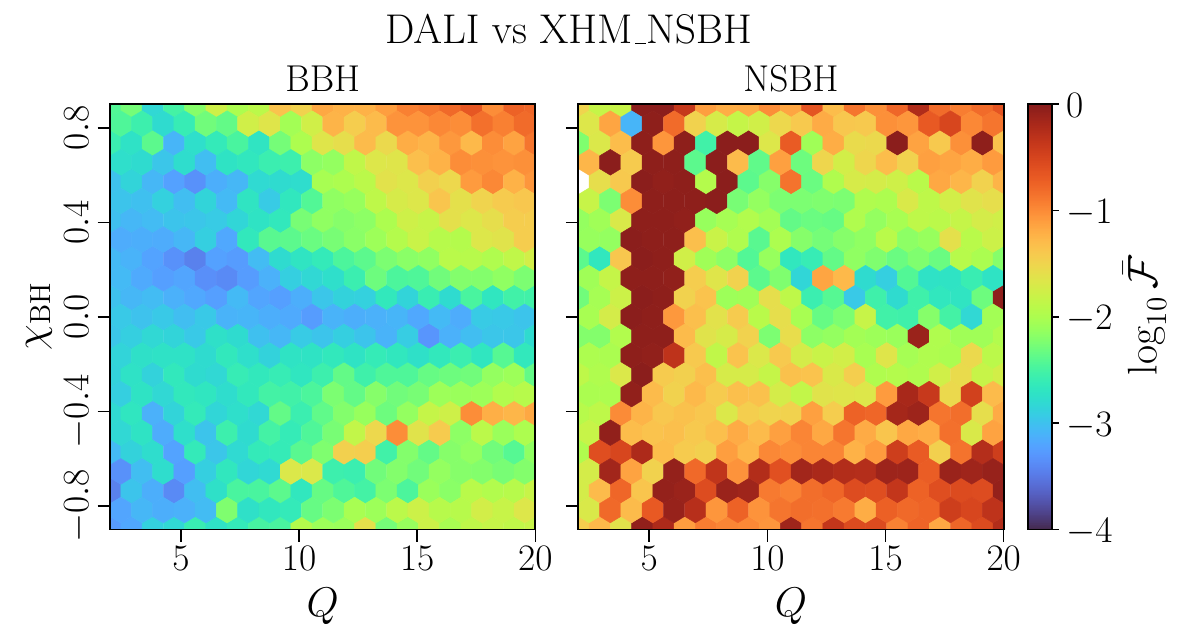}
		\caption{Mismatches between \alias{XNSBH}, \alias{v5HMROM_NSBH}, \alias{GIOTTO} and \alias{DALI} (right) compared to those between their respective BBH baseline models (left). Results are shown for the $(Q,\chiBHz)$ pair using the same conventions and sample as \cref{fig:as_nsbh_model_mismatches}. The range of the colorbars has been extended to accommodate the lower mismatches produced by the BBH comparisons.}
		\label{fig:bbh_nsbh_model_mismatches}
	\end{center}
\end{figure}

A further comparison including the underlying BBH models over the same sample (\cref{fig:bbh_nsbh_model_mismatches}) reveals that the mismatch trends between these BBH models are quite similar across the different models, with matches degrading primarily towards more asymmetric binaries and high positive BH spins. These trends are largely inherited by the NSBH models. However, the middle and bottom row of the figure illustrate that the NSBH extensions of \alias{GIOTTO} and \alias{DALI} do introduce significant difference with respect to other models. This is not the case between XHM and SEOB, where mismatches remain essentially unaltered with respect to the BBH comparison. 

The good agreement observed between \alias{XNSBH} and \alias{v5HMROM_NSBH} in these mismatches is consistent with the fact that both models share the tidal phase description and have their tidal amplitudes calibrated to the same NR simulations. Therefore, we can expect the two models to deliver largely equivalent results in most parts of parameter space at current detector sensitivities, as confirmed by our injections studies (see \cref{sec:pe}).

\begin{figure}[htp]
	\begin{center}
		\includegraphics[width=1\linewidth]{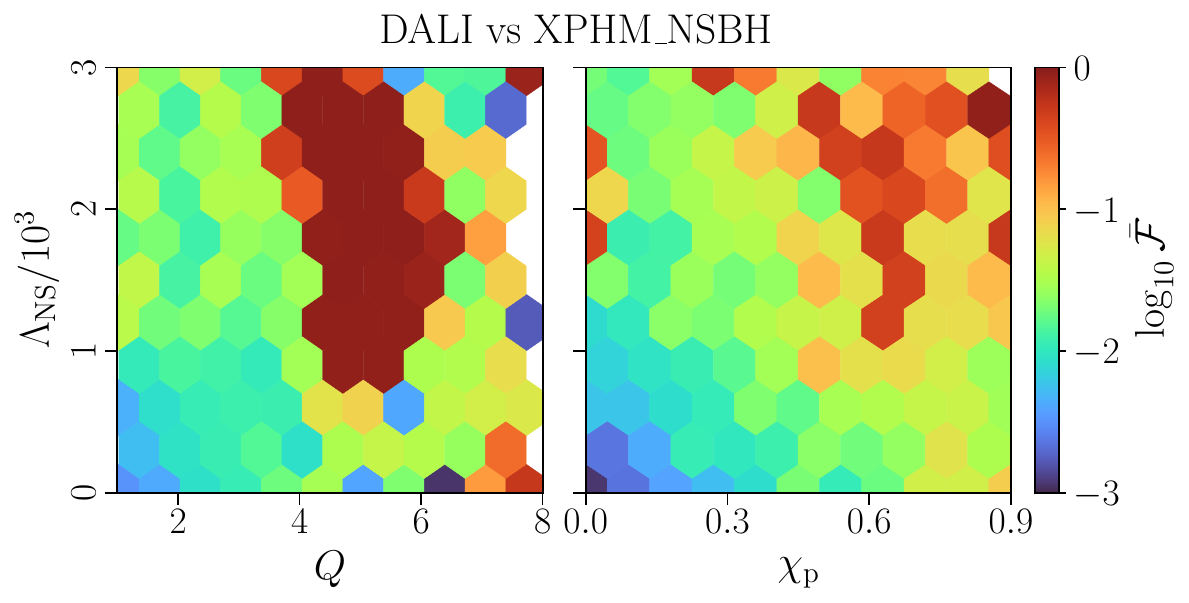}
		\caption{Mismatches between \alias{DALI} and \alias{XPNSBH} shown for the $(Q, \lambdaNS)$ and $(\chip, \lambdaNS)$ pairs using the precessing sample described in the main text.}
		\label{fig:prec_model_mismatches}
	\end{center}
\end{figure}

Finally, in \cref{fig:prec_model_mismatches}, we show mismatches between \alias{XPNSBH} and \alias{DALI} computed over a sample of 5000 precessing NSBH configurations with component masses uniformly sampled in $\MBH \in [3,8]\,\Msun$, $\MNS \in [1,3]\,\Msun$, spins isotropically distributed and magnitudes constrained to $\chiBH \leq 0.9$ and $\chiNS \leq 0.1$, and a uniform prior on the NS tidal deformability $\lambdaNS \in [0,3000]$. With this choice, the explored parameter space is closer to the one currently covered by NR simulations and more representative of astrophysically likely configurations. In this case, the mismatches are shown as a function of the tidal deformability $\lambdaNS$ and mass ratio $Q$ or precessing spin parameter $\chip$. In the left panel, we observe a localized region of very large mismatches, reflecting the same issues already identified for \alias{DALI} in the aligned-spin comparisons. Excluding these outliers, the overall trend is as expected: agreement is best for mildly precessing systems and smaller tidal deformabilities, with degrading values as we move away from these limits. 

\subsection{Timing Tests}\label{subsec:benchmarks}
In this subsection, we compare the evaluation times of the new models against those of their BBH baselines and other FD waveform models for NSBH systems. In particular, we consider \alias{phNSBH} and \alias{v4ROM_NSBH}. We do not include \model{TEOB} in this comparison, as it is a TD model with a substantially higher computational cost, making a direct comparison with FD models unfair.

For this test, we use a random sample of 5000 NSBH configurations with uniform sampling in mass ratios $Q \in [1,20]$, NS masses $\MNS \in [1,3]\,M_\odot$, spin magnitudes $\chiBH \leq 0.9$ and $\chiNS \leq 0.7$, and tidal deformabilities of the NS $\lambdaNS \leq 5000$. The spin components are obtained by isotropically distributing the spin directions for precessing models, and by randomly choosing the sign of the aligned component for aligned-spin models. Finally, we adopt uniform priors with standard bounds on other source parameters (e.g., polarization angle, reference phase, inclination), without averaging over them.

We then evaluate\footnote{The waveform calls are performed through the \texttt{SimInspiralFD} interface of LALSimulation and recorded on a 12-core Apple M3 Pro processor. In order to better reflect evaluation times in PE runs, OpenMP parallelism is disabled.} the different waveform models over each of these configurations, measuring the evaluation time. Each model is called with its default mode content over a frequency range of $20\text{--}4096\ \mathrm{Hz}$ with a frequency resolution $\Delta f = 1/128\ \mathrm{Hz}$, whose inverse corresponds to the typical duration of low mass signals in current detectors. Additionally, to enable direct comparison with previous models not including HMs, we also report the evaluation times of the new models when including only the $(2,2)$ mode.

\begin{figure*}[htp]
	\begin{center}
		\includegraphics[width=1\linewidth]{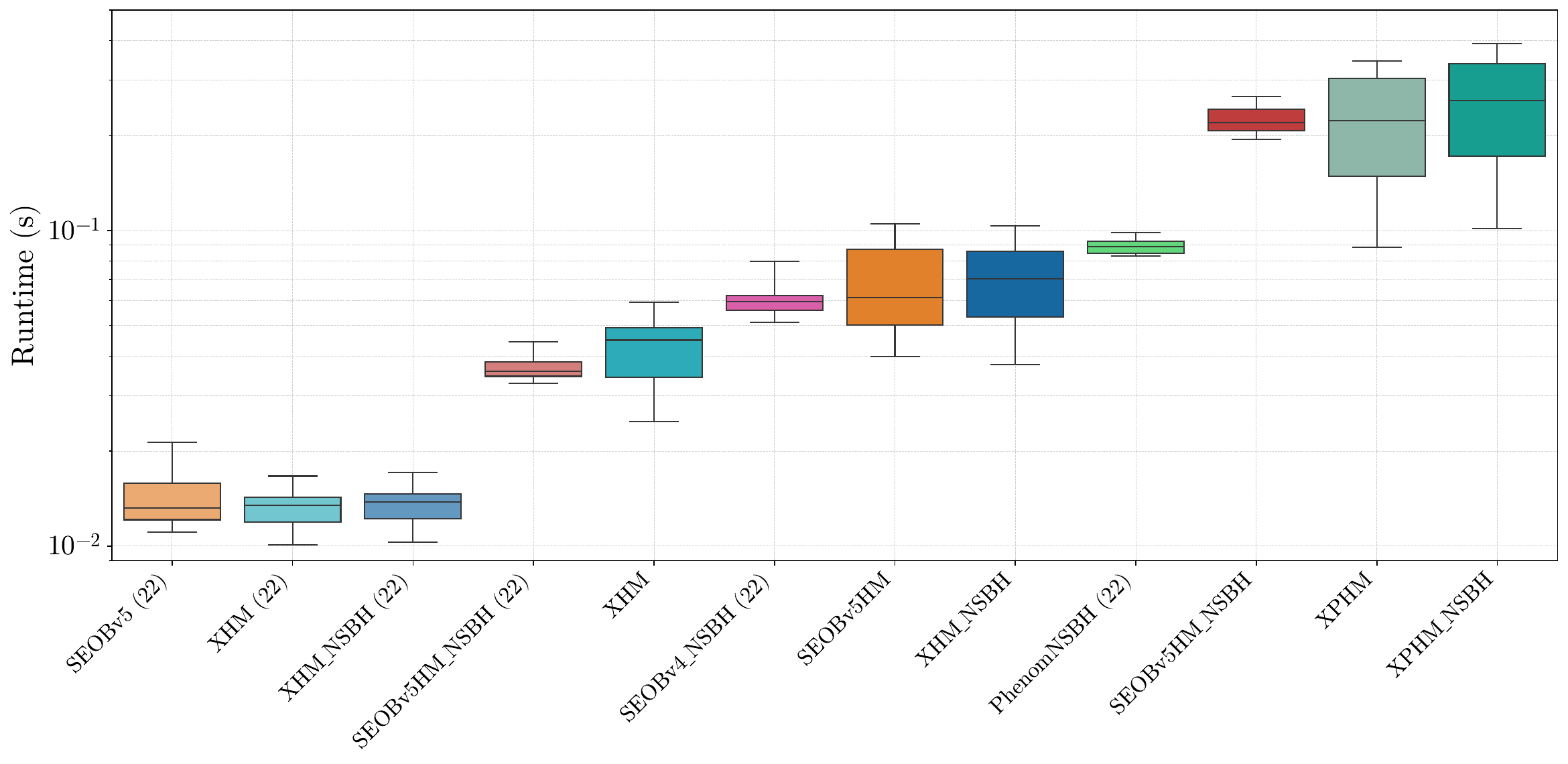}
		\caption{Evaluation times of \model{XNSBH}, \model{v5HMROM_NSBH} and \model{XPNSBH} compared to their BBH baselines and the FD models \model{phNSBH} and \model{v4ROM_NSBH}. The plot shows the distribution of evaluation times for each model across a sample of 5000 randomly generated NSBH configurations using a box to indicate the interquartile range, with a horizontal line within it marking the median of the distribution, and whiskers extending to the 1st and 99th percentiles. Models including ``(22)'' in the legend are restricted to the $(\ell, m) = (2,\pm 2)$ modes. All timings are recorded on a 12-core Apple M3 Pro processor.}
		\label{fig:timings}
	\end{center}
\end{figure*}

The results of this test are shown in \cref{fig:timings}. In the plot, we can observe how the full \alias{XNSBH} is comparable in speed with \alias{phNSBH} and \alias{v4ROM_NSBH}, despite these models including only the $(2,2)$ mode, and becoming roughly 6.4 and 4.3 times faster, respectively, when similarly restricted to that mode (with speed-up factors computed as ratios between the median evaluation times). On the other hand, \alias{v5HMROM_NSBH} is roughly 3.7 times slower than \alias{v4ROM_NSBH}, which is reasonable given that the former includes 6 additional modes. When similarly restricted to the $(2,2)$ mode, \alias{v5HMROM_NSBH} is also faster than \alias{phNSBH} and \alias{v4ROM_NSBH}, with speed-up factors of 2.5 and 1.7, respectively.

Compared to their baseline models, \alias{XNSBH} and \alias{v5HMROM_NSBH} exhibit a moderate increase in evaluation times due to the additional computational overheads associated with the inclusion of tidal effects. Specifically, \alias{XNSBH} is approximately 1.6 times slower than \alias{XHM}, while \alias{v5HMROM_NSBH} is about 3.6 times slower than \alias{v5HMROM}. Finally, the new precessing model, \alias{XPNSBH}, shows only a moderate increase in runtime of 1.2 times relative to \alias{XPHM}, due to the precessing routines dominating the evaluation time.

\section{Parameter Estimation}\label{sec:pe}
In this section, we present PE studies comparing the NSBH models introduced in this paper with their BBH baselines and \alias{phNSBH}. In particular, we analyse aligned-spin injections of NR signals in \cref{subsec:aligned_spin_injections}, an injection and recovery of a precessing signal using \alias{XPNSBH} in \cref{subsec:precessing_injection}, and the real events GW200105, GW200115, GW230518, and GW230529 in \cref{subsec:pe_real_events}.

The PE studies presented in this section have been carried out using \texttt{Bilby}~\cite{bilby_paper} (v2.6.0) together with \texttt{bilby\_pipe} (v1.6.0), the \texttt{MBGravitational\-Wave\-Transient} likelihood~\cite{Morisaki:2021ngj}, and the \texttt{dynesty} sampler~\cite{dynesty_paper} (v2.1.4). Unless otherwise stated, we employ the \texttt{acceptance-walk}\footnote{In the \texttt{acceptance-walk} method, at each iteration, the length of the Markov chain Monte Carlo (MCMC) chains used to evolve \texttt{nlive} live points is fixed to obtain, on average, \texttt{naccept} accepted proposals.} method with \texttt{nlive = 1000} and \texttt{naccept = 60}, which correspond to the LVK reviewed \texttt{dynesty} settings.

The prior distributions are set as follows, except where explicitly noted. We sample over detector-frame chirp mass and inverse mass ratio using priors uniform in the component masses. Spin magnitudes are assigned uniform priors up to $a_{\mathrm{BH},\mathrm{NS}}=0.9$ (similar to the object-agnostic high-spin priors used by the LVK~\cite{LIGOScientific:2025yae}), with sine priors on the tilt angles and uniform priors on the azimuthal angles. For aligned-spin models, we employ an \texttt{AlignedSpin} prior with the same upper bound on the spin magnitudes. The luminosity distance is sampled using a prior uniform in commoving volume and source-frame time, assuming a flat $\mathrm{\Lambda CDM}$ cosmology with Planck15 parameters~\cite{Planck:2015fie}, while standard priors are adopted for the remaining extrinsic parameters~\cite{Romero-Shaw:2020owr}. For NSBH analyses, we sample the secondary’s tidal deformability from a uniform prior over $\lambdaNS\in[0,5000]$.

In the injection studies presented in \cref{subsec:aligned_spin_injections,subsec:precessing_injection}, the synthetic signals are projected into the H(anford)--L(ivingston)--V(irgo) detector network in the absence of noise, assuming the Advanced LIGO zero-detuned, high-power sensitivity and Advanced Virgo design sensitivity curves as provided in \texttt{Bilby}. The injected signals share a sky location given by $\mathrm{RA} = 5.5\,\mathrm{rad}$ and $\mathrm{DEC} = 0.1\,\mathrm{rad}$, an inclination of the orbital plane with respect to the line of sight of $\iota = \pi/3$, a polarization angle of $\psi = 0\,\mathrm{rad}$, a reference phase of $\phi = 0\,\mathrm{rad}$, and geocentric GPS time of $t_0 \simeq 1239082262.7\,\mathrm{s}$. In the recovery, we set the minimum and maximum frequencies for the likelihood evaluations to $20\,\mathrm{Hz}$ and $2048\,\mathrm{Hz}$, respectively, with a sampling rate of $4096\,\mathrm{Hz}$.

Finally, the computational costs of the PE runs presented in this section are summarized in \cref{table:CPUh} in terms of the total CPU hours needed for each analysis. These values are not only influenced by the single waveform evaluation time discussed in \cref{subsec:benchmarks}, but also by the complexity of the likelihood surface produced by each model and the sampling settings. 

\subsection{Injection study for aligned-spin systems}\label{subsec:aligned_spin_injections}
Following the method described in \cref{subsec:hybrids}, we construct two hybrid signals combining a \model{NRSurTidal} inspiral with two SXS waveforms probing two different regimes in the parameter space of NSBHs, and create the corresponding injection frames using PyCBC~\cite{pycbc}. The chosen simulations are SXS:BHNS:0001, which corresponds to a $Q=6$ non-spinning binary yielding a non-disruptive merger~\cite{Foucart:2013psa}, and SXS:BHNS:0002, a $Q=2$ non-spinning binary resulting in a disruptive merger~\cite{Foucart:2018lhe}. In order for the two injections to yield a comparable SNR, the higher mass system (BHNS:0001) is injected at a higher luminosity distance $D_\mathrm{L} \simeq 126.6$ Mpc, whereas the lighter one is placed at $D_\mathrm{L} \simeq 87.8$ Mpc, resulting in both cases recovering a network matched filter $\mathrm{SNR} \simeq 26$. The mode content of the injected signals comprise the modes $(2,\pm2)$, $(2,\pm1)$, $(3,\pm2)$, $(3,\pm3)$, $(4,\pm4)$ and $(5,\pm5)$, where we explicitly hybridize only the positive $m$ modes and account for the negative ones via equatorial symmetry.

\begin{figure*}[tp]
  \centering
  \subfloat[][SXS:BHNS:0001]{
    \includegraphics[width=0.995\columnwidth]{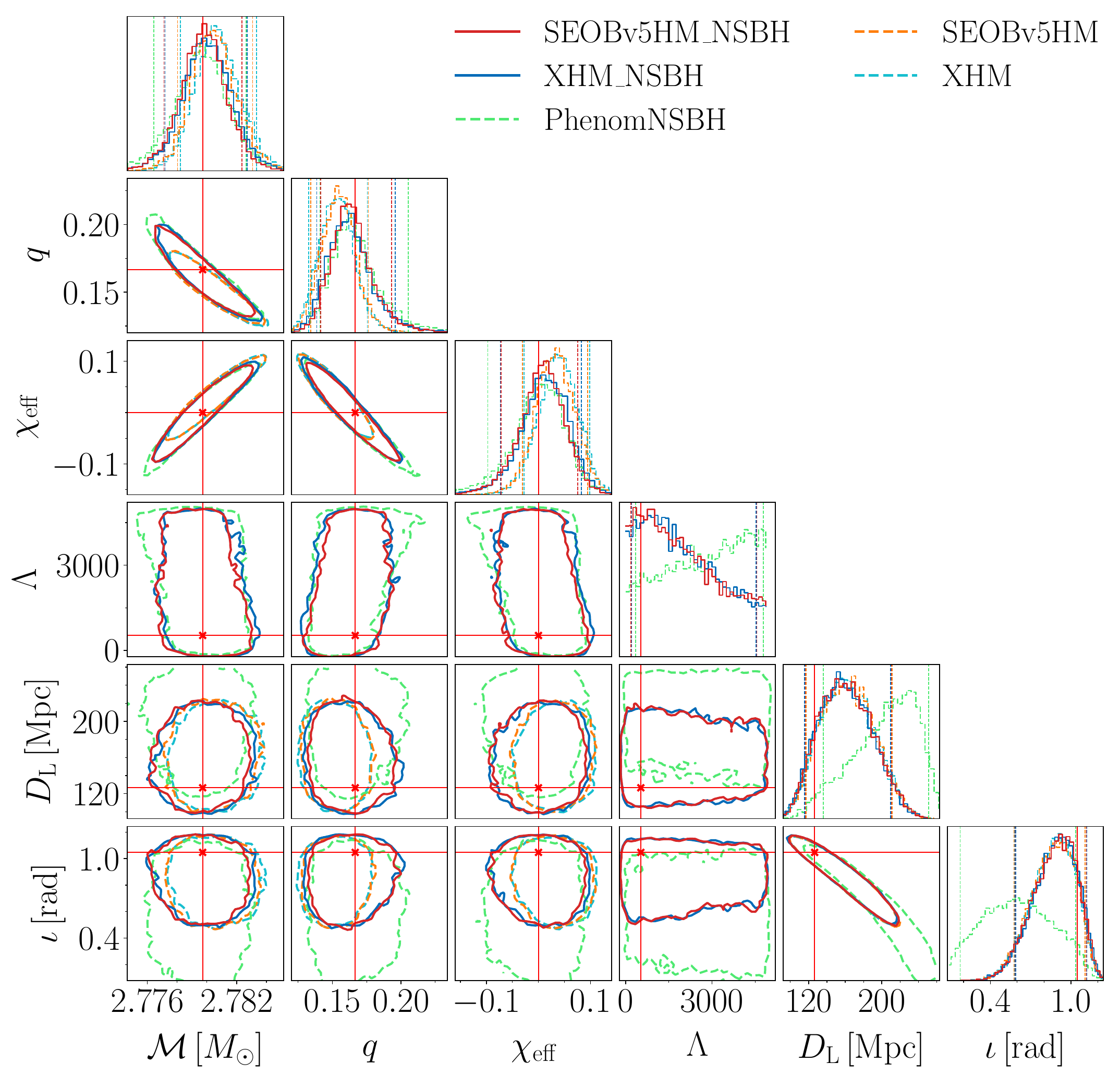}
    \label{fig:pe_sxs_0001}
  }
  \hfill
  \subfloat[][SXS:BHNS:0002]{
    \includegraphics[width=0.995\columnwidth]{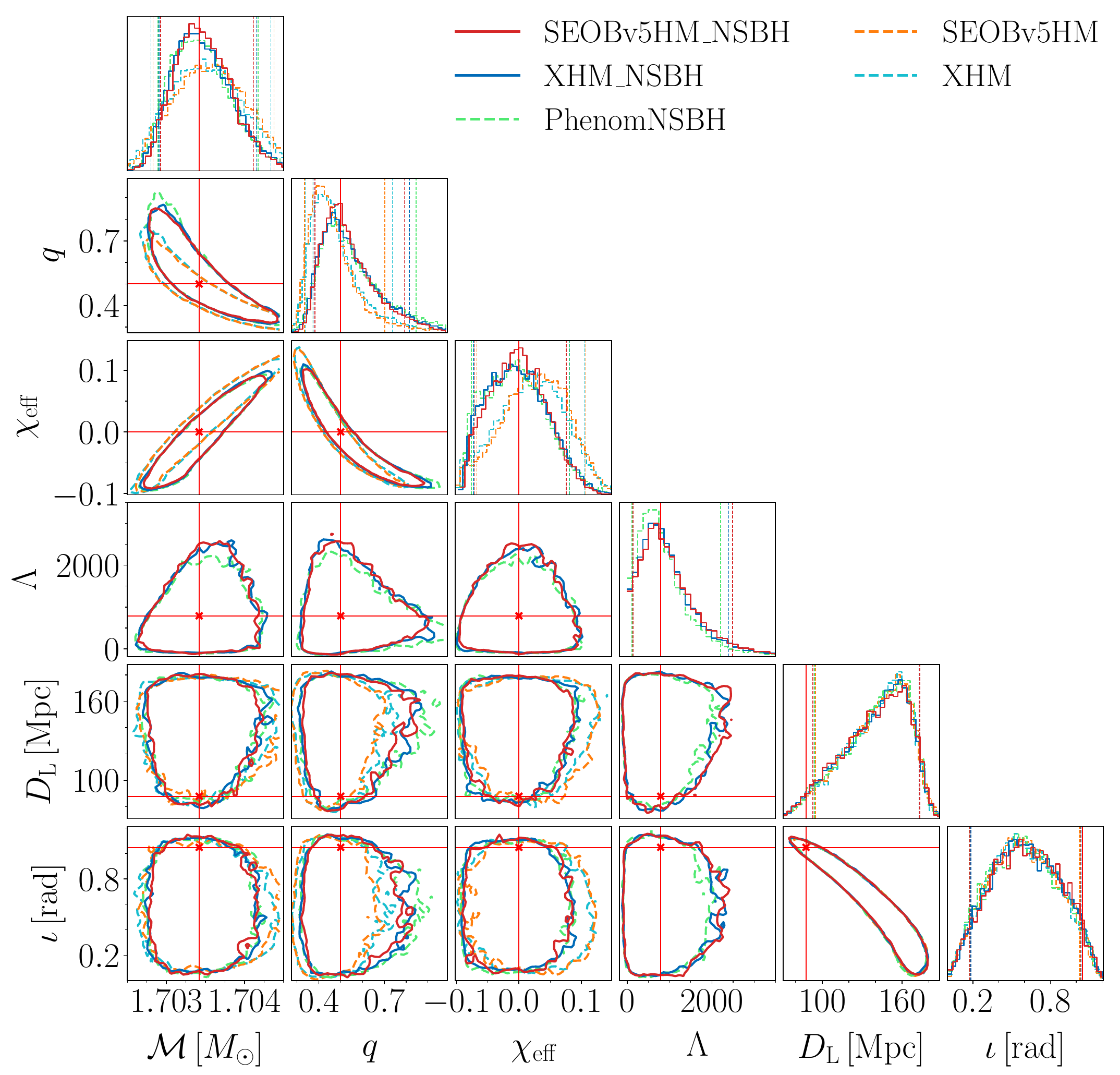}
    \label{fig:pe_sxs_0002}
  }
  \caption{1D and 2D posterior distributions for the injection and recovery study for the (a)~\textmd{SXS:BHNS:0001} and (b)~\textmd{SXS:BHNS:0002} hybrid waveforms. Vertical dashed lines in the 1D panels indicate 90\% credible intervals, while the 2D contours enclose 90\% credible regions, with injected values shown as red solid lines.}
  \label{fig:pe_as_injections}
\end{figure*}

The posterior distributions resulting from the recovery of these injections with different models are displayed in \cref{fig:pe_as_injections}, showing how the new models are very consistent with each other and show some improvements with respect to their BBH models and \alias{phNSBH}.

In the case of SXS:BHNS:0001 (\cref{fig:pe_sxs_0001}), where the contribution of the higher harmonics to the total SNR is higher, \alias{phNSBH} yields biased estimates for the luminosity distance and inclination, with the injected values falling outside its $90\%$ credible intervals for these parameters. At the same time, all NSBH models, regardless of the inclusion of HMs, provide a better recovery of the intrinsic parameters, with the BBH models showing biases due to their lack of tidal effects. For example, the mass ratio recovered by \alias{XHM}, $Q = 6.5^{+1.0}_{-0.8}$, exhibits a normalized deviation\footnote{We define the normalized deviation $\delta_{90}\theta$ of a parameter $\theta$ as $\delta_{90}\theta = (\theta_{50} - \theta_\mathrm{inj}) / (\theta_{95} - \theta_{5})$, where $\theta_{50}$ is the median of the posterior distribution for that parameter, $\theta_\mathrm{inj}$ is the injected value, and $\theta_{95}-\theta_5$ is the width of the 90\% credible interval.}
of $\delta_{90}Q \simeq 0.28$ from the injected value, whereas \alias{XNSBH} recovers $Q = 6.2^{+1.0}_{-1.1}$, producing a value of $\delta_{90}Q \simeq -0.08$.
Regarding the tidal deformability, although all NSBH models produce largely uninformative posteriors with broad 90\% credible intervals for this parameter due to matter effects being suppressed by the low mass ratio of this simulation, the posteriors of \alias{XNSBH} and \alias{v5HMROM_NSBH} peak closer to the true value. Despite the improved recovery of the parameters by the NSBH models, the marginal likelihoods show no preference for the inclusion of tides when comparing the models with their BBH baselines, with Bayes factors\footnote{The Bayes factor $\mathcal{B}_{ij}$ between two competing models $\mathcal{H}_i$ and $\mathcal{H}_j$ models is defined as the ratio of Bayesian evidences, $\mathcal{B}_{ij} = Z_i / Z_j$, where each evidence is the likelihood averaged over the prior. It quantifies how much more the data supports model $\mathcal{H}_i$ compared to $\mathcal{H}_j$, automatically penalizing more complex models through the prior volume. We report it here in terms of its decimal logarithm $\log_{10}{\mathcal{B}_{ij}}$.}
$\abs{\log_{10}{\mathcal{B}}} \sim 0.1$. This can be explained by the relatively low SNR of the injected signal, which implies tidal imprints in the waveform are weak. Consequently, models with extra parameters are automatically penalized in the Bayesian evidence, even if tidal effects are physically present. Therefore, the Bayes-factor result should not be interpreted as evidence against tidal effects, but rather as a reflection of the limited information contained in the signal. On the contrary, the inclusion of HMs leads to \alias{XNSBH} and \alias{v5HMROM_NSBH} being strongly preferred over \alias{phNSBH}, with Bayes factors of $\log_{10}{\mathcal{B}} \simeq 3.7$ and $\log_{10}{\mathcal{B}} \simeq 4.0$, respectively.

In contrast, in the case of SXS:BHNS:0002 (\cref{fig:pe_sxs_0002}), the lower mass ratio suppresses the contribution of HMs while enhancing matter effects. As a result, the inclusion of HMs does not lead to significant improvements in the recovery of the source parameters, while differences between BBH and NSBH models are still pronounced. Using again the mass ratio as an example, \alias{v5HMROM} recovers $Q = 2.25^{+0.74}_{-0.83}$, whereas \alias{v5HMROM_NSBH} recovers $Q = 1.95^{+0.67}_{-0.69}$, with normalized deviations of $\delta_{90}Q \simeq 0.16$ and $\delta_{90}Q \simeq -0.04$, respectively. Regarding the tidal deformability, in this case the NSBH models manage to recover informative posteriors that peak near the true value of the simulation ($\lambdaNS = 791$), with median values of $\lambdaNS = 838^{+1557}_{-703}$ for \alias{XNSBH}, $\lambdaNS = 870^{+1619}_{-734}$ for \alias{v5HMROM_NSBH} and $\lambdaNS = 738^{+1466}_{-622}$ for \alias{phNSBH}. Despite the above, no preference is found for the inclusion of tides when comparing NSBH models and their BBH baselines, nor for the inclusion of HMs when comparing \alias{XNSBH} and \alias{v5HMROM_NSBH} with \alias{phNSBH}, with Bayes factors $\abs{\log_{10}{\mathcal{B}}} \sim 0.1$ in both cases. Finally, all models yield biased estimates for the inclination and luminosity distance, with the injected values lying outside all 90\% credible intervals. These biases have been seen in previous studies \cite{Abac:2023ujg,Abac:2025brd}, and found to progressively diminish as the signal is injected at higher SNRs.

\subsection{Injection study for a precessing system}\label{subsec:precessing_injection}
In this subsection, we inject a synthetic signal generated with \alias{XPNSBH} and perform PE using the same model and \alias{XPHM}. The source parameters are specified at a reference frequency of $f_\mathrm{ref} = 20\,\mathrm{Hz}$ corresponding to component masses $\MBH = 5 \Msun$ and $\MNS = 1.4 \Msun$, dimensionless spins $\chiBHvec = (0.2,0.2,0.2)$ and $\chiNSvec = (0.0,0.0,0.02)$, and a NS tidal deformability of $\lambdaNS = 400$. The extrinsic parameters (sky location, orientation angles, and coalescence time) are set to the common values specified above, with the system being placed at a luminosity distance $D_\mathrm{L} \simeq 178.8\,\mathrm{Mpc}$, yielding a network matched filter $\mathrm{SNR} \simeq 18$. 

The prior and sampling settings are otherwise the same as those described at the beginning of the section, including the number of live points, $\texttt{nlive}=1000$, which was found to provide adequate sampling convergence for this self-recovery injection in zero-noise.
The only difference concerns the luminosity distance prior, which was set in this case to the uniform-in-Euclidean-volume prior used in older LVK publications~\cite{LIGOScientific:2018mvr,LIGOScientific:2020ibl}. This difference in prior is not expected to affect the intrinsic parameters in the detector frame.

\begin{figure}[tp]
  \centering
  \includegraphics[width=\columnwidth]{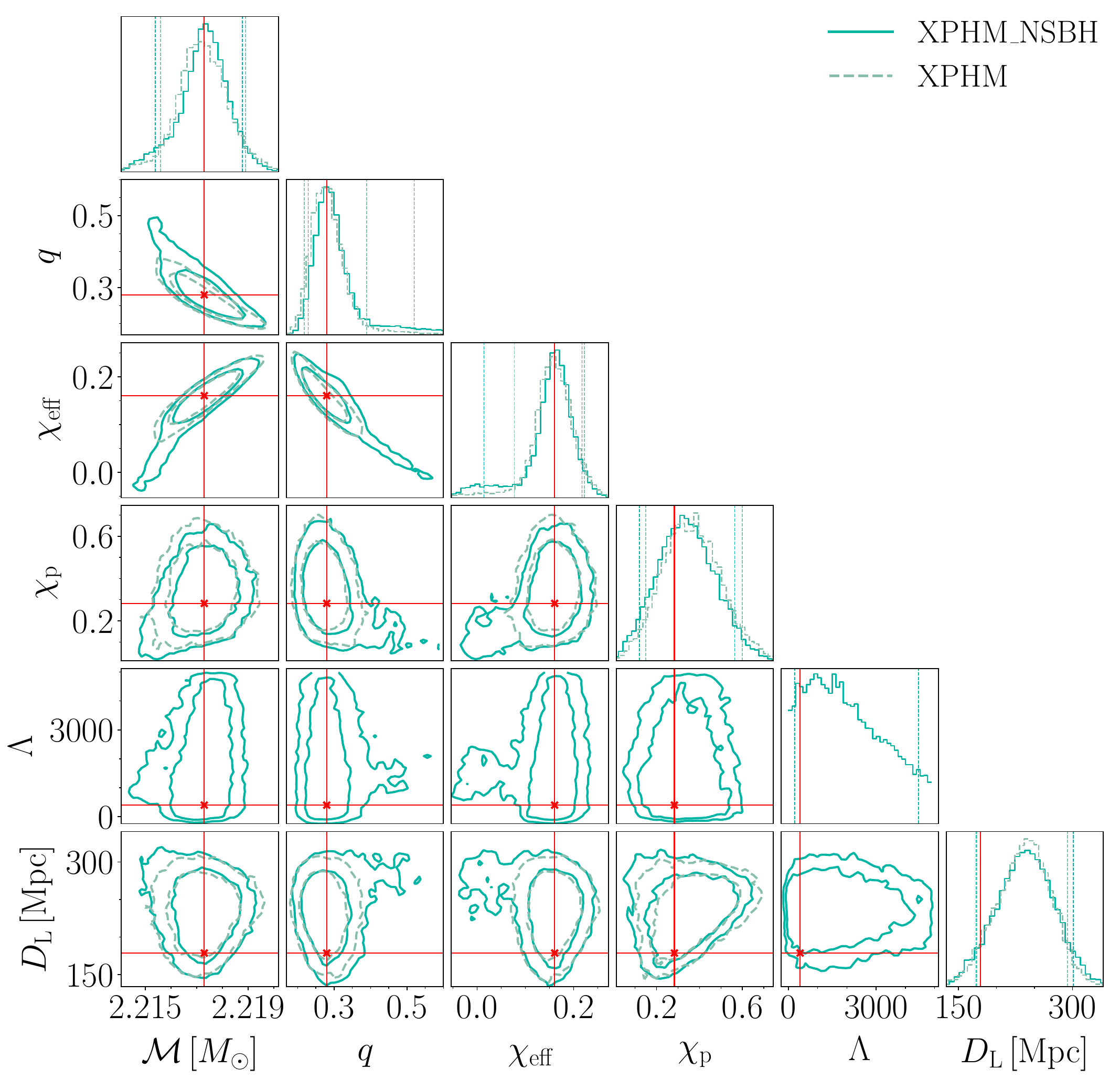}
	\caption{1D and 2D posterior distributions for the injection and recovery study of a signal simulated with \alias{XPNSBH}, including the recovery with \alias{XPHM}. Vertical dashed lines in the 1D panels indicate 90\% credible intervals, while the 2D contours enclose 90\% and 68\% credible regions, with injected values shown as red solid lines.}
  \label{fig:pe_prec_injection}
\end{figure}

The results of this injection and recovery study are shown in \cref{fig:pe_prec_injection}, where we see how both models yield consistent posterior distributions that peak close to the injected values except for the luminosity distance. Relative to \alias{XPHM}, \alias{XPNSBH} exhibits increased support for lower values of the mass ratio and effective spin across its 90\% credible intervals. For \alias{XPHM}, we obtain $Q = 3.57^{+1.00}_{-1.00}$ and $\chieff = 0.158^{+0.064}_{-0.081}$, while \alias{XPNSBH} yields $Q = 3.43^{+0.94}_{-1.50}$ and $\chieff = 0.159^{+0.058}_{-0.145}$. Consequently, the median mass ratio recovered by \alias{XPNSBH} deviates more from the injected values $Q = 3.57$ and $\chieff = 0.161$. The recovery of the tidal deformability with \alias{XPNSBH} is largely consistent with that obtained for the SXS:BHNS:0001 aligned-spin injection shown in \cref{fig:pe_sxs_0001}, with a posterior distribution that leans towards the true value despite its broad support. At this SNR, no preference if found for the inclusion of tidal effects, with a Bayes factor between \alias{XPNSBH} and \alias{XPHM} of $\log_{10}{\mathcal{B}} \simeq -0.09$.

\subsection{Parameter estimation of real events}\label{subsec:pe_real_events}
In this subsection we present a re-analysis of the NSBH candidate events GW200105 (GW200105\_162426), GW200115 (GW200115\_042309), GW230518 (GW230518\_125908), and GW230529 (GW230529\_181500) using the NSBH models introduced in this paper, their respective BBH baselines, and \alias{phNSBH}, as done in the injection studies of \cref{subsec:aligned_spin_injections}.

The astrophysical context and general properties of these systems are discussed in \cref{sec:intro} and in their discovery and catalogue papers~\cite{LIGOScientific:2021qlt,KAGRA:2021vkt,LIGOScientific:2025slb,LIGOScientific:2024elc}. All four events have moderate network SNRs (${\sim}\,11$--14), spins consistent with zero, and no conclusive evidence for spin-precession. GW200105, GW200115, and GW230518 probe more asymmetric mass-ratio configurations ($Q \sim 4$--6), where HMs may carry nonnegligible power, but tidal effects are suppressed, whereas the lower mass ratio of GW230529 ($Q \sim 2.6$) suppresses HMs but increases the potential imprint of tidal interactions. At these SNRs, however, statistical uncertainties dominate over the contribution of either effect, so we expect all waveform models to yield broadly consistent posteriors and only weak constraints on the tidal deformability.

In all cases, we use the latest publicly available data frames from the Gravitational Wave Open Science Center (GWOSC)~\cite{KAGRA:2021vkt,LIGOScientific:2025slb} and analyse the events using the sampler settings detailed at the beginning of this section. Exceptions are made for GW200105 and GW200115, where we adopt the uniform-in-Euclidean-volume prior for the luminosity distance employed in the original LVK analyses prior to a cosmological reweighing of the samples~\cite{KAGRA:2021vkt}. Additionally, for the precessing models \alias{XPHM} and \alias{XPNSBH}, we increase \texttt{nlive} to 2000 for a more conservative sampling of the higher-dimensional parameter space.

\begin{figure*}[tp]
  \centering

  \subfloat[][GW200105]{
    \includegraphics[width=0.995\columnwidth]{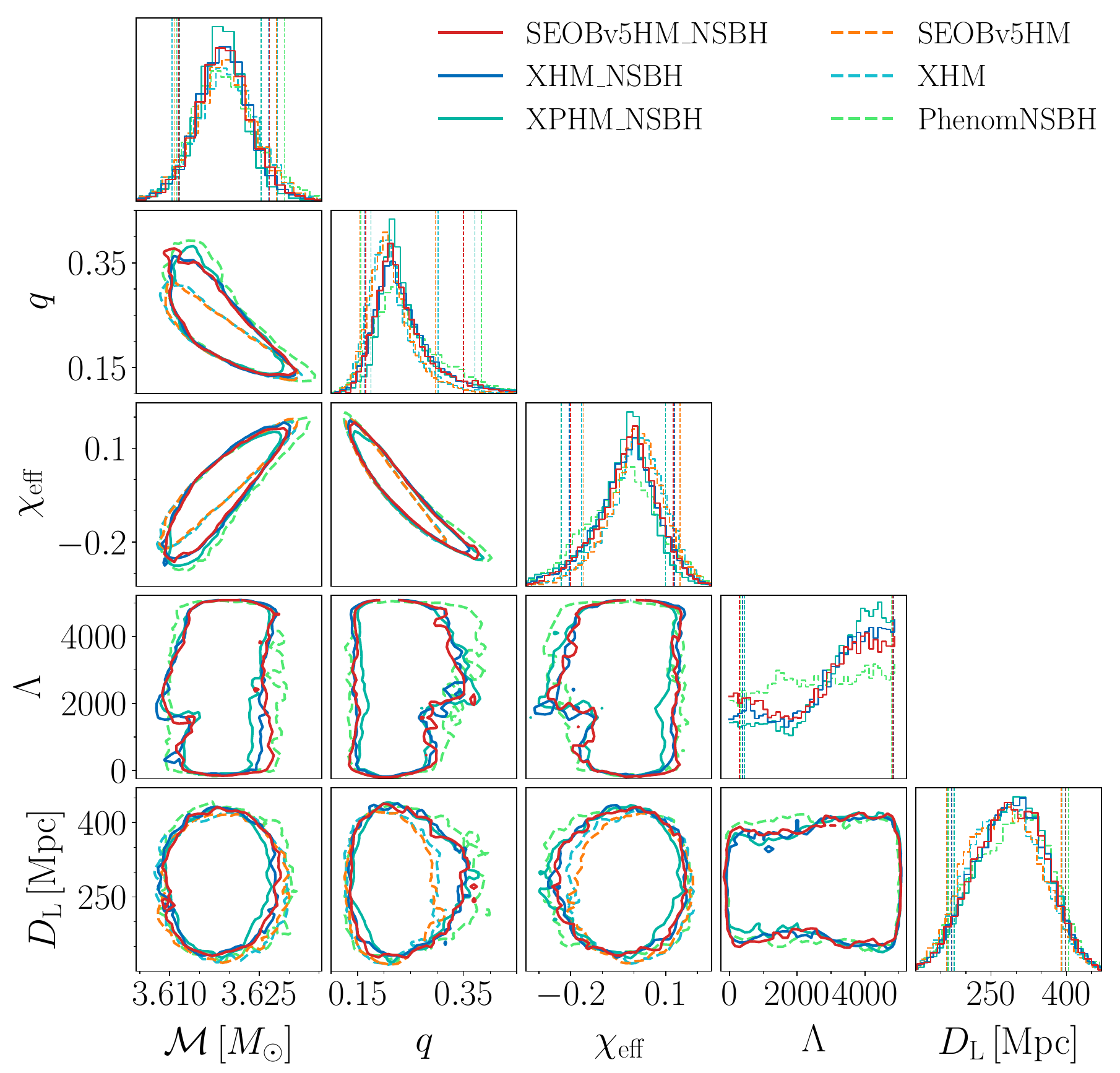}
    \label{fig:gw200105}
  }
  \hfill
  \subfloat[][GW200115]{
    \includegraphics[width=0.995\columnwidth]{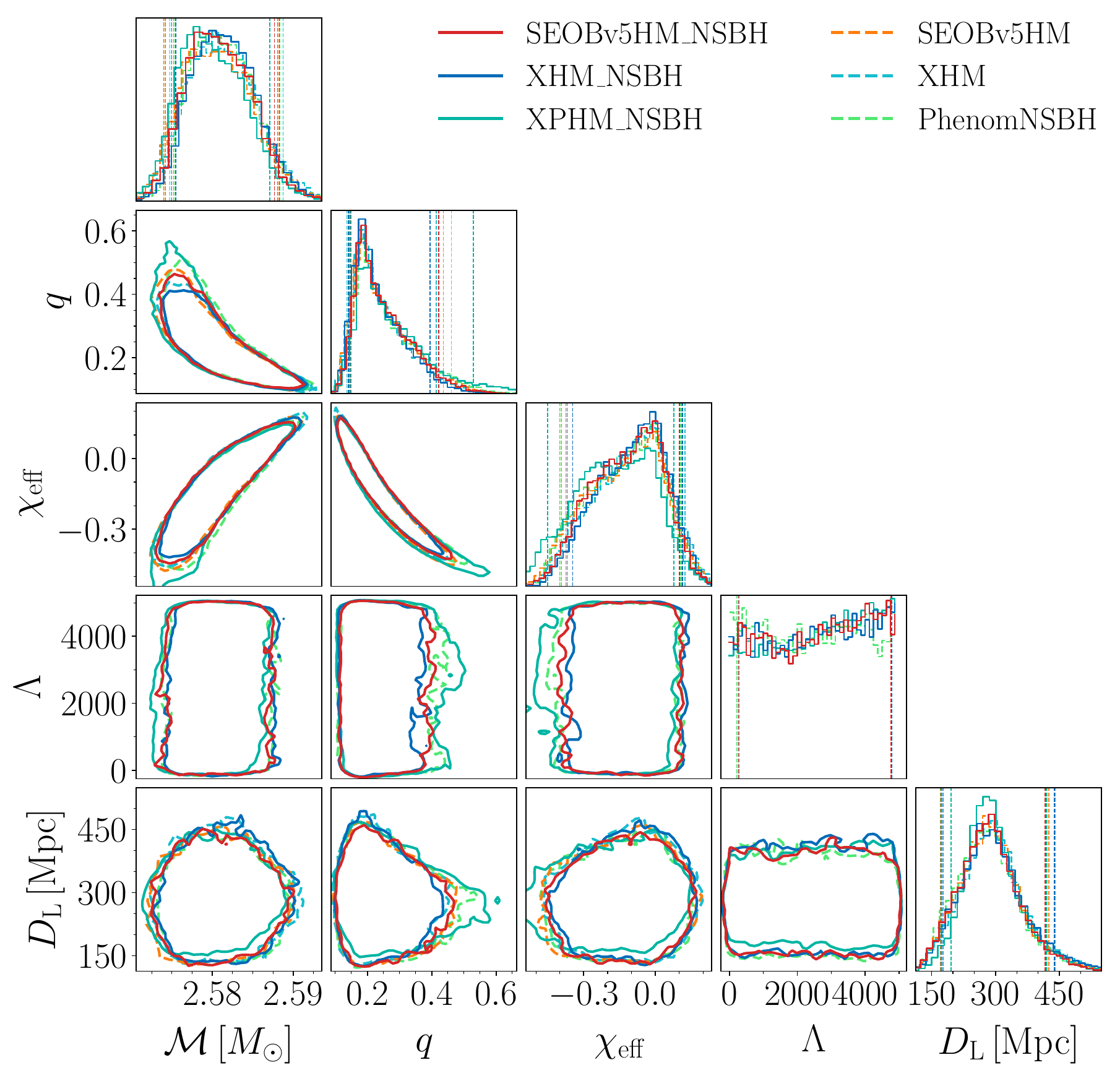}
    \label{fig:gw200115}
  }

  \subfloat[][GW230518]{
    \includegraphics[width=0.995\columnwidth]{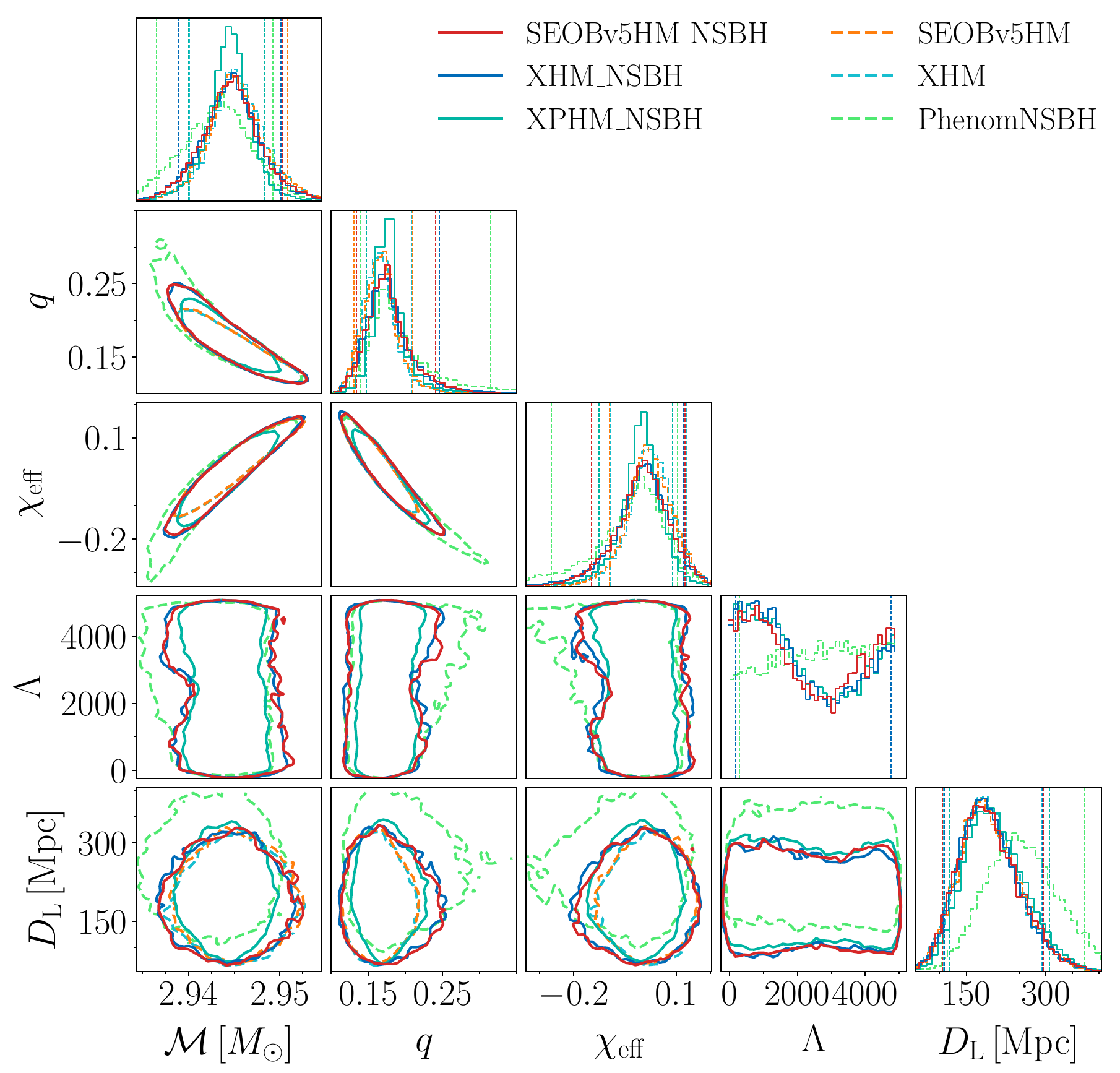}
    \label{fig:gw230518}
  }
  \hfill
  \subfloat[][GW230529]{
    \includegraphics[width=0.995\columnwidth]{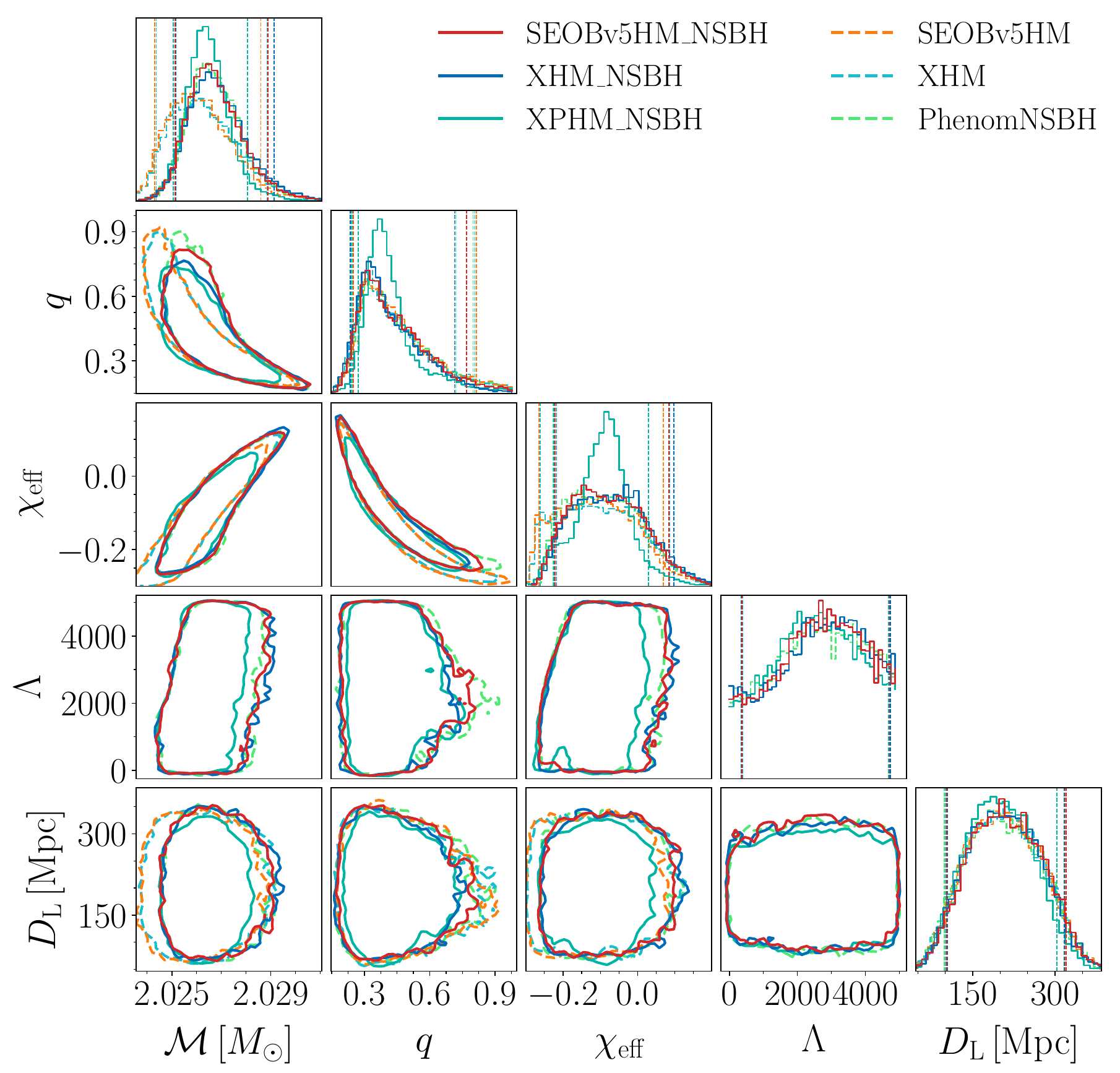}
    \label{fig:gw230529}
  }

  \caption{1D and 2D posterior distributions of the PE for (a)~GW200105, (b)~GW200115, (c)~GW230518, and (d)~GW230529. Vertical dashed lines in the 1D panels mark 90\% credible intervals, whereas 2D contours delimit 90\% credible regions.}
  \label{fig:pe_real_events}
\end{figure*}

The obtained posterior distributions are summarized in the corner plots shown in \cref{fig:pe_real_events}. Overall, the inferred parameters are broadly consistent across all waveform models, particularly between \alias{XNSBH} and \alias{v5HMROM_NSBH}, and are in agreement with previous analyses of these events~\cite{KAGRA:2021vkt,LIGOScientific:2025slb}. Consistent with those results, tidal models do not yield meaningful constraints on the tidal deformability parameter owing to the moderate SNRs of these signals.

Despite this general consistency, some differences can be observed when subdominant effects are included in the waveform models.

In the case of GW200105 (\cref{fig:gw200105}), the inclusion of tides leads to a moderate shift and broadening of the mass ratio and effective spin posteriors, with the largest difference being in the mass ratio changing from $Q = 4.9^{+1.6}_{-1.5}$ (\alias{XHM}) to $Q = 4.4^{+1.7}_{-1.6}$ (\alias{XNSBH}). Additionally, the waveform models introduced in this work show a preference for larger tidal deformabilities over \alias{phNSBH}, with median values of $\lambdaNS = 3300^{+1500}_{-2900}$ for \alias{XNSBH} and $\lambdaNS = 3200^{+1700}_{-2900}$ for \alias{v5HMROM_NSBH}, compared to $\lambdaNS = 2600^{+2100}_{-2300}$ for \alias{phNSBH}.

For GW200115 (\cref{fig:gw200115}), all waveform models produce very similar posterior distributions, with no significant differences observed between NSBH and BBH models or between models with and without HMs.

The situation is different for GW230518 (\cref{fig:gw230518}), with noticeable shifts in both the chirp mass ($\mathcal{M} = 2.9433^{+0.0060}_{-0.0068}\,\Msun$ for \alias{phNSBH} versus $\mathcal{M} = 2.9447^{+0.0055}_{-0.0057}\,\Msun$ for \alias{XNSBH}) and inferred luminosity distance ($D_\mathrm{L} = 250^{+120}_{-100}\,\mathrm{Mpc}$ for \alias{phNSBH} versus $D_\mathrm{L} = 190^{+110}_{-80}\,\mathrm{Mpc}$ for \alias{XNSBH}). Some smaller differences of the same character as those seen in the SXS:BHNS:0001 injection of \cref{subsec:aligned_spin_injections} are also observed when comparing NSBH models with their BBH baselines for this event.

Finally, the posterior distributions of GW230529 (\cref{fig:gw230529}) exhibit a shift in the chirp mass parameter between BBH and NSBH models, with \alias{XHM} recovering $\mathcal{M} = 2.0262^{+0.0027}_{-0.0018}\,\Msun$ and \alias{XNSBH} recovering $\mathcal{M} = 2.0267^{+0.0024}_{-0.0016}\,\Msun$, while the various NSBH models yield mutually consistent results.

Among these four events, the precessing model \alias{XPNSBH} stands out by having more pronounced peaks on its 1D posterior distributions for GW230518 and GW230529. However, these peaks are similar to those produced by \alias{XPHM} on previous analyses of these events~\cite{LIGOScientific:2025slb}.

Looking at the Bayes factors between the different models, no significant preference is found across the four events, except for \alias{XNSBH} and \alias{v5HMROM_NSBH} being preferred over \alias{phNSBH} for GW230518, with Bayes factors of $\log_{10}{\mathcal{B}} \simeq 2.3$ and $\log_{10}{\mathcal{B}} \simeq 2.4$, respectively.

\begin{table*}
  \caption{Computational cost, measured in CPU hours, of the PE runs of \cref{subsec:pe_real_events}. Values with an asterisk indicate runs using an increased number of live points, and thus not directly comparable to the rest. The runs were performed on either AMD EPYC 7H12 (2.6GHz) or AMD EPYC 9754 (2.25GHz) CPUs.}
  \label{table:CPUh}
  \centering
  \scriptsize
  \begin{tblr}{
    width   = \textwidth,
    colspec = {l | *{4}{X[c,si={table-format=4.0}]} | *{2}{X[c,si={table-format=4.0}]} | X[1.5,c,si={table-format=4.0}]},
    rowsep  = 0.3mm,
    row{1}  = {font=\bfseries, belowsep=0.5mm},
    row{2}  = {abovesep=1mm},
    row{6}  = {belowsep=0.5mm},
  }
    \SetCell{c} \text{Model} & \text{GW200105} & \text{GW200115} & \text{GW230529} & \text{GW230518} & \text{SXS1} & \text{SXS2} & \text{Prec. Injection} \\
    \midrule
    \alias{XPHM}         & \text{---} & \text{---} & \text{---} & \text{---} & \text{---} & \text{---} & 4369       \\
    \alias{XPNSBH}       & 8365*      & 5826*      & 4235*      & 9259*      & \text{---} & \text{---} & 6104       \\
    \alias{v5HMROM}      & 2210       & 5628       & 1643       & 2786       & 4523       & 4403       & \text{---} \\
    \alias{v5HMROM_NSBH} & 2268       & 2633       & 1752       & 3373       & 3076       & 3541       & \text{---} \\
    \alias{XHM}          & 836        & 781        & 267        & 2078       & 766        & 804        & \text{---} \\
    \alias{XNSBH}        & 1020       & 900        & 324        & 1058       & 966        & 1041       & \text{---} \\
    \alias{phNSBH}       & 450        & 478        & 132        & 440        & 348        & 431        & \text{---} \\
\end{tblr}
\end{table*}

\section{Conclusions}\label{sec:conclusions}
In this work, we have presented the first gravitational waveform models native to the FD for aligned-spin quasi-circular NSBH binaries including HMs and tidal effects: the phenomenological model \model{XNSBH} and the effective-one-body model \model{v5HMROM_NSBH}. We have also introduced the extension of the former to the case of precessing binaries, \model{XPNSBH}.

These models have been constructed by augmenting existing BBH waveform models with the \model{NRT3} tidal phase model calibrated to BNS simulations, and two new amplitude models tuned to NSBH simulations, including HMs. The inclusion of HMs and spin-precession is expected to be particularly important for NSBH systems, as these effects become more prominent in configurations with asymmetric masses, as are expected of these binaries. On the other hand, the calibrated amplitudes allow capturing the large tidal effects that take place in disruptive configurations, the accurate modelling of which is crucial for reliable PE and astrophysical interpretations. The combination of these effects in computationally efficient models enables fully consistent PE of GW signals rather than resorting to models missing some of these effects.

Comparison with hybrid NR waveforms shows that the new models including HMs yield lower mismatches and exhibit less variation across different choices of extrinsic parameters than previous models restricted to the dominant $(2,2)$ mode. The new models also show reasonable agreement with each other across a wide region of the parameter space, with differences largely inherited from their respective BBH baselines. Timing tests display the computational efficiency of the models, with \model{XNSBH} being comparable in speed with \model{phNSBH} despite the inclusion of HMs, and \model{v5HMROM_NSBH} exhibiting a moderate increase in computational cost compared to \model{v4ROM_NSBH} given the inclusion of HMs.

Parameter estimation studies using both injections and real GW events demonstrate the practical relevance of including HMs and tidal effects for NSBH signals in the context of current detectors. For hybrid aligned-spin NR injections at $\mathrm{SNR}{\;\sim\;}26$, the new models recover the injected parameters more accurately than both BBH models and NSBH models restricted to the dominant harmonic mode, with very good mutual agreement. For the analysis of real signals, we obtain results that are consistent with previous studies, with posteriors that are largely similar across different waveform models due to the low SNRs, but with some noticeable shifts in the posterior distributions when including HMs and/or tidal effects.

The impact of sub-dominant physical effects on PE is expected to become increasingly significant as detector sensitivities improve, enabling the observation of longer and higher-SNR signals. It is therefore crucial to incorporate all relevant physical effects into waveform models in order to avoid systematic biases in our scientific inferences. In particular, beyond refining the effects already included in the models presented in this work, future efforts could aim to incorporate orbital eccentricity. Eccentricity has been neglected in numerous analyses of GW events under the assumption that it is efficiently radiated away before binaries enter the sensitive band of ground-based detectors. However, several astrophysical formation channel, such as dynamical interactions in dense stellar environments~\cite{Sedda:2020wzl} or hierarchical triples in the field~\cite{Stegmann:2025clo, Dorozsmai:2025jlu}, could produce systems retaining measurable eccentricity in the tens-of-Hz range. There have been indications that GW200105\_162426 might have been emitted by an eccentric inspiralling binary~\cite{Morras:2025xfu, Jan:2025fps, Kacanja:2025kpr, Planas:2025plq}. If confirmed, such events could point to a population of eccentric mergers involving neutron stars. Hence, further generalisations of existing waveform models will be extremely useful to characterise the properties and potential origin of mergers detected by current and future instruments. 

\section{Acknowledgements}\label{sec:acknowledgements}
The authors would like to thank Quentin Henry, Joan Llobera-Queról, Eleanor Hamilton and Antoni Ramos-Baudes for useful discussions. A special thanks goes to Rossella Gamba for her assistance with the TEOBResumS interface, and to Jonathan Thompson for his comments on the draft as the PNP reviewer within the LIGO Collaboration.

This work was supported by the Universitat de les Illes Balears (UIB); the Spanish Agencia Estatal de Investigación grants PID2022-138626NB-I00, RED2024-153978-E, RED2024-153735-E, funded by MICIU/AEI/10.13039/501100011033 and the ERDF/EU; and the Comunitat Autònoma de les Illes Balears through the Conselleria d'Educació i Universitats with funds from the European Union - NextGenerationEU/PRTR-C17.I1 (SINCO2022/6719) and from the European Union - European Regional Development Fund (ERDF) (SINCO2022/18146).

F.\,A.\,R.\,V.\ is supported through the Conselleria d'Educació i Universitats del Govern de les Illes Balears via an FPI-CAIB doctoral grant (FPI\_2022\_092) with funds from the European Social Fund+ in the framework of the Balearic Islands ESF+ Program 2021-2027.
M.\,C.\ acknowledges support through the Ministry of Education and Universities through the Vicenç Mut program for the incorporation of research personnel into the science, technology and innovation ecosystem of the Balearic Islands, with reference number POSTDOC2024\_52, funded by the Ministry of Education and Universities and by the European Union (ESF+), within the framework of the ESF+ Balearic Islands Program for the period 2021-2027.
T.\,D.\ and A.\,A.\ acknowledge funding from the EU Horizon under ERC Starting Grant, no.\ SMArt-101076369. Views and opinions expressed are however those of the authors only and do not necessarily reflect those of the European Union or the European Research Council. Neither the European Union nor the granting authority can be held responsible for them.
I.\,M.\ and T.\,D.\ gratefully acknowledge support from the Deutsche Forschungsgemeinschaft (DFG) under the project 504148597 (DI 2553/7). A.\,A.\ acknowledges support from the IN2P3 Master Project MAC.

The authors thankfully acknowledge the computer resources at Picasso and the technical support provided by Barcelona Supercomputing Center (BSC) through grant
No.\ AECT-2025-2-0025 from the Red Española Supercomputación (RES). The authors thank the Supercomputing and Bioinnovation Center (SCBI) of the University of Malaga for their provision of computational resources and technical support (www.scbi.uma.es/site).

This research has made use of data or software obtained from the Gravitational Wave Open Science Center (gwosc.org), a service of the LIGO Scientific Collaboration, the Virgo Collaboration, and KAGRA. This material is based upon work supported by NSF's LIGO Laboratory which is a major facility fully funded by the National Science Foundation, as well as the Science and Technology Facilities Council (STFC) of the United Kingdom, the Max-Planck-Society (MPS), and the State of Niedersachsen/Germany for support of the construction of Advanced LIGO and construction and operation of the GEO600 detector. Additional support for Advanced LIGO was provided by the Australian Research Council. Virgo is funded, through the European Gravitational Observatory (EGO), by the French Centre National de Recherche Scientifique (CNRS), the Italian Istituto Nazionale di Fisica Nucleare (INFN) and the Dutch Nikhef, with contributions by institutions from Belgium, Germany, Greece, Hungary, Ireland, Japan, Monaco, Poland, Portugal, Spain. KAGRA is supported by Ministry of Education, Culture, Sports, Science and Technology (MEXT), Japan Society for the Promotion of Science (JSPS) in Japan; National Research Foundation (NRF) and Ministry of Science and ICT (MSIT) in Korea; Academia Sinica (AS) and National Science and Technology Council (NSTC) in Taiwan.

\appendix

\crefalias{section}{appendix}
\crefalias{subsection}{appendix}
\crefalias{subsubsection}{appendix}

\section{\texorpdfstring{Amplitude Model of IMRPhenomXHM{\undr}NSBH}{Amplitude Model of IMRPhenomXHM_NSBH}}\label{app:phenomxnsbh-amplitude}
As introduced in \cref{subsec:amplitudes}, the amplitude model of \model{XNSBH} is constructed by augmenting the amplitude of \model{XHM} with PN tidal corrections over the inspiral and a calibrated suppressing function over the merger-ringdown, with a smooth transition between the two regions. This appendix provides further details on the construction of this amplitude model, including the selection of frequencies for the calibration of the merger-ringdown, the fitting of the amplitudes at those frequencies across the parameter space, and the final expression of the amplitude informed by these fits as a function of frequency. 

\subsection{Collocation Points}\label{app:collocation_points}

For the calibration of the merger–ringdown amplitude, we select five or six log-spaced frequencies constructed from combinations of characteristic frequencies delimiting this regime: the frequency of the MECO, $\fmeco[\lm]$, which approximately marks the end of the inspiral; the frequency at which the NS is expected to be tidally disrupted by its BH companion, $\ftide[\lm]$; the fundamental ringdown frequency of the remnant BH, $\fring[\lm]$; and its associated damping frequency, $\fdamp[\lm]$. The first of these frequencies, $\fmrd[\lm]{1}$, is placed at
\begin{equation}
    \fmrd[\lm]{1} = 0.9\fmeco[\lm]\,,
\end{equation}
where the $0.9$ factor is added to allow for a better connection with the inspiral ansatz on those cases where tidal effects are already significant around $\fmeco[\lm]$. Towards the high-frequency end, the fifth collocation point, $\fmrd[\lm]{5}$, is set as:
\begin{equation}\label{eq:Mf5mrd}
    \fmrd[\lm]{5} = \frac{3}{4}\min(\ftide[\lm], \fcutRD[\lm]) + \frac{1}{4}\fcutRD[\lm]\,,
\end{equation}
where $\fcutRD[\lm] = \fring[\lm] - \fdamp[\lm]$. For a better characterization of the ringdown amplitude, when $\ftide[\lm] > \fcutRD[\lm]$, one extra collocation point, $\fmrd[\lm]{6}$, is placed at the ringdown frequency,
\begin{equation}
    \fmrd[\lm]{6} = \fring[\lm]\,.
\end{equation}

Finally, the remaining collocation points between $\fmrd[\lm]{1}$ and $\fmrd[\lm]{5}$ are given by
\begin{equation}
    \fmrd[\lm]{i} = \fmrd[\lm]{1}\cdot \del{\frac{\fmrd[\lm]{5}}{\fmrd[\lm]{1}}}^{\frac{i-1}{4}}
\end{equation}
for $i \in \{1,\dots,5\}$.

\subsubsection{MECO Frequency}
The MECO frequency used in the expressions above is computed by evaluating a fit of the stationary points of the energy function including tidal contributions provided in Ref.~\cite{Henry:2019xhg} up to 7PN order. The fit for the $(2,2)$ mode is given by
\begin{equation}
    \fmeco[22] = \del{1 - w\xi}\!\fmeco[22,\mathrm{BBH}]\,,
\end{equation}
where $\fmeco[22,\mathrm{BBH}]$ is a phenomenological fit for the MECO frequency of the BBH case as a function of the symmetric mass ratio and aligned spins of the components provided in Ref.~\cite{Cabero:2016ayq}\footnote{In the placing of the collocation points, we assume by construction a non-spinning NS; that is, we evaluate $\fmeco[22,\mathrm{BBH}]$ with $\chiBHz = 0$.}, and the functions $w$ and  $\xi$ are given by
\begin{align}
    w &= \frac{1}{2}\sbr{1 + \tanh\del{a_0 + a_1\eta + a_2 |\chiBHz|}}\,,\\
    \xi &= 1 - \frac{1 - \eta \lambdaNS \del{b_0 + b_1\eta + b_2\eta^2}}{1 + c_0\,\lambdaNS^{c_1}\,\exp\!\del{d_1\chiBHz + d_2\chiBHz^2}}\,,
\end{align}
with parameters:
\begin{equation*}
    \begin{aligned}
    a_{0} &= -8.0458412                 \,, \quad & b_{2} &= 6.3235640\!\times\!10^{-3}\,,\\
    a_{1} &= 50                         \,, \quad & c_{0} &= 3.4417565\!\times\!10^{-8}\,,\\
    a_{2} &= 16.508398\vphantom{10^{-6}}\,, \quad & c_{1} &= 1.5976586\,,                 \\
    b_{0} &= 5.8048225\!\times\!10^{-6} \,, \quad & d_{1} &= 1.1070675\,,                 \\
    b_{1} &= -1.1333823\!\times\!10^{-3}\,, \quad & d_{2} &= 1.0829260\,.                 \\
    \end{aligned}
\end{equation*}
For the HMs, $\fmeco[\lm]$ is obtained by scaling $\fmeco[22]$ according to
\begin{equation}
    \fmeco[\lm] = \frac{m}{2}\fmeco[22]\,.
\end{equation}

\subsubsection{Frequency of Tidal Disruption}
The frequency of the quadrupolar mode associated with the tidal disruption of the NS, denoted as $\ftide[22]$, is estimated as the frequency at which the BH's tidal force acting on the NS becomes comparable in magnitude to its self-gravitating force including relativistic corrections as computed in the Appendix A of Ref.~\cite{Thompson:2020nei}.
For the HMs, in the same way as done with the MECO frequency, $\ftide[\lm]$ is obtained by scaling $\ftide[22]$ according to
\begin{equation}
    \ftide[\lm] = \frac{m}{2}\ftide[22]\,.
\end{equation}

\subsubsection{Ringdown and Damping Frequencies}\label{app:xnsbh_fring_fdamp}
The ringdown frequency, $\fring[\lm]$, and damping frequency, $\fdamp[\lm]$, are computed by evaluating fits that express these quantities as functions of the final BH's mass and spin, where the final mass and spin are obtained from the fits provided in Ref.~\cite{Gonzalez:2022prs}, and the fits for $\fring[\lm]$ and $\fdamp[\lm]$ as functions of the remnant properties are the same ones used in \model{XHM}~\cite{Garcia-Quiros:2020qpx} in order to leave unaltered the BBH limit.

\subsection{Parameter Space Fits}\label{app:parameter-space-fits}
Having defined our scheme for the collocation point frequencies, we perform the calibration by fitting the amplitude of NR simulations at those frequencies as a function of the intrinsic parameters of the system: the mass ratio, $Q$, the dimensionless spin magnitude of the BH, $\chiBHz$, and the tidal deformability of the NS, $\lambdaNS$. However, instead of fitting the amplitudes directly, we fit the amplitude ratio with respect to the BBH model,
\begin{equation}
    \Ymrd[\lm]{i} \equiv \frac{\ANSBH[\lm](\fmrd[\lm]{i})}{\ABBH[\lm](\fmrd[\lm]{i})}\,.
\end{equation}
Using this quantity, we model the deviations from the BBH case instead of the full amplitude, inheriting the accuracy of the underlying model, and allowing us to use a single fit for the merger and ringdown regions. Furthermore, this quantity has the advantage of being bounded between 0 and 1 due to the extra dissipation introduced by the tides, and has a more predictable behaviour in the high mass ratio and low tidal deformability limits, where we expect tidal effects to be negligible, and thus the amplitude of the NSBH waveform to approach that of a BBH waveform.

For the fitting of these amplitude ratios, we use the following functions of the intrinsic parameters
\begin{widetext}
    \begin{equation}\label{eq:PS_ansatz}
        \Yfit[\lm]{i}(Q,\chiBHz,\lambdaNS) = \sbr{1 + \frac{\lambdaNS^{a_3}}{(Q + a_5)^{a_4}} \exp\del{- a_0 + a_1 \chiBHz + a_2 \del{\frac{1-Q}{1+Q}}^2 \chiBHz \abs{\chiBHz}}}^{- a_6}
        \hspace{-1em}\raisebox{-1.2ex}{,}
\end{equation}
\end{widetext}
where $a_0$ and $a_1,\dots,a_6 \geq 0$ are the fitted coefficients\footnote{Due to their degeneracy, only one of $a_0$ or $a_5$ is used at a time, choosing the one providing the best fit.}, which depend on $i$. This phenomenological ansatz was obtained by identifying a common trend for the qualitative dependence on the tidal parameter, and then parametrizing the remaining observed dependencies on the mass ratio and spin, while enforcing the desired analytical limits and a smooth variation with each intrinsic parameter. We also avoided overfitting by keeping a limited number of degrees of freedom.

The fits are obtained through a least-squares minimization using residuals normalized by the square root of the fitting values to prevent the fit from underweighting regions with low value of the amplitude ratio. Furthermore, a term penalizing points above (below) the fit for the preceding (subsequent) collocation point over a grid is added to ensure monotonically decreasing values of $\Yfit[\lm]{i}$, which is necessary to ensure a good reconstruction using the ansatz of \cref{sec:merger-ringdown}.

When producing these fits for the HMs, we weight the SXS simulations with a factor of 4 relative to the BAM simulations, reflecting that the former are longer, cleaner, and with fewer issues than the latter, which make up most of the dataset.

\subsection{Amplitude Ansatz}\label{app:amplitude-ansatz}

The amplitude ansatz is split in two regions, one for the inspiral, and one for the merger-ringdown, connected together via a windowing function that ensures a smooth transition between the two regions. Specifically, the ansatz is given by
\begin{equation}\label{eq:IMR_ansatz}
    \begin{aligned}
        \ANSBH[\lm](\Mf) &= \sbr[0]{1 - W_{\lm}(\Mf)}\Ains^\lm(\Mf)\\ &\qquad + W_{\lm}(\Mf)\Amrd[\lm]{}(\Mf)\,,
    \end{aligned}
\end{equation}
where $\Ains^\lm$ and $\Amrd[\lm]{}$ are, respectively, the amplitudes in the inspiral and merger-ringdown regions, and $W_{\lm}$ is the windowing function that smoothly transitions between the two. In the following subsections we describe in detail these frequencies and component functions.

\subsubsection{Inspiral}
The amplitude in the inspiral region, defined by $\Mf \leq \ftransI[\lm]$, is constructed by adding tidal corrections derived from PN results to the amplitude of \model{XHM}. That is, the amplitude of the inspiral region is given by
\begin{equation}\label{eq:inspiral}
    \Ains^\lm(\Mf) = A_\mathrm{XHM}^\lm(\Mf) + A_\mathrm{TidalPN}^\lm(\Mf)\,,
\end{equation}
where $A_\mathrm{XHM}^\lm$ is the amplitude of \model{XHM}, and $A_\mathrm{TidalPN}^\lm$ are the added tidal corrections. These tidal corrections result from applying the stationary phase approximation (SPA), as described in Appendix E of Ref.~\cite{Garcia-Quiros:2020qpx}, to the TD adiabatic tidal corrections up to 7.5PN order provided in Refs.~\cite{Henry:2022ccf,Dones:2024odv}, using the expression for the time derivative of the PN parameter $x$ provided in the same references. In order to account for tidal effects in the spin-induced sector, the corresponding PN terms are subtracted with coefficients fixed to their BBH values (since these are already included in $A_\mathrm{XHM}^{\lm}$ for the BBH case) and reinstated in their general form as a function of spin-induced parameters.

The corrections specified above depend on a number of tidal Love numbers and spin-induced multipole parameters, all of which are computed from the dimensionless quadrupolar tidal deformability of the NS, $\lambdaNS$, via quasi-universal relations. In particular, we obtain the magnetic tidal love number (for an irrotational fluid) from Ref.~\cite{JimenezForteza:2018rwr}, the octupolar tidal love number from Ref.~\cite{Yagi:2013sva}, the spin-induced quadrupole parameter from Ref.~\cite{Yagi:2013awa}, and the spin-induced octupole parameter from Ref.~\cite{Yagi:2016bkt}.

\subsubsection{Merger-Ringdown}\label{sec:merger-ringdown}
The amplitude of the merger-ringdown region, covering the frequencies $\Mf \geq \ftransM[\lm]$, is constructed by modulating the amplitude of the BBH model with a function $0 \leq S_{\lm}(\Mf) \leq 1$ representing the ratio between the NSBH and BBH amplitudes, i.e.,
\begin{equation}
    \Amrd[\lm]{}(\Mf) = S_{\lm}(\Mf)A_\mathrm{XHM}^\lm(\Mf)\,.
\end{equation}
This function is modelled as an inverse power law of the form
\begin{equation}\label{eq:merger-ringdown}
    S_{\lm}(\Mf) = \sbr[1]{1 + \alpha(\Mf)^{\beta}}^{-\gamma}, 
\end{equation}
where $\alpha$, $\beta$, and $\gamma$ are the mode-dependent free parameters of the ansatz. Out of these parameters, $\alpha$ and $\beta$ are obtained through a least-squares minimization with respect to the collocation points values $\Ymrd[\lm]{i}$ provided by the parameter-space fits of \cref{app:parameter-space-fits}. On the other hand, $\gamma$ is pre-determined for each mode via a direct Nelder-Mead optimization against the same NR dataset used for the calibration of the collocation points in \cref{app:parameter-space-fits}.

Fixing $\gamma$ enables the determination of the remaining parameters through a linear (hence more efficient and stable) least-squares minimization. This fit is performed using a weighted scheme that compensates for the change in objective function introduced by the linearization of the problem. The weighting scheme is further modified by incorporating the equivalent of applying the conditioning function $y \mapsto \sqrt{y}$ before performing the linearized fit, and by including a small additive term that prevents the weights from vanishing. This additional weighting ensures that configurations with very small amplitude ratios are not effectively ignored by the fit and therefore remain properly represented in the optimization.

\subsubsection{Transition Region}
The transition region, defined over the frequencies $\ftransI[\lm] \leq \Mf \leq \ftransM[\lm]$, smoothly blends the inspiral and merger–ringdown regions of the model using the function
\begin{equation}
    W_{\lm}(\Mf) = s\!\del{\dfrac{\Mf - \ftransI[\lm]}{\ftransM[\lm] - \ftransI[\lm]}}\,,
\end{equation}
where $s(x)$ is a smooth step function given by
\begin{equation}\label{eq:smooth-step}
    s(x) =
    \left\{
    \begin{aligned}
        &\,0                                                        &&\text{for}\:\:\: x \leq 0  \\
        &\sbr{1 + \exp\del{\dfrac{1}{x} - \dfrac{1}{1-x}}}^{-1}\!\! &&\text{for}\:\:\: 0 < x < 1 \\
        &\,1                                                        &&\text{for}\:\:\: x \geq 1
    \end{aligned}
    \right..
\end{equation}
This analytic form continuously transitions from zero to unity, ensuring smoothness with vanishing derivatives of all orders at the boundaries.

For the transition frequency $\ftransM[\lm]$, beyond which the amplitude is given entirely by the merger-ringdown contribution, we adopt\footnote{Although the use of \cref{eq:ftransM} generally places the transition region \emph{entirely} before the point where the PN corrections added to the inspiral start to diverge, we also implement a safeguard that enforces $\abs[0]{A_\mathrm{TidalPN}^\lm}/\ABBH[\lm] < 1/2$ at $\ftransM[\lm]$. This ensures the difference between the inspiral and merger-ringodwn regions is not too large over the transition region.}
\begin{equation}\label{eq:ftransM}
    \ftransM[\lm] = \min\!{\del[0]{\fzero[\lm]\!,\:\fmrd[\lm]{1}}}\,,
\end{equation}
where $\fmrd[\lm]{1}$ is the first collocation frequency introduced in \cref{app:collocation_points}. $\fzero[22]$ is the midpoint of the inspiral window of \model{phNSBH} for disruptive mergers with a torus remnant, listed as $\tilde{f}_0$ in Table I of Ref.~\cite{Thompson:2020nei}. For the higher-order modes, we obtain $\fzero[\lm]$ by scaling $\fzero[22]$ according to,
\begin{equation}
    \fzero[\lm] = \frac{m}{2}\fzero[22]\,.
\end{equation}

The transition frequency $\ftransI[\lm]$, below which the amplitude is fully described by the inspiral contribution, is simply set as half the value of $\ftransM[\lm]$, avoiding sharp transitions between the inspiral and merger-ringdown contributions, namely,
\begin{equation}
    \ftransI[\lm] = \frac{1}{2}\ftransM[\lm]\,.
\end{equation} 
\section{\texorpdfstring{Amplitude Model of SEOBNRv5HM{\Undr}ROM{\Undr}NRTidalv3{\Undr}NSBH}{Amplitude Model of SEOBNRv5HM_ROM_NRTidalv3_NSBH}}\label{app:seobv5nsbh-amplitude}
In this appendix, for consistency with the notation used in Ref.~\cite{Matas:2020wab}, the dimensionless gravitational-wave frequency $\Mf$ is denoted simply by $f$.

The amplitude model of \model{v5HMROM_NSBH} is defined as
\begin{equation}
    \ANSBH[\lm] = w^{\lm}_\mathrm{NSBH} \ABBH[\lm]\,,
\end{equation}
where $\ABBH[\lm]$ is the amplitude of \model{v5HMROM}, and $w^{\lm}_\mathrm{NSBH}$ is the introduced amplitude correction. This correction is given for the $(2,2)$-mode by
\begin{equation}\label{eq:EOB_amplitude_correction_22}
    w^{22}_\mathrm{NSBH}(f) = 1 - s_a(f)[1 - w_\mathrm{corr}(f)]\,,
\end{equation}
where $s_a(f)$ is defined in terms of the smooth step function of \cref{eq:smooth-step} as
\begin{equation}
    s_a(f) = s\del{\frac{f - f_1}{f_2-f_1}}\,.
\end{equation}
Here, $f_1$ and $f_2$ denote the transition frequencies, which are specified in terms of a pivot frequency $f_\mathrm{piv}$ as $[f_1, f_2] = [0.9f_\mathrm{piv}, 1.1f_\mathrm{piv}]$. The pivot frequency itself is written in terms of the effective tidal deformability of Ref.~\cite{Wade:2014vqa}\footnote{For an NSBH binary, assuming $\MBH \geq \MNS$, the expression for the effective tidal deformability reduces to \[\tilde{\Lambda} = \frac{16}{13} \frac{1 + 12Q}{(1 + Q)^5}\lambdaNS\,.\]} as
\begin{equation}
    f_\mathrm{piv} = pC\tilde{\Lambda}^{\alpha}\,,
\end{equation}
where $p = 0.9$ is empirically chosen to allow for a smoother transition, while $C = 0.101206$ and $\alpha = -0.314966$ are fitted parameters. The modified form of \cref{eq:EOB_amplitude_correction_22} ensures that the amplitude corrections reduce to the BNS amplitude at low enough frequencies, i.e., $f \le 0.9 f_\mathrm{piv}$.

The amplitude correction function $w_\mathrm{corr}(f)$ takes the same form as in Ref.~\cite{Matas:2020wab},
\begin{equation}
    w_\mathrm{corr}(f) = w^{-}(f; f_0, \sigma) + \epsilon w^{+}(f;f_0, \sigma)\,,
\end{equation}
where
\begin{equation}
    w^{\pm} (f; f_0, \sigma) = \frac{1}{2}\sbr{1 \pm \tanh{\del{4\frac{f-f_0}{\sigma}}}}\,.
\end{equation}
In these expressions, the parameters $f_0$, $\sigma$ and $\epsilon$ are fitted to NR simulations using different functional forms depending on whether the coalescence is classified as disruptive or not. This classification is established by comparing the ringdown and tidal disruption frequencies, $f_\mathrm{RD}$ and $f_\mathrm{tide}$\footnote{As in \model{v4ROM_NSBH}, the tidal disruption frequency is computed as a function of the final, rather than initial, BH mass and spin.}, with disruptive mergers being further classified according to the expected torus remnant mass $M_\mathrm{torus}$. The different cases of this classification are detailed here:
\begin{enumerate}
    \item Nondisruptive (ND) merger: $f_\mathrm{RD} < f_\mathrm{tide}$, and $M_\mathrm{torus} = 0$. In this case, the parameters are set to $f_0 = f_\mathrm{ND}$, $\sigma_0 = \sigma_\mathrm{ND}$, $\epsilon = \epsilon_\mathrm{ND}$, resulting in
    \begin{equation}
        \begin{aligned}
            w_\mathrm{ND}(f) &= w^{-}(f;f_\mathrm{ND}, \sigma_\mathrm{ND}) \\
            &\qquad + \epsilon_\mathrm{ND}w^+(f; f_\mathrm{ND}, \sigma_\mathrm{ND})\,.
        \end{aligned}
    \end{equation}
    
    \item Disruptive (D) merger with a torus remnant: $f_\mathrm{RD} > f_\mathrm{tide}$, and $M_\mathrm{torus} > 0$. In this case, the parameters are set to $f_0 = f_\mathrm{D}$, $\sigma_0 = \sigma_\mathrm{D}$, $\epsilon = 0$, resulting in
    \begin{equation}
        w_\mathrm{D}(f) = w^-(f; f_\mathrm{D}, \sigma_\mathrm{D})\,.
    \end{equation}
    
    \item Mildly disruptive merger with no torus remnant: $f_\mathrm{RD} > f_\mathrm{tide}$, $M_\mathrm{torus} = 0$. In this case, the parameters are set to $f_0 = (1- Q^{-1})f_\mathrm{ND}+Q^{-1}f_\mathrm{tide}$, $\sigma_0 = (\sigma_\mathrm{D}+\sigma_\mathrm{ND})/2$, and $\epsilon = 0$.
    
    \item Mildly disruptive merger with a torus remnant: $f_\mathrm{RD} < f_\mathrm{tide}$, $M_\mathrm{torus} > 0$. Here, the information from cases 1 and 2 is combined by setting $f_0 = f_\mathrm{D}$, $\sigma_0 = \sigma_\mathrm{ND}$, and $\epsilon = \epsilon_\mathrm{ND}$.
\end{enumerate}

The forms of the fitting formulae for $f_\mathrm{D/ND}$, $\sigma_\mathrm{D/ND}$, and $\epsilon_\mathrm{D/ND}$ are discussed in detail in the Appendix of Ref.~\cite{Matas:2020wab}. For completeness, we also provide them here:
\begin{align}
    f_\mathrm{ND} &= f_\mathrm{RD}\,, \label{eq:EOB_D_ND_parameters_first}\\
    \sigma_\mathrm{ND} &= \bar{\sigma}_\mathrm{ND} + 2w^-(x;x_0, \sigma_x)\,,\\
    \epsilon_\mathrm{ND} &= w^+(y; y_0, \sigma_y)\,,\\
    x &= \del[3]{\frac{f_\mathrm{RD}-f_\mathrm{tide}}{f_\mathrm{RD}}}^2 + x_CC_\mathrm{NS} + x_{\chi}\chiBHz\,,\\
    y &= \del[3]{\frac{f_\mathrm{RD}-f_\mathrm{tide}}{f_\mathrm{RD}}}^2 + y_CC_\mathrm{NS} + y_{\chi}\chiBHz\,,\\
    f_\mathrm{D} &= \del[3]{a_0 + a_M\frac{M_{b, \rm torus}}{M_{b, \rm NS}} + a_CC_\mathrm{NS} + a_{\nu}\sqrt{\nu} + a_{\chi}\chi}\!f_\mathrm{tide}\,,\\
    \sigma_\mathrm{D} &= b_0 + b_M\frac{M_{b, \rm torus}}{M_{b, \rm NS}}+ b_CC_\mathrm{NS} + b_{\nu}\sqrt{\nu} + \sum_{k = 1}^3 b_\chi^{(k)}\chi^k \label{eq:EOB_D_ND_parameters_last}\,.
\end{align}
In our case, using \cref{eq:EOB_amplitude_correction_22}, we obtain a new set of values for the free parameters in \crefrange{eq:EOB_D_ND_parameters_first}{eq:EOB_D_ND_parameters_last} by calibrating to the set of NR simulations discussed in \cref{sec:nr_dataset}, resulting in the values:
\begin{center}
    \begin{tblr}{
        width = \columnwidth,
        colspec = {c c S[table-format=1.8,mode=math] l c c c S[table-format=1.8,mode=math] l},
        rowsep  = 0pt,
        row{0-Z} = {ht=5mm, valign=m},
        column{1} = {rightsep=0pt},
        column{3} = {rightsep=0pt},
        column{4} = {leftsep=0.166em},
        column{6} = {rightsep=0pt},
        column{8} = {rightsep=0pt},
        column{9} = {leftsep=0.166em, rightsep=0pt},
    }
        $\bar{\sigma}_\mathrm{ND}$ & = & -1.18567265 &, & & $a_0$             & = &  0.88253874 &, \\
        $x_0$                      & = & -0.28554656 &, & & $a_M$             & = & -3.45459063 &, \\
        $\sigma_x$                 & = &  0.13817884 &, & & $a_C$             & = & -2.63742394 &, \\
        $x_C$                      & = & -1.79416592 &, & & $a_{\nu}$         & = &  1.14154394 &, \\
        $x_{\chi}$                 & = & -0.00347289 &, & & $a_{\chi}$        & = &  0.36136213 &, \\
        $y_0$                      & = &  0.01143411 &, & & $b_0$             & = &  0.30219480 &, \\
        $\sigma_y$                 & = &  1.56110486 &, & & $b_M$             & = & -0.17397504 &, \\
        $y_C$                      & = & -0.91501537 &, & & $b_C$             & = &  0.14752663 &, \\
        $y_{\chi}$                 & = & -0.01035293 &, & & $b_{\nu}$         & = & -0.49273232 &, \\
                                &   &             &  & & $b_{\chi}^{(1)}$  & = & -0.01012444 &, \\
                                &   &             &  & & $b_{\chi}^{(2)}$  & = &  0.02599226 &, \\
                                &   &             &  & & $b_{\chi}^{(3)}$  & = &  0.00391012 &. \\
    \end{tblr}
\end{center}

For the higher-mode corrections, we simply scale the $(2,2)$ mode correction via:
\begin{equation}
    w^{\lm}_\mathrm{NSBH}(f) = w^{22}_\mathrm{NSBH}(a_{\lm} f)\,,
\end{equation}
for $\lm \neq 22$, where $a_{\lm}$ are fitted constants that scale the frequency for each mode (analogous to the scaling used for the phase in \cref{subsec:phases}), taking the values:
\begin{equation*}
    \begin{aligned}
    a_{21} &= 1.11512249\,, \qquad & a_{43} &= 0.84205758\,,\\
    a_{32} &= 0.91153288\,, \qquad & a_{44} &= 0.66971269\,,\\
    a_{33} &= 0.84131887\,, \qquad & a_{55} &= 0.60794726\,.
    \end{aligned}
\end{equation*} 

\bibliography{references}

@ARTICLE{bilby_paper,
       author = {{Ashton}, Gregory and {H{\"u}bner}, Moritz and {Lasky}, Paul D. and {Talbot}, Colm and {Ackley}, Kendall and {Biscoveanu}, Sylvia and {Chu}, Qi and {Divakarla}, Atul and {Easter}, Paul J. and {Goncharov}, Boris and {Hernandez Vivanco}, Francisco and {Harms}, Jan and {Lower}, Marcus E. and {Meadors}, Grant D. and {Melchor}, Denyz and {Payne}, Ethan and {Pitkin}, Matthew D. and {Powell}, Jade and {Sarin}, Nikhil and {Smith}, Rory J.~E. and {Thrane}, Eric},
        title = "{BILBY: A User-friendly Bayesian Inference Library for Gravitational-wave Astronomy}",
      journal = {\apjs},
         year = 2019,
        month = apr,
       volume = {241},
       number = {2},
          eid = {27},
        pages = {27},
          doi = {10.3847/1538-4365/ab06fc},
archivePrefix = {arXiv},
       eprint = {1811.02042},
 primaryClass = {astro-ph.IM},
       adsurl = {https://ui.adsabs.harvard.edu/abs/2019ApJS..241...27A}
}

@article{Romero-Shaw:2020owr,
    author = "Romero-Shaw, I. M. and others",
    title = "{Bayesian inference for compact binary coalescences with bilby: validation and application to the first LIGO{\textendash}Virgo gravitational-wave transient catalogue}",
    eprint = "2006.00714",
    archivePrefix = "arXiv",
    primaryClass = "astro-ph.IM",
    doi = "10.1093/mnras/staa2850",
    journal = "Mon. Not. Roy. Astron. Soc.",
    volume = "499",
    number = "3",
    pages = "3295--3319",
    year = "2020"
}

@ARTICLE{dynesty_paper,
       author = {{Speagle}, Joshua S.},
        title = "{DYNESTY: a dynamic nested sampling package for estimating Bayesian posteriors and evidences}",
      journal = {\mnras},
         year = 2020,
        month = apr,
       volume = {493},
       number = {3},
        pages = {3132-3158},
          doi = {10.1093/mnras/staa278},
archivePrefix = {arXiv},
       eprint = {1904.02180},
 primaryClass = {astro-ph.IM},
       adsurl = {https://ui.adsabs.harvard.edu/abs/2020MNRAS.493.3132S}
}

@misc{SpEC,
  note		= {SpEC - Spectral Einstein Code},
  howpublished = "\url{http://www.black-holes.org/SpEC.html}"
}

@article{Barkett:2015wia,
    author = "Barkett, Kevin and others",
    title = "{Gravitational waveforms for neutron star binaries from binary black hole simulations}",
    eprint = "1509.05782",
    archivePrefix = "arXiv",
    primaryClass = "gr-qc",
    doi = "10.1103/PhysRevD.93.044064",
    journal = "Phys. Rev. D",
    volume = "93",
    number = "4",
    pages = "044064",
    year = "2016"
}

@article{Campanelli:2005dd,
    author = "Campanelli, Manuela and Lousto, C. O. and Marronetti, P. and Zlochower, Y.",
    title = "{Accurate evolutions of orbiting black-hole binaries without excision}",
    eprint = "gr-qc/0511048",
    archivePrefix = "arXiv",
    doi = "10.1103/PhysRevLett.96.111101",
    journal = "Phys. Rev. Lett.",
    volume = "96",
    pages = "111101",
    year = "2006"
}

@article{Bruegmann:2006ulg,
    author = "Bruegmann, Bernd and Gonzalez, Jose A. and Hannam, Mark and Husa, Sascha and Sperhake, Ulrich and Tichy, Wolfgang",
    title = "{Calibration of Moving Puncture Simulations}",
    eprint = "gr-qc/0610128",
    archivePrefix = "arXiv",
    doi = "10.1103/PhysRevD.77.024027",
    journal = "Phys. Rev. D",
    volume = "77",
    pages = "024027",
    year = "2008"
}

@article{Thierfelder:2011yi,
    author = "Thierfelder, Marcus and Bernuzzi, Sebastiano and Bruegmann, Bernd",
    title = "{Numerical relativity simulations of binary neutron stars}",
    eprint = "1104.4751",
    archivePrefix = "arXiv",
    primaryClass = "gr-qc",
    doi = "10.1103/PhysRevD.84.044012",
    journal = "Phys. Rev. D",
    volume = "84",
    pages = "044012",
    year = "2011"
}

@article{Lindblom:2005qh,
    author = "Lindblom, Lee and Scheel, Mark A. and Kidder, Lawrence E. and Owen, Robert and Rinne, Oliver",
    title = "{A New generalized harmonic evolution system}",
    eprint = "gr-qc/0512093",
    archivePrefix = "arXiv",
    doi = "10.1088/0264-9381/23/16/S09",
    journal = "Class. Quant. Grav.",
    volume = "23",
    pages = "S447--S462",
    year = "2006"
}

@ARTICLE{1982ApJ...260..838S,
       author = {{Shapiro}, S.~L. and {Wasserman}, I.},
        title = "{Gravitational radiation from nonspherical infall into black holes}",
      journal = {\apj},
         year = 1982,
        month = sep,
       volume = {260},
        pages = {838-848},
          doi = {10.1086/160302},
       adsurl = {https://ui.adsabs.harvard.edu/abs/1982ApJ...260..838S}
}

@ARTICLE{2010ApJ...725.1918O,
       author = {{{\"O}zel}, Feryal and {Psaltis}, Dimitrios and {Narayan}, Ramesh and {McClintock}, Jeffrey E.},
        title = "{The Black Hole Mass Distribution in the Galaxy}",
      journal = {\apj},
         year = 2010,
        month = dec,
       volume = {725},
       number = {2},
        pages = {1918-1927},
          doi = {10.1088/0004-637X/725/2/1918},
archivePrefix = {arXiv},
       eprint = {1006.2834},
 primaryClass = {astro-ph.GA},
       adsurl = {https://ui.adsabs.harvard.edu/abs/2010ApJ...725.1918O}
}

@ARTICLE{2012ApJ...749...91F,
       author = {{Fryer}, Chris L. and {Belczynski}, Krzysztof and {Wiktorowicz}, Grzegorz and {Dominik}, Michal and {Kalogera}, Vicky and {Holz}, Daniel E.},
        title = "{Compact Remnant Mass Function: Dependence on the Explosion Mechanism and Metallicity}",
      journal = {\apj},
         year = 2012,
        month = apr,
       volume = {749},
       number = {1},
          eid = {91},
        pages = {91},
          doi = {10.1088/0004-637X/749/1/91},
archivePrefix = {arXiv},
       eprint = {1110.1726},
 primaryClass = {astro-ph.SR},
       adsurl = {https://ui.adsabs.harvard.edu/abs/2012ApJ...749...91F}
}

@ARTICLE{2012ApJ...757...91B,
       author = {{Belczynski}, Krzysztof and {Wiktorowicz}, Grzegorz and {Fryer}, Chris L. and {Holz}, Daniel E. and {Kalogera}, Vassiliki},
        title = "{Missing Black Holes Unveil the Supernova Explosion Mechanism}",
      journal = {\apj},
         year = 2012,
        month = sep,
       volume = {757},
       number = {1},
          eid = {91},
        pages = {91},
          doi = {10.1088/0004-637X/757/1/91},
archivePrefix = {arXiv},
       eprint = {1110.1635},
 primaryClass = {astro-ph.GA},
       adsurl = {https://ui.adsabs.harvard.edu/abs/2012ApJ...757...91B}
}

@ARTICLE{2019Sci...366..637T,
       author = {{Thompson}, Todd A. and {Kochanek}, Christopher S. and {Stanek}, Krzysztof Z. and {Badenes}, Carles and {Post}, Richard S. and {Jayasinghe}, Tharindu and {Latham}, David W. and {Bieryla}, Allyson and {Esquerdo}, Gilbert A. and {Berlind}, Perry and {Calkins}, Michael L. and {Tayar}, Jamie and {Lindegren}, Lennart and {Johnson}, Jennifer A. and {Holoien}, Thomas W. -S. and {Auchettl}, Katie and {Covey}, Kevin},
        title = "{A noninteracting low-mass black hole-giant star binary system}",
      journal = {Science},
         year = 2019,
        month = nov,
       volume = {366},
       number = {6465},
        pages = {637-640},
          doi = {10.1126/science.aau4005},
archivePrefix = {arXiv},
       eprint = {1806.02751},
 primaryClass = {astro-ph.HE},
       adsurl = {https://ui.adsabs.harvard.edu/abs/2019Sci...366..637T}
}

@ARTICLE{2021MNRAS.504.2577J,
       author = {{Jayasinghe}, T. and {Stanek}, K.~Z. and {Thompson}, Todd A. and {Kochanek}, C.~S. and {Rowan}, D.~M. and {Vallely}, P.~J. and {Strassmeier}, K.~G. and {Weber}, M. and {Hinkle}, J.~T. and {Hambsch}, F. -J. and {Martin}, D.~V. and {Prieto}, J.~L. and {Pessi}, T. and {Huber}, D. and {Auchettl}, K. and {Lopez}, L.~A. and {Ilyin}, I. and {Badenes}, C. and {Howard}, A.~W. and {Isaacson}, H. and {Murphy}, S.~J.},
        title = "{A unicorn in monoceros: the 3 M$_{{\ensuremath{\odot}}}$ dark companion to the bright, nearby red giant V723 Mon is a non-interacting, mass-gap black hole candidate}",
      journal = {\mnras},
         year = 2021,
        month = jun,
       volume = {504},
       number = {2},
        pages = {2577-2602},
          doi = {10.1093/mnras/stab907},
archivePrefix = {arXiv},
       eprint = {2101.02212},
 primaryClass = {astro-ph.SR},
       adsurl = {https://ui.adsabs.harvard.edu/abs/2021MNRAS.504.2577J}
}

@article{Abac:2023ujg,
    author = "Abac, Adrian and Dietrich, Tim and Buonanno, Alessandra and Steinhoff, Jan and Ujevic, Maximiliano",
    title = "{New and robust gravitational-waveform model for high-mass-ratio binary neutron star systems with dynamical tidal effects}",
    eprint = "2311.07456",
    archivePrefix = "arXiv",
    primaryClass = "gr-qc",
    doi = "10.1103/PhysRevD.109.024062",
    journal = "Phys. Rev. D",
    volume = "109",
    number = "2",
    pages = "024062",
    year = "2024"
}

@article{Barkett:2019tus,
    author = "Barkett, Kevin and Chen, Yanbei and Scheel, Mark A. and Varma, Vijay",
    title = "{Gravitational waveforms of binary neutron star inspirals using post-Newtonian tidal splicing}",
    eprint = "1911.10440",
    archivePrefix = "arXiv",
    primaryClass = "gr-qc",
    doi = "10.1103/PhysRevD.102.024031",
    journal = "Phys. Rev. D",
    volume = "102",
    number = "2",
    pages = "024031",
    year = "2020"
}

@article{Barr:2024wwl,
    author = "Barr, Ewan D. and others",
    title = "{A pulsar in a binary with a compact object in the mass gap between neutron stars and black holes}",
    eprint = "2401.09872",
    archivePrefix = "arXiv",
    primaryClass = "astro-ph.HE",
    doi = "10.1126/science.adg3005",
    journal = "Science",
    volume = "383",
    number = "6680",
    pages = "275--279",
    year = "2024"
}

@article{Boccioli:2024kvw,
    author = "Boccioli, Luca and Fragione, Giacomo",
    title = "{Remnant masses from 1D+ core-collapse supernovae simulations: Bimodal neutron star mass distribution and black holes in the low-mass gap}",
    eprint = "2404.05927",
    archivePrefix = "arXiv",
    primaryClass = "astro-ph.HE",
    doi = "10.1103/PhysRevD.110.023007",
    journal = "Phys. Rev. D",
    volume = "110",
    number = "2",
    pages = "023007",
    year = "2024"
}

@article{Buonanno:1998gg,
    author = "Buonanno, A. and Damour, T.",
    title = "{Effective one-body approach to general relativistic two-body dynamics}",
    eprint = "gr-qc/9811091",
    archivePrefix = "arXiv",
    reportNumber = "IHES-P-98-74",
    doi = "10.1103/PhysRevD.59.084006",
    journal = "Phys. Rev. D",
    volume = "59",
    pages = "084006",
    year = "1999"
}

@article{Chatziioannou:2016ezg,
    author = "Chatziioannou, Katerina and Klein, Antoine and Cornish, Neil and Yunes, Nicolas",
    title = "{Analytic Gravitational Waveforms for Generic Precessing Binary Inspirals}",
    eprint = "1606.03117",
    archivePrefix = "arXiv",
    primaryClass = "gr-qc",
    doi = "10.1103/PhysRevLett.118.051101",
    journal = "Phys. Rev. Lett.",
    volume = "118",
    number = "5",
    pages = "051101",
    year = "2017"
}

@article{Chatziioannou:2017tdw,
    author = "Chatziioannou, Katerina and Klein, Antoine and Yunes, Nicol\'as and Cornish, Neil",
    title = "{Constructing Gravitational Waves from Generic Spin-Precessing Compact Binary Inspirals}",
    eprint = "1703.03967",
    archivePrefix = "arXiv",
    primaryClass = "gr-qc",
    doi = "10.1103/PhysRevD.95.104004",
    journal = "Phys. Rev. D",
    volume = "95",
    number = "10",
    pages = "104004",
    year = "2017"
}

@article{Chattopadhyay:2020lff,
    author = "Chattopadhyay, Debatri and Stevenson, Simon and Hurley, Jarrod R. and Bailes, Matthew and Broekgaarden, Floor",
    title = "{Modelling neutron star{\textendash}black hole binaries: future pulsar surveys and gravitational wave detectors}",
    eprint = "2011.13503",
    archivePrefix = "arXiv",
    primaryClass = "astro-ph.HE",
    doi = "10.1093/mnras/stab973",
    journal = "Mon. Not. Roy. Astron. Soc.",
    volume = "504",
    number = "3",
    pages = "3682--3710",
    year = "2021"
}

@article{Dexheimer:2020rlp,
    author = {Dexheimer, V. and Gomes, R. O. and Kl\"ahn, T. and Han, S. and Salinas, M.},
    title = "{GW190814 as a massive rapidly rotating neutron star with exotic degrees of freedom}",
    eprint = "2007.08493",
    archivePrefix = "arXiv",
    primaryClass = "astro-ph.HE",
    doi = "10.1103/PhysRevC.103.025808",
    journal = "Phys. Rev. C",
    volume = "103",
    number = "2",
    pages = "025808",
    year = "2021"
}

@article{Dorozsmai:2025jlu,
    author = "Dorozsmai, Andris and Romero-Shaw, Isobel M. and Vijaykumar, Aditya and Toonen, Silvia and Antonini, Fabio and Kremer, Kyle and Zevin, Michael and Grishin, Evgeni",
    title = "{Hierarchical triples versus globular clusters: binary black hole merger eccentricity distributions compete and evolve with redshift}",
    eprint = "2507.23212",
    archivePrefix = "arXiv",
    primaryClass = "astro-ph.GA",
    doi = "10.1093/mnras/staf1938",
    journal = "Mon. Not. Roy. Astron. Soc.",
    volume = "545",
    number = "2",
    pages = "staf1938",
    year = "2025"
}

@article{Henry:2022ccf,
    author = "Henry, Quentin",
    title = "{Complete gravitational-waveform amplitude modes for quasicircular compact binaries to the 3.5PN order}",
    eprint = "2210.15602",
    archivePrefix = "arXiv",
    primaryClass = "gr-qc",
    doi = "10.1103/PhysRevD.107.044057",
    journal = "Phys. Rev. D",
    volume = "107",
    number = "4",
    pages = "044057",
    year = "2023"
}

@article{Dones:2024odv,
    author = "Dones, Eve and Henry, Quentin and Bernard, Laura",
    title = "{Tidal contributions to the full gravitational waveform to the second-and-a-half post-Newtonian order}",
    eprint = "2412.14249",
    archivePrefix = "arXiv",
    primaryClass = "gr-qc",
    doi = "10.1103/PhysRevD.111.084043",
    journal = "Phys. Rev. D",
    volume = "111",
    number = "8",
    pages = "084043",
    year = "2025"
}

@article{East:2021spd,
    author = "East, William E. and Lehner, Luis and Liebling, Steven L. and Palenzuela, Carlos",
    title = "{Multimessenger Signals from Black Hole\textendash{}Neutron Star Mergers without Significant Tidal Disruption}",
    eprint = "2101.12214",
    archivePrefix = "arXiv",
    primaryClass = "astro-ph.HE",
    doi = "10.3847/2041-8213/abf566",
    journal = "Astrophys. J. Lett.",
    volume = "912",
    number = "1",
    pages = "L18",
    year = "2021"
}

@article{Essick:2020ghc,
    author = "Essick, Reed and Landry, Philippe",
    title = "{Discriminating between Neutron Stars and Black Holes with Imperfect Knowledge of the Maximum Neutron Star Mass}",
    eprint = "2007.01372",
    archivePrefix = "arXiv",
    primaryClass = "astro-ph.HE",
    doi = "10.3847/1538-4357/abbd3b",
    journal = "Astrophys. J.",
    volume = "904",
    number = "1",
    pages = "80",
    year = "2020"
}

@article{Fernandez:2016sbf,
    author = "Fern\'andez, Rodrigo and Foucart, Francois and Kasen, Daniel and Lippuner, Jonas and Desai, Dhruv and Roberts, Luke F.",
    title = "{Dynamics, nucleosynthesis, and kilonova signature of black hole\textemdash{}neutron star merger ejecta}",
    eprint = "1612.04829",
    archivePrefix = "arXiv",
    primaryClass = "astro-ph.HE",
    doi = "10.1088/1361-6382/aa7a77",
    journal = "Class. Quant. Grav.",
    volume = "34",
    number = "15",
    pages = "154001",
    year = "2017"
}

@article{Foucart:2013psa,
    author = "Foucart, Francois and Buchman, Luisa and Duez, Matthew D. and Grudich, Michael and Kidder, Lawrence E. and MacDonald, Ilana and Mroue, Abdul and Pfeiffer, Harald P. and Scheel, Mark A. and Szilagyi, Bela",
    title = "{First direct comparison of nondisrupting neutron star-black hole and binary black hole merger simulations}",
    eprint = "1307.7685",
    archivePrefix = "arXiv",
    primaryClass = "gr-qc",
    doi = "10.1103/PhysRevD.88.064017",
    journal = "Phys. Rev. D",
    volume = "88",
    number = "6",
    pages = "064017",
    year = "2013"
}

@article{Foucart:2012nc,
    author = "Foucart, Francois",
    title = "{Black Hole-Neutron Star Mergers: Disk Mass Predictions}",
    eprint = "1207.6304",
    archivePrefix = "arXiv",
    primaryClass = "astro-ph.HE",
    doi = "10.1103/PhysRevD.86.124007",
    journal = "Phys. Rev. D",
    volume = "86",
    pages = "124007",
    year = "2012"
}

@article{Foucart:2018lhe,
    author = "Foucart, Francois and others",
    title = "{Gravitational waveforms from spectral Einstein code simulations: Neutron star-neutron star and low-mass black hole-neutron star binaries}",
    eprint = "1812.06988",
    archivePrefix = "arXiv",
    primaryClass = "gr-qc",
    doi = "10.1103/PhysRevD.99.044008",
    journal = "Phys. Rev. D",
    volume = "99",
    number = "4",
    pages = "044008",
    year = "2019"
}

@article{Foucart:2020xkt,
    author = "Foucart, Francois and others",
    title = "{High-accuracy waveforms for black hole-neutron star systems with spinning black holes}",
    eprint = "2010.14518",
    archivePrefix = "arXiv",
    primaryClass = "gr-qc",
    doi = "10.1103/PhysRevD.103.064007",
    journal = "Phys. Rev. D",
    volume = "103",
    number = "6",
    pages = "064007",
    year = "2021"
}

@article{Fryer:2022lla,
    author = "Fryer, Chris L. and Olejak, Aleksandra and Belczynski, Krzysztof",
    title = "{The Effect of Supernova Convection On Neutron Star and Black Hole Masses}",
    eprint = "2204.13025",
    archivePrefix = "arXiv",
    primaryClass = "astro-ph.HE",
    reportNumber = "LA-UR-21-32205",
    doi = "10.3847/1538-4357/ac6ac9",
    journal = "Astrophys. J.",
    volume = "931",
    number = "2",
    pages = "94",
    year = "2022"
}

@article{Garcia-Quiros:2020qpx,
    author = "Garc\'\i{}a-Quir\'os, Cecilio and Colleoni, Marta and Husa, Sascha and Estell\'es, H\'ector and Pratten, Geraint and Ramos-Buades, Antoni and Mateu-Lucena, Maite and Jaume, Rafel",
    title = "{Multimode frequency-domain model for the gravitational wave signal from nonprecessing black-hole binaries}",
    eprint = "2001.10914",
    archivePrefix = "arXiv",
    primaryClass = "gr-qc",
    doi = "10.1103/PhysRevD.102.064002",
    journal = "Phys. Rev. D",
    volume = "102",
    number = "6",
    pages = "064002",
    year = "2020"
}

@article{Gonzalez:2022mgo,
    author = "Gonzalez, Alejandra and others",
    title = "{Second release of the CoRe database of binary neutron star merger waveforms}",
    eprint = "2210.16366",
    archivePrefix = "arXiv",
    primaryClass = "gr-qc",
    doi = "10.1088/1361-6382/acc231",
    journal = "Class. Quant. Grav.",
    volume = "40",
    number = "8",
    pages = "085011",
    year = "2023"
}

@article{Gonzalez:2022prs,
    author = "Gonzalez, Alejandra and Gamba, Rossella and Breschi, Matteo and Zappa, Francesco and Carullo, Gregorio and Bernuzzi, Sebastiano and Nagar, Alessandro",
    title = "{Numerical-relativity-informed effective-one-body model for black-hole\textendash{}neutron-star mergers with higher modes and spin precession}",
    eprint = "2212.03909",
    archivePrefix = "arXiv",
    primaryClass = "gr-qc",
    doi = "10.1103/PhysRevD.107.084026",
    journal = "Phys. Rev. D",
    volume = "107",
    number = "8",
    pages = "084026",
    year = "2023"
}

@article{Gonzalez:2025xba,
    author = "Gonzalez, Alejandra and Bernuzzi, Sebastiano and Rashti, Alireza and Brandoli, Francesco and Gamba, Rossella",
    title = "{Black-hole - neutron-star mergers: new numerical-relativity simulations and multipolar effective-one-body model with spin precession and eccentricity}",
    eprint = "2507.00113",
    archivePrefix = "arXiv",
    journal = {arXiv e-prints},
    primaryClass = "gr-qc",
    month = "6",
    year = "2025"
}

@article{Gottlieb:2023est,
    author = "Gottlieb, Ore and others",
    title = "{Large-scale Evolution of Seconds-long Relativistic Jets from Black Hole\textendash{}Neutron Star Mergers}",
    eprint = "2306.14947",
    archivePrefix = "arXiv",
    primaryClass = "astro-ph.HE",
    doi = "10.3847/2041-8213/aceeff",
    journal = "Astrophys. J. Lett.",
    volume = "954",
    number = "1",
    pages = "L21",
    year = "2023"
}

@article{Hayashi:2022cdq,
    author = "Hayashi, Kota and Kiuchi, Kenta and Kyutoku, Koutarou and Sekiguchi, Yuichiro and Shibata, Masaru",
    title = "{General-relativistic neutrino-radiation magnetohydrodynamics simulation of seconds-long black hole-neutron star mergers: Dependence on the initial magnetic field strength, configuration, and neutron-star equation of state}",
    eprint = "2211.07158",
    archivePrefix = "arXiv",
    primaryClass = "astro-ph.HE",
    doi = "10.1103/PhysRevD.107.123001",
    journal = "Phys. Rev. D",
    volume = "107",
    number = "12",
    pages = "123001",
    year = "2023"
}

@article{Henry:2019xhg,
    author = "Henry, Quentin and Faye, Guillaume and Blanchet, Luc",
    title = "{Tidal effects in the equations of motion of compact binary systems to next-to-next-to-leading post-Newtonian order}",
    eprint = "1912.01920",
    archivePrefix = "arXiv",
    primaryClass = "gr-qc",
    doi = "10.1103/PhysRevD.101.064047",
    journal = "Phys. Rev. D",
    volume = "101",
    number = "6",
    pages = "064047",
    year = "2020"
}

@article{Huang:2020pba,
    author = "Huang, Yiwen and Haster, Carl-Johan and Vitale, Salvatore and Varma, Vijay and Foucart, Francois and Biscoveanu, Sylvia",
    title = "{Statistical and systematic uncertainties in extracting the source properties of neutron star - black hole binaries with gravitational waves}",
    eprint = "2005.11850",
    archivePrefix = "arXiv",
    primaryClass = "gr-qc",
    doi = "10.1103/PhysRevD.103.083001",
    journal = "Phys. Rev. D",
    volume = "103",
    number = "8",
    pages = "083001",
    year = "2021"
}

@article{Kyutoku:2010zd,
    author = "Kyutoku, Koutarou and Shibata, Masaru and Taniguchi, Keisuke",
    title = "{Gravitational waves from nonspinning black hole-neutron star binaries: dependence on equations of state}",
    eprint = "1008.1460",
    archivePrefix = "arXiv",
    primaryClass = "astro-ph.HE",
    doi = "10.1103/PhysRevD.82.044049",
    journal = "Phys. Rev. D",
    volume = "82",
    pages = "044049",
    year = "2010",
    note = "[Erratum: Phys.Rev.D 84, 049902 (2011)]"
}

@article{Kyutoku:2011vz,
    author = "Kyutoku, Koutarou and Okawa, Hirotada and Shibata, Masaru and Taniguchi, Keisuke",
    title = "{Gravitational waves from spinning black hole-neutron star binaries: dependence on black hole spins and on neutron star equations of state}",
    eprint = "1108.1189",
    archivePrefix = "arXiv",
    primaryClass = "astro-ph.HE",
    doi = "10.1103/PhysRevD.84.064018",
    journal = "Phys. Rev. D",
    volume = "84",
    pages = "064018",
    year = "2011"
}

@article{Kyutoku:2013wxa,
    author = "Kyutoku, Koutarou and Ioka, Kunihito and Shibata, Masaru",
    title = "{Anisotropic mass ejection from black hole-neutron star binaries: Diversity of electromagnetic counterparts}",
    eprint = "1305.6309",
    archivePrefix = "arXiv",
    primaryClass = "astro-ph.HE",
    reportNumber = "KEK-TH-1655, KEK-COSMO-125",
    doi = "10.1103/PhysRevD.88.041503",
    journal = "Phys. Rev. D",
    volume = "88",
    number = "4",
    pages = "041503",
    year = "2013"
}

@article{Kyutoku:2015gda,
    author = "Kyutoku, Koutarou and Ioka, Kunihito and Okawa, Hirotada and Shibata, Masaru and Taniguchi, Keisuke",
    title = "{Dynamical mass ejection from black hole-neutron star binaries}",
    eprint = "1502.05402",
    archivePrefix = "arXiv",
    primaryClass = "astro-ph.HE",
    reportNumber = "KEK-TH-1796, KEK-COSMO-163",
    doi = "10.1103/PhysRevD.92.044028",
    journal = "Phys. Rev. D",
    volume = "92",
    pages = "044028",
    year = "2015"
}

@article{Lackey:2013axa,
    author = "Lackey, Benjamin D. and Kyutoku, Koutarou and Shibata, Masaru and Brady, Patrick R. and Friedman, John L.",
    title = "{Extracting equation of state parameters from black hole-neutron star mergers: aligned-spin black holes and a preliminary waveform model}",
    eprint = "1303.6298",
    archivePrefix = "arXiv",
    primaryClass = "gr-qc",
    doi = "10.1103/PhysRevD.89.043009",
    journal = "Phys. Rev. D",
    volume = "89",
    number = "4",
    pages = "043009",
    year = "2014"
}

@article{Levin:2018mzg,
    author = "Levin, Janna and D'Orazio, Daniel J. and Garcia-Saenz, Sebastian",
    title = "{Black Hole Pulsar}",
    eprint = "1808.07887",
    archivePrefix = "arXiv",
    primaryClass = "astro-ph.HE",
    doi = "10.1103/PhysRevD.98.123002",
    journal = "Phys. Rev. D",
    volume = "98",
    number = "12",
    pages = "123002",
    year = "2018"
}

@article{LIGOScientific:2020zkf,
    author = "Abbott, R. and others",
    collaboration = "LIGO Scientific, Virgo",
    title = "{GW190814: Gravitational Waves from the Coalescence of a 23 Solar Mass Black Hole with a 2.6 Solar Mass Compact Object}",
    eprint = "2006.12611",
    archivePrefix = "arXiv",
    primaryClass = "astro-ph.HE",
    reportNumber = "LIGO-P190814",
    doi = "10.3847/2041-8213/ab960f",
    journal = "Astrophys. J. Lett.",
    volume = "896",
    number = "2",
    pages = "L44",
    year = "2020"
}

@article{LIGOScientific:2021qlt,
    author = "Abbott, R. and others",
    collaboration = "LIGO Scientific, KAGRA, VIRGO",
    title = "{Observation of Gravitational Waves from Two Neutron Star\textendash{}Black Hole Coalescences}",
    eprint = "2106.15163",
    archivePrefix = "arXiv",
    primaryClass = "astro-ph.HE",
    reportNumber = "LIGO Document P2000357",
    doi = "10.3847/2041-8213/ac082e",
    journal = "Astrophys. J. Lett.",
    volume = "915",
    number = "1",
    pages = "L5",
    year = "2021"
}

@article{LIGOScientific:2021usb,
    author = "Abbott, R. and others",
    collaboration = "LIGO Scientific, VIRGO",
    title = "{GWTC-2.1: Deep extended catalog of compact binary coalescences observed by LIGO and Virgo during the first half of the third observing run}",
    eprint = "2108.01045",
    archivePrefix = "arXiv",
    primaryClass = "gr-qc",
    reportNumber = "LIGO-P2100063",
    doi = "10.1103/PhysRevD.109.022001",
    journal = "Phys. Rev. D",
    volume = "109",
    number = "2",
    pages = "022001",
    year = "2024"
}

@article{LIGOScientific:2020ibl,
    author = "Abbott, R. and others",
    collaboration = "LIGO Scientific, Virgo",
    title = "{GWTC-2: Compact Binary Coalescences Observed by LIGO and Virgo During the First Half of the Third Observing Run}",
    eprint = "2010.14527",
    archivePrefix = "arXiv",
    primaryClass = "gr-qc",
    reportNumber = "P2000061",
    doi = "10.1103/PhysRevX.11.021053",
    journal = "Phys. Rev. X",
    volume = "11",
    pages = "021053",
    year = "2021"
}

@article{LIGOScientific:2018mvr,
    author = "Abbott, B. P. and others",
    collaboration = "LIGO Scientific, Virgo",
    title = "{GWTC-1: A Gravitational-Wave Transient Catalog of Compact Binary Mergers Observed by LIGO and Virgo during the First and Second Observing Runs}",
    eprint = "1811.12907",
    archivePrefix = "arXiv",
    primaryClass = "astro-ph.HE",
    reportNumber = "LIGO-P1800307",
    doi = "10.1103/PhysRevX.9.031040",
    journal = "Phys. Rev. X",
    volume = "9",
    number = "3",
    pages = "031040",
    year = "2019"
}

@article{LIGOScientific:2024elc,
    author = "Abac, A. G. and others",
    collaboration = "LIGO Scientific, Virgo,, KAGRA, VIRGO",
    title = "{Observation of Gravitational Waves from the Coalescence of a 2.5\textendash{}4.5 M $_{⊙}$ Compact Object and a Neutron Star}",
    eprint = "2404.04248",
    archivePrefix = "arXiv",
    primaryClass = "astro-ph.HE",
    reportNumber = "LIGO-P2300352",
    doi = "10.3847/2041-8213/ad5beb",
    journal = "Astrophys. J. Lett.",
    volume = "970",
    number = "2",
    pages = "L34",
    year = "2024"
}

@article{McWilliams:2011zi,
    author = "McWilliams, Sean T. and Levin, Janna",
    title = "{Electromagnetic extraction of energy from black hole-neutron star binaries}",
    eprint = "1101.1969",
    archivePrefix = "arXiv",
    primaryClass = "astro-ph.HE",
    doi = "10.1088/0004-637X/742/2/90",
    journal = "Astrophys. J.",
    volume = "742",
    pages = "90",
    year = "2011"
}

@article{Most:2020bba,
    author = "Most, Elias R. and Papenfort, L. Jens and Weih, Lukas R. and Rezzolla, Luciano",
    title = "{A lower bound on the maximum mass if the secondary in GW190814 was once a rapidly spinning neutron star}",
    eprint = "2006.14601",
    archivePrefix = "arXiv",
    primaryClass = "astro-ph.HE",
    doi = "10.1093/mnrasl/slaa168",
    journal = "Mon. Not. Roy. Astron. Soc.",
    volume = "499",
    number = "1",
    pages = "L82--L86",
    year = "2020"
}

@article{Planck:2015fie,
    author = "Ade, P. A. R. and others",
    collaboration = "Planck",
    title = "{Planck 2015 results. XIII. Cosmological parameters}",
    eprint = "1502.01589",
    archivePrefix = "arXiv",
    primaryClass = "astro-ph.CO",
    doi = "10.1051/0004-6361/201525830",
    journal = "Astron. Astrophys.",
    volume = "594",
    pages = "A13",
    year = "2016"
}

@article{Pratten:2020fqn,
    author = "Pratten, Geraint and Husa, Sascha and Garcia-Quiros, Cecilio and Colleoni, Marta and Ramos-Buades, Antoni and Estelles, Hector and Jaume, Rafel",
    title = "{Setting the cornerstone for a family of models for gravitational waves from compact binaries: The dominant harmonic for nonprecessing quasicircular black holes}",
    eprint = "2001.11412",
    archivePrefix = "arXiv",
    primaryClass = "gr-qc",
    reportNumber = "LIGO-P2000018",
    doi = "10.1103/PhysRevD.102.064001",
    journal = "Phys. Rev. D",
    volume = "102",
    number = "6",
    pages = "064001",
    year = "2020"
}

@article{Cabero:2016ayq,
    author = "Cabero, Miriam and Nielsen, Alex B. and Lundgren, Andrew P. and Capano, Collin D.",
    title = "{Minimum energy and the end of the inspiral in the post-Newtonian approximation}",
    eprint = "1602.03134",
    archivePrefix = "arXiv",
    primaryClass = "gr-qc",
    doi = "10.1103/PhysRevD.95.064016",
    journal = "Phys. Rev. D",
    volume = "95",
    number = "6",
    pages = "064016",
    year = "2017"
}

@article{Mandel:2020cig,
    author = {Mandel, Ilya and M\"uller, Bernhard and Riley, Jeff and de Mink, Selma E. and Vigna-G\'omez, Alejandro and Chattopadhyay, Debatri},
    title = "{Binary population synthesis with probabilistic remnant mass and kick prescriptions}",
    eprint = "2007.03890",
    archivePrefix = "arXiv",
    primaryClass = "astro-ph.HE",
    doi = "10.1093/mnras/staa3390",
    journal = "Mon. Not. Roy. Astron. Soc.",
    volume = "500",
    number = "1",
    pages = "1380--1384",
    year = "2020"
}

@article{Matas:2020wab,
    author = "Matas, Andrew and others",
    title = "{Aligned-spin neutron-star\textendash{}black-hole waveform model based on the effective-one-body approach and numerical-relativity simulations}",
    eprint = "2004.10001",
    archivePrefix = "arXiv",
    primaryClass = "gr-qc",
    doi = "10.1103/PhysRevD.102.043023",
    journal = "Phys. Rev. D",
    volume = "102",
    number = "4",
    pages = "043023",
    year = "2020"
}

@article{Morisaki:2021ngj,
    author = "Morisaki, Soichiro",
    title = "{Accelerating parameter estimation of gravitational waves from compact binary coalescence using adaptive frequency resolutions}",
    eprint = "2104.07813",
    archivePrefix = "arXiv",
    primaryClass = "gr-qc",
    doi = "10.1103/PhysRevD.104.044062",
    journal = "Phys. Rev. D",
    volume = "104",
    number = "4",
    pages = "044062",
    year = "2021"
}

@article{Olejak:2022zee,
    author = "Olejak, Aleksandra and Fryer, Chris L. and Belczynski, Krzysztof and Baibhav, Vishal",
    title = "{The role of supernova convection for the lower mass gap in the isolated binary formation of gravitational wave sources}",
    eprint = "2204.09061",
    archivePrefix = "arXiv",
    primaryClass = "astro-ph.HE",
    doi = "10.1093/mnras/stac2359",
    journal = "Mon. Not. Roy. Astron. Soc.",
    volume = "516",
    number = "2",
    pages = "2252--2271",
    year = "2022"
}

@article{Pannarale:2015jka,
    author = "Pannarale, Francesco and Berti, Emanuele and Kyutoku, Koutarou and Lackey, Benjamin D. and Shibata, Masaru",
    title = "{Aligned spin neutron star-black hole mergers: a gravitational waveform amplitude model}",
    eprint = "1509.00512",
    archivePrefix = "arXiv",
    primaryClass = "gr-qc",
    reportNumber = "LIGO-P1500135",
    doi = "10.1103/PhysRevD.92.084050",
    journal = "Phys. Rev. D",
    volume = "92",
    number = "8",
    pages = "084050",
    year = "2015"
}

@article{Paschalidis:2014qra,
    author = "Paschalidis, Vasileios and Ruiz, Milton and Shapiro, Stuart L.",
    title = "{Relativistic Simulations of Black Hole\textendash{}neutron Star Coalescence: the jet Emerges}",
    eprint = "1410.7392",
    archivePrefix = "arXiv",
    primaryClass = "astro-ph.HE",
    doi = "10.1088/2041-8205/806/1/L14",
    journal = "Astrophys. J. Lett.",
    volume = "806",
    number = "1",
    pages = "L14",
    year = "2015"
}

@article{PhysRevD.10.1680,
  title = {Black hole in a uniform magnetic field},
  author = {Wald, Robert M.},
  journal = {Phys. Rev. D},
  volume = {10},
  issue = {6},
  pages = {1680--1685},
  numpages = {0},
  year = {1974},
  month = {Sep},
  publisher = {American Physical Society},
  doi = {10.1103/PhysRevD.10.1680},
  url = {https://link.aps.org/doi/10.1103/PhysRevD.10.1680}
}

@misc{Pillas:2025pfc,
    author = "Pillas, M. and others",
    title = "{Limits on the Ejecta Mass During the Search for Kilonovae Associated with Neutron Star-Black Hole Mergers: A case study of S230518h, GW230529, S230627c and the Low-Significance Candidate S240422ed}",
    eprint = "2503.15422",
    archivePrefix = "arXiv",
    primaryClass = "astro-ph.HE",
    month = "3",
    year = "2025"
}

@article{Pompili:2023tna,
    author = "Pompili, Lorenzo and others",
    title = "{Laying the foundation of the effective-one-body waveform models SEOBNRv5: Improved accuracy and efficiency for spinning nonprecessing binary black holes}",
    eprint = "2303.18039",
    archivePrefix = "arXiv",
    primaryClass = "gr-qc",
    doi = "10.1103/PhysRevD.108.124035",
    journal = "Phys. Rev. D",
    volume = "108",
    number = "12",
    pages = "124035",
    year = "2023"
}

@article{Sedda:2020wzl,
    author = "Sedda, Manuel Arca",
    title = "{Dissecting the properties of neutron star - black hole mergers originating in dense star clusters}",
    eprint = "2003.02279",
    archivePrefix = "arXiv",
    primaryClass = "astro-ph.GA",
    doi = "10.1038/s42005-020-0310-x",
    journal = "Commun. Phys.",
    volume = "3",
    pages = "43",
    year = "2020"
}

@article{Stegmann:2025clo,
    author = "Stegmann, Jakob and Klencki, Jakub",
    title = "{Orbital Eccentricity and Spin{\textendash}Orbit Misalignment Are Evidence that Neutron Star{\textendash}Black Hole Mergers Form through Triple Star Evolution}",
    eprint = "2506.09121",
    archivePrefix = "arXiv",
    primaryClass = "astro-ph.HE",
    doi = "10.3847/2041-8213/ae055b",
    journal = "Astrophys. J. Lett.",
    volume = "991",
    number = "2",
    pages = "L54",
    year = "2025"
}

@article{Steppohn:2025kbh,
    author = {Steppohn, Oliver and V{\"o}lkel, Sebastian H. and Dietrich, Tim},
    title = "{Black hole spectroscopy of collapsing and merging neutron stars}",
    eprint = "2508.15534",
    archivePrefix = "arXiv",
    primaryClass = "gr-qc",
    doi = "10.1103/hnhc-m4lw",
    journal = "Phys. Rev. D",
    volume = "113",
    number = "2",
    pages = "024011",
    year = "2026"
}

@misc{            SXS:catalog,
  title         = {{SXS Gravitational Waveform Database}},
  howpublished = "\url{https://data.black-holes.org/simulations/index.html}"
}

@article{Tews:2020ylw,
    author = "Tews, Ingo and Pang, Peter T. H. and Dietrich, Tim and Coughlin, Michael W. and Antier, Sarah and Bulla, Mattia and Heinzel, Jack and Issa, Lina",
    title = "{On the Nature of GW190814 and Its Impact on the Understanding of Supranuclear Matter}",
    eprint = "2007.06057",
    archivePrefix = "arXiv",
    primaryClass = "astro-ph.HE",
    reportNumber = "LA-UR-20-24959",
    doi = "10.3847/2041-8213/abdaae",
    journal = "Astrophys. J. Lett.",
    volume = "908",
    number = "1",
    pages = "L1",
    year = "2021"
}

@article{Thompson:2020nei,
    author = "Thompson, Jonathan E. and Fauchon-Jones, Edward and Khan, Sebastian and Nitoglia, Elisa and Pannarale, Francesco and Dietrich, Tim and Hannam, Mark",
    title = "{Modeling the gravitational wave signature of neutron star black hole coalescences}",
    eprint = "2002.08383",
    archivePrefix = "arXiv",
    primaryClass = "gr-qc",
    reportNumber = "LIGO-P2000059",
    doi = "10.1103/PhysRevD.101.124059",
    journal = "Phys. Rev. D",
    volume = "101",
    number = "12",
    pages = "124059",
    year = "2020"
}

@article{Topolski:2024jva,
    author = "Topolski, Konrad and Tootle, Samuel D. and Rezzolla, Luciano",
    title = "{Black hole-neutron star binaries with high spins and large mass asymmetries. II. Properties of dynamical simulations}",
    eprint = "2409.06777",
    archivePrefix = "arXiv",
    primaryClass = "gr-qc",
    doi = "10.1103/PhysRevD.111.064023",
    journal = "Phys. Rev. D",
    volume = "111",
    number = "6",
    pages = "064023",
    year = "2025"
}

@article{Tsang:2012PhysRevLett.108.011102,
  title = {Resonant Shattering of Neutron Star Crusts},
  author = {Tsang, David and Read, Jocelyn S. and Hinderer, Tanja and Piro, Anthony L. and Bondarescu, Ruxandra},
  journal = {Phys. Rev. Lett.},
  volume = {108},
  issue = {1},
  pages = {011102},
  numpages = {5},
  year = {2012},
  month = {Jan},
  publisher = {American Physical Society},
  doi = {10.1103/PhysRevLett.108.011102},
  url = {https://link.aps.org/doi/10.1103/PhysRevLett.108.011102}
}

@article{Vigna-Gomez:2021oqy,
    author = "Vigna-G{\'o}mez, Alejandro and Schr{\o}der, Sophie L. and Ramirez-Ruiz, Enrico and Aguilera-Dena, David R. and Batta, Aldo and Langer, Norbert and Willcox, Reinhold",
    title = "{Fallback Supernova Assembly of Heavy Binary Neutron Stars and Light Black Hole{\textendash}Neutron Star Pairs and the Common Stellar Ancestry of GW190425 and GW200115}",
    eprint = "2106.12381",
    archivePrefix = "arXiv",
    primaryClass = "astro-ph.HE",
    doi = "10.3847/2041-8213/ac2903",
    journal = "Astrophys. J. Lett.",
    volume = "920",
    number = "1",
    pages = "L17",
    year = "2021"
}

@misc{Xing:2024ydg,
    author = "Xing, Zepei and others",
    title = "{Mass-gap Black Holes in Coalescing Neutron Star Black Hole Binaries}",
    eprint = "2410.20415",
    archivePrefix = "arXiv",
    primaryClass = "astro-ph.HE",
    month = "10",
    year = "2024"
}

@article{Ye:2024wqj,
    author = "Ye, Claire S. and Kremer, Kyle and Ransom, Scott M. and Rasio, Frederic A.",
    title = "{Lower-mass-gap Black Holes in Dense Star Clusters}",
    eprint = "2408.00076",
    archivePrefix = "arXiv",
    primaryClass = "astro-ph.HE",
    doi = "10.3847/1538-4357/ad76a0",
    journal = "Astrophys. J.",
    volume = "975",
    number = "1",
    pages = "77",
    year = "2024"
}

@misc{Zackay:2018qdy,
    author = "Zackay, Barak and Dai, Liang and Venumadhav, Tejaswi",
    title = "{Relative Binning and Fast Likelihood Evaluation for Gravitational Wave Parameter Estimation}",
    eprint = "1806.08792",
    archivePrefix = "arXiv",
    primaryClass = "astro-ph.IM",
    month = "6",
    year = "2018"
}

@article{Pratten:2020ceb,
    author = "Pratten, Geraint and others",
    title = "{Computationally efficient models for the dominant and subdominant harmonic modes of precessing binary black holes}",
    eprint = "2004.06503",
    archivePrefix = "arXiv",
    primaryClass = "gr-qc",
    doi = "10.1103/PhysRevD.103.104056",
    journal = "Phys. Rev. D",
    volume = "103",
    number = "10",
    pages = "104056",
    year = "2021"
}

@techreport{SpinTaylor_TechNote,
	author = {Sturani, R.},
	institution = {{LIGO} Project},
	note = {\url{https://dcc.ligo.org/T1500554/public}},
	number = {{LIGO}-T1500554},
	title = {Note on the derivation of the angular momentum and spin precessing equations in SpinTaylor codes},
	url = {https://dcc.ligo.org/T1500554/public},
	year = {2021}
}

@article{Colleoni:2025aoh,
    author = "Colleoni, Marta and Ramis Vidal, Felip A. and Johnson-McDaniel, Nathan K. and Dietrich, Tim and Haney, Maria and Pratten, Geraint",
    title = "{New gravitational waveform model for precessing binary neutron stars with double-spin effects}",
    doi = "10.1103/PhysRevD.111.064025",
    journal = "Phys. Rev. D",
    volume = "111",
    number = "6",
    pages = "064025",
    year = "2025"
}

@misc{lalsuite,
       author         = "{LIGO Scientific Collaboration} and {Virgo Collaboration} and {KAGRA Collaboration}",
       title          = "{LVK} {A}lgorithm {L}ibrary - {LALS}uite",
       howpublished   = "Free software (GPL)",
       doi            = "10.7935/GT1W-FZ16",
       year           = "2018"
 }

@article{swiglal,
      title     = "{SWIGLAL: Python and Octave interfaces to the LALSuite gravitational-wave data analysis libraries}",
      author    = "Karl Wette",
      journal   = "SoftwareX",
      volume    = "12",
      pages     = "100634",
      year      = "2020",
      doi       = "10.1016/j.softx.2020.100634"
}

@misc{Abac:2025brd,
      title={Leveraging NRTidalv3 to develop gravitational waveform models with higher-order modes for binary neutron star systems}, 
      author={Adrian Abac and Felip Ramis Vidal and Marta Colleoni and Anna Puecher and Alejandra Gonzalez and Tim Dietrich},
      year={2025},
      eprint={2507.15426},
      archivePrefix={arXiv},
      primaryClass={gr-qc},
      url={https://arxiv.org/abs/2507.15426}, 
}

@article{Dietrich:2019kaq,
    author = "Dietrich, Tim and Samajdar, Anuradha and Khan, Sebastian and Johnson-McDaniel, Nathan K. and Dudi, Reetika and Tichy, Wolfgang",
    title = "{Improving the NRTidal model for binary neutron star systems}",
    eprint = "1905.06011",
    archivePrefix = "arXiv",
    primaryClass = "gr-qc",
    doi = "10.1103/PhysRevD.100.044003",
    journal = "Phys. Rev. D",
    volume = "100",
    number = "4",
    pages = "044003",
    year = "2019"
}

@article{Yamamoto:2008js,
    author = "Yamamoto, Tetsuro and Shibata, Masaru and Taniguchi, Keisuke",
    title = "{Simulating coalescing compact binaries by a new code SACRA}",
    eprint = "0806.4007",
    archivePrefix = "arXiv",
    primaryClass = "gr-qc",
    doi = "10.1103/PhysRevD.78.064054",
    journal = "Phys. Rev. D",
    volume = "78",
    pages = "064054",
    year = "2008"
}

@article{KAGRA:2021vkt,
    author = "Abbott, R. and others",
    collaboration = "KAGRA, VIRGO, LIGO Scientific",
    title = "{GWTC-3: Compact Binary Coalescences Observed by LIGO and Virgo during the Second Part of the Third Observing Run}",
    eprint = "2111.03606",
    archivePrefix = "arXiv",
    primaryClass = "gr-qc",
    reportNumber = "LIGO-P2000318",
    doi = "10.1103/PhysRevX.13.041039",
    journal = "Phys. Rev. X",
    volume = "13",
    number = "4",
    pages = "041039",
    year = "2023"
}

@article{LIGOScientific:2025slb,
       author = "Abac, A. G. and others",
collaboration = "LIGO Scientific, VIRGO, KAGRA",
        title = "{GWTC-4.0: Updating the Gravitational-Wave Transient Catalog with Observations from the First Part of the Fourth LIGO-Virgo-KAGRA Observing Run}",
          doi = {10.48550/arXiv.2508.18082},
archivePrefix = {arXiv},
      journal = {arXiv e-prints},
       eprint = {2508.18082},
 primaryClass = {gr-qc},
       adsurl = {https://ui.adsabs.harvard.edu/abs/2025arXiv250818082T}
}

@article{LIGOScientific:2025yae,
    author = "Abac, A. G. and others",
    collaboration = "LIGO Scientific, VIRGO, KAGRA",
    title = "{GWTC-4.0: Methods for Identifying and Characterizing Gravitational-wave Transients}",
    journal = {arXiv e-prints},
    eprint = "2508.18081",
    archivePrefix = "arXiv",
    primaryClass = "gr-qc",
    reportNumber = "LIGO-P2400300",
    month = "8",
    year = "2025"
}

@article{Morras:2025xfu,
    author = "Morras, Gonzalo and Pratten, Geraint and Schmidt, Patricia",
    title = "{Orbital eccentricity in a neutron star - black hole binary}",
    eprint = "2503.15393",
    journal = {arXiv e-prints},
    archivePrefix = "arXiv",
    primaryClass = "astro-ph.HE",
    month = "3",
    year = "2025"
}

@article{Planas:2025plq,
    author = "Planas, Maria de Lluc and Husa, Sascha and Ramos-Buades, Antoni and Valencia, Jorge",
    title = "{First Eccentric Inspiral{\textendash}Merger{\textendash}Ringdown Analysis of Neutron Star{\textendash}Black Hole Mergers}",
    eprint = "2506.01760",
    archivePrefix = "arXiv",
    primaryClass = "astro-ph.HE",
    doi = "10.3847/1538-4357/ae1d7d",
    journal = "Astrophys. J.",
    volume = "995",
    number = "1",
    pages = "47",
    year = "2025"
}

@article{Jan:2025fps,
    author = "Jan, Aasim and Tsao, Bing-Jyun and O'Shaughnessy, Richard and Shoemaker, Deirdre and Laguna, Pablo",
    title = "{GW200105: A detailed study of eccentricity in the neutron star-black hole binary}",
    eprint = "2508.12460",
    archivePrefix = "arXiv",
    journal = {arXiv e-prints},
    primaryClass = "gr-qc",
    month = "8",
    year = "2025"
}

@article{Kacanja:2025kpr,
    author = "Kacanja, Keisi and Soni, Kanchan and Nitz, Alexander Harvey",
    title = "{Eccentricity signatures in LIGO-Virgo-KAGRA{\textquoteright}s binary neutron star and neutron-star black holes}",
    eprint = "2508.00179",
    archivePrefix = "arXiv",
    primaryClass = "gr-qc",
    doi = "10.1103/jnsc-783p",
    journal = "Phys. Rev. D",
    volume = "112",
    number = "12",
    pages = "122007",
    year = "2025"
}

@article{JimenezForteza:2018rwr,
    author = "Jim{\'e}nez Forteza, Xisco and Abdelsalhin, Tiziano and Pani, Paolo and Gualtieri, Leonardo",
    title = "{Impact of high-order tidal terms on binary neutron-star waveforms}",
    eprint = "1807.08016",
    archivePrefix = "arXiv",
    primaryClass = "gr-qc",
    doi = "10.1103/PhysRevD.98.124014",
    journal = "Phys. Rev. D",
    volume = "98",
    number = "12",
    pages = "124014",
    year = "2018"
}

@article{Yagi:2013sva,
    author = "Yagi, Kent",
    title = "{Multipole Love Relations}",
    eprint = "1311.0872",
    archivePrefix = "arXiv",
    primaryClass = "gr-qc",
    doi = "10.1103/PhysRevD.89.043011",
    journal = "Phys. Rev. D",
    volume = "89",
    number = "4",
    pages = "043011",
    year = "2014",
    note = "[Erratum: Phys.Rev.D 96, 129904 (2017), Erratum: Phys.Rev.D 97, 129901 (2018)]"
}

@article{Yagi:2013awa,
    author = "Yagi, Kent and Yunes, Nicolas",
    title = "{I-Love-Q Relations in Neutron Stars and their Applications to Astrophysics, Gravitational Waves and Fundamental Physics}",
    eprint = "1303.1528",
    archivePrefix = "arXiv",
    primaryClass = "gr-qc",
    doi = "10.1103/PhysRevD.88.023009",
    journal = "Phys. Rev. D",
    volume = "88",
    number = "2",
    pages = "023009",
    year = "2013"
}

@article{Yagi:2016bkt,
    author = "Yagi, Kent and Yunes, Nicol{\'a}s",
    title = "{Approximate Universal Relations for Neutron Stars and Quark Stars}",
    eprint = "1608.02582",
    archivePrefix = "arXiv",
    primaryClass = "gr-qc",
    doi = "10.1016/j.physrep.2017.03.002",
    journal = "Phys. Rept.",
    volume = "681",
    pages = "1--72",
    year = "2017"
}

@article{Woodford:2019tlo,
    author = "Woodford, Charles J. and Boyle, Michael and Pfeiffer, Harald P.",
    title = "{Compact Binary Waveform Center-of-Mass Corrections}",
    eprint = "1904.04842",
    archivePrefix = "arXiv",
    primaryClass = "gr-qc",
    doi = "10.1103/PhysRevD.100.124010",
    journal = "Phys. Rev. D",
    volume = "100",
    number = "12",
    pages = "124010",
    year = "2019"
}

@article{Schmidt:2010it,
    author = "Schmidt, Patricia and Hannam, Mark and Husa, Sascha and Ajith, P.",
    title = "{Tracking the precession of compact binaries from their gravitational-wave signal}",
    eprint = "1012.2879",
    archivePrefix = "arXiv",
    primaryClass = "gr-qc",
    doi = "10.1103/PhysRevD.84.024046",
    journal = "Phys. Rev. D",
    volume = "84",
    pages = "024046",
    year = "2011"
}

@software{pycbc,
  author       = {Alex Nitz and
                  Ian Harry and
                  Duncan Brown and
                  Christopher M. Biwer and
                  Josh Willis and
                  Tito Dal Canton and
                  Collin Capano and
                  Thomas Dent and
                  Larne Pekowsky and
                  Gareth S Cabourn Davies and
                  Soumi De and
                  Miriam Cabero and
                  Shichao Wu and
                  Andrew R. Williamson and
                  Bernd Machenschalk and
                  Duncan Macleod and
                  Francesco Pannarale and
                  Prayush Kumar and
                  Steven Reyes and
                  dfinstad and
                  Sumit Kumar and
                  Márton Tápai and
                  Leo Singer and
                  Praveen Kumar and
                  veronica-villa and
                  maxtrevor and
                  Bhooshan Uday Varsha Gadre and
                  Sebastian Khan and
                  Stephen Fairhurst and
                  Arthur Tolley},
  title        = {gwastro/pycbc: v2.3.3 release of PyCBC},
  month        = jan,
  year         = 2024,
  publisher    = {Zenodo},
  version      = {v2.3.3},
  doi          = {10.5281/zenodo.10473621},
  url          = {https://doi.org/10.5281/zenodo.10473621},
}

@article{Markin:2026eyc,
    author = "Markin, Ivan and Bulla, Mattia and Dietrich, Tim",
    title = "{Numerical simulations of black hole-neutron star mergers with equal and near-equal mass ratios}",
    eprint = "2601.19405",
    archivePrefix = "arXiv",
    primaryClass = "astro-ph.HE",
    month = "1",
    year = "2026",
    journal = "",
}

@article{Khan:2015jqa,
    author = {Khan, Sebastian and Husa, Sascha and Hannam, Mark and Ohme, Frank and P{\"u}rrer, Michael and Jim{\'e}nez Forteza, Xisco and Boh{\'e}, Alejandro},
    title = "{Frequency-domain gravitational waves from nonprecessing black-hole binaries. II. A phenomenological model for the advanced detector era}",
    eprint = "1508.07253",
    archivePrefix = "arXiv",
    primaryClass = "gr-qc",
    doi = "10.1103/PhysRevD.93.044007",
    journal = "Phys. Rev. D",
    volume = "93",
    number = "4",
    pages = "044007",
    year = "2016"
}

@article{Garcia-Quiros:2020qlt,
    author = "Garc{\'\i}a-Quir{\'o}s, Cecilio and Husa, Sascha and Mateu-Lucena, Maite and Borchers, Angela",
    title = "{Accelerating the evaluation of inspiral{\textendash}merger{\textendash}ringdown waveforms with adapted grids}",
    eprint = "2001.10897",
    archivePrefix = "arXiv",
    primaryClass = "gr-qc",
    doi = "10.1088/1361-6382/abc36e",
    journal = "Class. Quant. Grav.",
    volume = "38",
    number = "1",
    pages = "015006",
    year = "2021"
}

@article{Bohe:2016gbl,
    author = "Boh{\'e}, Alejandro and others",
    title = "{Improved effective-one-body model of spinning, nonprecessing binary black holes for the era of gravitational-wave astrophysics with advanced detectors}",
    eprint = "1611.03703",
    archivePrefix = "arXiv",
    primaryClass = "gr-qc",
    reportNumber = "LIGO-P1600315",
    doi = "10.1103/PhysRevD.95.044028",
    journal = "Phys. Rev. D",
    volume = "95",
    number = "4",
    pages = "044028",
    year = "2017"
}

@article{Wade:2014vqa,
    author = "Wade, Leslie and Creighton, Jolien D. E. and Ochsner, Evan and Lackey, Benjamin D. and Farr, Benjamin F. and Littenberg, Tyson B. and Raymond, Vivien",
    title = "{Systematic and statistical errors in a bayesian approach to the estimation of the neutron-star equation of state using advanced gravitational wave detectors}",
    eprint = "1402.5156",
    archivePrefix = "arXiv",
    primaryClass = "gr-qc",
    doi = "10.1103/PhysRevD.89.103012",
    journal = "Phys. Rev. D",
    volume = "89",
    number = "10",
    pages = "103012",
    year = "2014"
}

@article{LIGOScientific:2014pky,
    author = "Aasi, J. and others",
    collaboration = "LIGO Scientific",
    title = "{Advanced LIGO}",
    eprint = "1411.4547",
    archivePrefix = "arXiv",
    primaryClass = "gr-qc",
    doi = "10.1088/0264-9381/32/7/074001",
    journal = "Class. Quant. Grav.",
    volume = "32",
    pages = "074001",
    year = "2015"
}

@article{VIRGO:2014yos,
    author = "Acernese, F. and others",
    collaboration = "VIRGO",
    title = "{Advanced Virgo: a second-generation interferometric gravitational wave detector}",
    eprint = "1408.3978",
    archivePrefix = "arXiv",
    primaryClass = "gr-qc",
    doi = "10.1088/0264-9381/32/2/024001",
    journal = "Class. Quant. Grav.",
    volume = "32",
    number = "2",
    pages = "024001",
    year = "2015"
}

@article{KAGRA:2018plz,
    author = "Akutsu, T. and others",
    collaboration = "KAGRA",
    title = "{KAGRA: 2.5 Generation Interferometric Gravitational Wave Detector}",
    eprint = "1811.08079",
    archivePrefix = "arXiv",
    primaryClass = "gr-qc",
    reportNumber = "JGW-P1809243",
    doi = "10.1038/s41550-018-0658-y",
    journal = "Nature Astron.",
    volume = "3",
    number = "1",
    pages = "35--40",
    year = "2019"
}

@article{Santamaria:2010yb,
    author = "Santamaria, L. and others",
    title = "{Matching post-Newtonian and numerical relativity waveforms: systematic errors and a new phenomenological model for non-precessing black hole binaries}",
    eprint = "1005.3306",
    archivePrefix = "arXiv",
    primaryClass = "gr-qc",
    reportNumber = "LIGO-P1000048, AEI-2010-122",
    doi = "10.1103/PhysRevD.82.064016",
    journal = "Phys. Rev. D",
    volume = "82",
    pages = "064016",
    year = "2010"
}

\end{document}